\let\mypdfximage\pdfximage
\def\pdfximage{\immediate\mypdfximage}

\documentclass[letterpaper, 11pt]{article}

\usepackage[ruled]{algorithm2e}
\usepackage{amsbsy}
\usepackage{amsfonts}
\usepackage{amsmath}
\usepackage{amssymb}
\usepackage{amsthm}
\usepackage[page]{appendix}
\usepackage{bm}
\usepackage{color}
\usepackage{epsfig}
\usepackage{geometry}
\usepackage{graphicx}
\usepackage[colorlinks=true, citecolor=blue, filecolor=black, linkcolor=red, urlcolor=blue]{hyperref}
\usepackage{soul}
\usepackage{subfigure}
\usepackage{xspace}

\usepackage[capitalise,nameinlink,noabbrev]{cleveref} 

\usepackage{pgfplots}
\usepackage{tikz}

\usetikzlibrary{calc, shapes, arrows, positioning, angles, quotes,
  patterns, fit}

\pgfplotsset{compat=newest} 
\pgfplotsset{plot coordinates/math parser=false} 
\newlength\figureheight 
\newlength\figurewidth

\newcommand{\iid}{\textit{i.i.d.}\xspace}

\newcommand{\myinvisible}[1]{ }

\newcommand\blr{\begin{scriptsize}\color{blue}}
\newcommand\elr{\normalcolor\end{scriptsize}}

\newcommand{\bi}{\begin{itemize}}
\newcommand{\ei}{\end{itemize}}
\newcommand{\be}{\begin{equation}}
\newcommand{\ee}{\end{equation}}
\newcommand{\bea}{\begin{eqnarray}}
\newcommand{\eea}{\end{eqnarray}}
\newcommand{\bean}{\begin{eqnarray*}}
\newcommand{\eean}{\end{eqnarray*}}
\newcommand{\ben}{\begin{equation*}}        \newcommand{\een}{\end{equation*}}

\newenvironment{greenItemize} { \begin{list}
{\color{green}$\bullet$\color{black}}
{ \setlength{\rightmargin}{\leftmargin} \setlength{\itemsep}{0in} } }{ \end{list} }

\newenvironment{redItemize} { \begin{list}
{\color{red}$\bullet$\color{black}}
{ \setlength{\rightmargin}{\leftmargin} \setlength{\itemsep}{0in} } }{ \end{list} }

\newenvironment{blueItemize} { \begin{list}
{\color{blue}$\bullet$\color{black}}
{ \setlength{\rightmargin}{\leftmargin} \setlength{\itemsep}{0in} } }{ \end{list} }

\newenvironment{dashItemize} { \begin{list}
{\color{black}--\color{black}}
{ \setlength{\rightmargin}{\leftmargin} \setlength{\itemsep}{0in} } }{ \end{list} }

\newcommand{\bgi}{\begin{greenItemize}}
\newcommand{\egi}{\end{greenItemize}}
\newcommand{\bri}{\begin{redItemize}}
\newcommand{\eri}{\end{redItemize}}
\newcommand{\bbi}{\begin{blueItemize}}
\newcommand{\ebi}{\end{blueItemize}}
\newcommand{\bdi}{\begin{dashItemize}}
\newcommand{\edi}{\end{dashItemize}}

\newcounter{Enumi}

\definecolor{babyblueeyes}{rgb}{0.63, 0.79, 0.95}

\def\1{$_\mathrm{1}$}
\def\2{$_\mathrm{2}$}
\def\3{$_\mathrm{3}$}
\def\4{$_\mathrm{4}$}
\def\5{$_\mathrm{5}$}
\def\6{$_\mathrm{6}$}
\def\7{$_\mathrm{7}$}
\def\8{$_\mathrm{8}$}

\def\leftpar{\raggedright\leftskip=0.0cm\hsize=8.0cm}
\def\rightpar{\raggedright\leftskip=12cm}

\def\cap#1#2
{\tenpoint\vskip-0.5cm\parindent=0pt
{\hskip-0.0cm\leftpar {#1} \par}
\vskip -\prevgraf \baselineskip
{\hskip-3.0cm\rightpar {#2} \par}
}

\newcommand\br{\par\vspace*{-2mm}\begin{scriptsize}\color{blue}\begin{flushright}}
\newcommand\er{\end{flushright}\normalcolor\end{scriptsize}}
\newcommand{\bp}{ \begin{flushright}\color{blue}\vspace*{-4mm}\begin{large} }
\newcommand{\ep}{ \end{large}\color{black}\end{flushright} }

\newcommand{\gae}{\lower 8pt \hbox{$\, \buildrel {\scriptstyle >}\over {\scriptstyle \sim}\,$}}
\newcommand{\lae}{\lower 8pt \hbox{$\, \buildrel {\scriptstyle <}\over {\scriptstyle \sim}\,$}}

\def\<{\left< }
\def\>{ \right>}

\def\co2{{\rm CO}_2}

\usetikzlibrary{positioning,arrows,shapes,shadows}
\tikzstyle{format} = [draw, very thick, fill=blue!40]
\tikzstyle{medium} = [ellipse, draw, thin, fill=green!20]
\tikzstyle{qoi}=[rectangle, draw=black, rounded corners, fill=green, drop shadow]
\tikzstyle{cblue}=[circle, draw, thin, fill=blue!10, scale=0.8]
\tikzstyle{decision} = [diamond, draw, fill=blue!20,
    text width=5.5em, text badly centered, node distance=3cm, inner sep=0pt]
\tikzstyle{biggray}=[rectangle, draw=black, thin, fill=gray!20,
                                 anchor=north west, text width=10em,
                                 text height=5em, minimum width=12.5em,
                                 minimum height=5.5em, rounded
                                 corners=5pt]

\def\eid{{\epsilon_i^{\textrm{d}}}}

\DeclareMathOperator*{\argmin}{arg\,min}
\DeclareMathOperator*{\argmax}{arg\,max}

\def\LL{{\cal L}}

\def\VV{\mathbb{V}}
\def\EE{\mathbb{E}}
\def\RR{\mathbb{R}}

\newcommand{\CC}{\mathbb{C}}

\newcommand{\tf}{\tilde{f}}

\def\fmp{f_{\textrm{MPF}}}

\def\vy{\Vector y}

\def\Vector#1{\mbox{\boldmath $#1$}}

\def\vH{{\Vector H}}

\def\vz{{\Vector z}}
\def\vy{{\Vector y}}

\def\vc{{\Vector c}}
\def\vC{{\Vector C}}

\def\vp{{\Vector p}}

\def\vlam{{\bm \lambda}}
\def\vLam{{\bm \Lambda}}

\def\vh{{\Vector h}}

\def\vtheta{{\Vector \theta}}
\def\vxi{{\Vector \xi}}

\def\vtheta{{\Vector \theta}}
\def\vsigma{{\Vector \sigma}}

\def\vxi{{\Vector \xi}}
\def\vmu{{\Vector \mu}}

\def\vmu{{\Vector \mu}}
\def\vc{{\Vector c}}
\def\vc{{\Vector c}}
\def\vlam{{\Vector \lambda}}
\def\vtheta{{\Vector \theta}}

\def\vC{{\Vector C}}

\def\vW{{\Vector W}}

\def\vz{{\Vector z}}
\def\vmu{{\Vector \mu}}

\def\vsigma{{\Vector \sigma}}
\def\valpha{{\Vector \alpha}}

\def\vy{{\Vector y}}

\newcommand{\vf}{{\bm f}}

\newcommand{\ed}{{\epsilon}}
\newcommand{\ved}{{\bm \ed}}
\newcommand{\vdel}{{\bm \delta}}
\newcommand{\tvalpha}{{\tilde{\valpha}}}
\newcommand{\SSS}{{\cal S}}

\providecommand{\ie}{i.e.,\;}
\providecommand{\eg}{e.g.,\;}

\newcommand{\reb}[1]{\textcolor{black}{#1}}

\usepackage{lineno}
\usepackage{color}
\usepackage{bm}
\usepackage{xspace}
\usepackage{amsmath}
\usepackage{amssymb}
\usepackage{amsbsy}
\usepackage{soul}

\usepackage{tikz}

\usepackage[ruled]{algorithm2e}
\providecommand{\SetAlgoLined}{\SetLine} \SetKw{KwReq}{Key Algorithmic Decisions Required:}

\title{Embedded Model Error Representation\\for Bayesian Model
  Calibration}

\author{Khachik Sargsyan\footnote{Corresponding author:
    \href{mailto:ksargsy@sandia.gov}{ksargsy@sandia.gov}}~\footnote{Sandia
    National Laboratories, Livermore, CA 94550, USA.}, Xun
  Huan\footnote{Department of Mechanical Engineering, University of
    Michigan, Ann Arbor, MI 48109, USA.}~\footnotemark[2], and Habib N. Najm\footnotemark[2]}
 
\begin{document}

\maketitle

\begin{abstract}
Model error estimation remains one of the key challenges in
uncertainty quantification and predictive science. For
computational models of complex physical systems,
model error, also known
as structural error or model inadequacy, is often the largest
contributor to the overall predictive uncertainty.
This work builds on a recently developed framework of embedded,
internal model correction, in order to represent and quantify
structural errors, together with model parameters, within a
Bayesian inference context.  We focus specifically on a Polynomial Chaos
representation with additive modification of existing model parameters,
enabling a non-intrusive procedure for efficient approximate
likelihood construction, model error estimation, and disambiguation of
model and data errors' contributions to predictive uncertainty.  The
framework is demonstrated on several synthetic examples, as well as on
a chemical ignition problem.
\end{abstract}


\section{Introduction} \label{sec:intro}

Both computational capabilities
and observational data availability have seen rapid advances in the past decade.  While
these improvements catalyzed and enabled the development of new,
sophisticated methods for analyzing and assimilating data with complex
computational models, considerable challenges remain. One such challenge involves
the quantification of uncertainty in model predictions, and its accurate
attribution to different uncertainty sources, all of which is important for
targeted uncertainty reduction and improved predictability.
While methods for quantifying data noise/error and associated parametric uncertainty
have grown relatively mature, the quantification of \emph{model
  error} (also known as structural error and model inadequacy) and associated
  predictive uncertainty is
less developed. All models involve assumptions, and none is perfect.
Models of complex physical systems are also often in error due to lack of
proper understanding of underlying phenomena.
Accurate attribution and quantification of model
error in the process of model calibration is crucial for ensuring reliable
predictions with meaningful estimates of predictive uncertainty,
but yet this is rarely done.
This critical gap is widely acknowledged in the
literature~\cite{Campbell:2006,OHagan:2013,Bojke:2009,Curry:2011,Gupta:2012}.

Conventional model
calibration seeks best-fit estimates of model parameters
according to
a penalty on the mismatch
between model predictions and observational data.  Statistical
inference techniques for model calibration and parameter estimation
typically ignore errors associated with the model itself---i.e., the
model is assumed to be correct. However, every model has assumptions and
therefore model error. If unaccounted for, model error can introduce
bias in calibration and handicap predictive utility of the
model~\cite{Kennedy:2001, Strong:2014, Smith:2013}. Conventional
statistical approaches for capturing model error typically estimate
error correlations from observational data employing \emph{ad hoc} covariance
structures on quantities of interest (QoIs)~\cite{Kennedy:2001,
  Higdon:2008, Brynjarsdottir:2013}. This subsequently burdens the
model with additive statistical mismatch terms that are QoI-specific,
and lead to predictions that can violate the underlying physical
constraints imposed through the
models~\cite{Strong:2014,Sargsyan:2015}.

A class of methods for model error quantification,
inspired by the seminal work of Kennedy and
O'Hagan~\cite{Kennedy:2001}, augments model output quantities of
interest (QoIs) with \emph{external} statistical terms for bias
correction. For example, Wang \emph{et. al.}~\cite{Wang:2009}
developed a two-step
variation of the external discrepancy approach with sequential inference of model bias and the true output, while Joseph and Melkote~\cite{Joseph:2009}
proposed a compromise between engineering and statistical approaches
for explicit correction by adjusting the model at hand incrementally and sequentially using empirical Bayes methods retaining both the fitting power of statistical methods and the predictive power of engineering models. Higdon \textit{et al.}~\cite{Higdon:2004}
further developed the model correction framework and demonstrated it
for physical systems. However, it
is well-recognized that corrections at the QoI level produce
predictions that could potentially violate physical laws in the model,
unless ameliorated with special prior
constructions~\cite{Brynjarsdottir:2013}. More
crucially, external, QoI-specific corrections do not provide any model error augmentation
for extrapolative scenarios or for prediction of
other QoIs~\cite{Bayarri:2007a,Oliver:2015,Pernot:2017a}. Indeed, the resulting calibration information cannot be
easily carried to making predictions of other model outputs. Besides,
the model errors evaluated as explicit bias corrections are often entangled with data/measurement errors.

As a response to these limitations, \emph{internal} model correction
approaches have been gaining popularity, where select internal model
components are augmented with statistical bias correction terms. He
and Xiu~\cite{He:2016} described a general framework for both internal embedding and external correction under a constrained optimization
setting without quantifying the full Bayesian uncertainty. Strong \textit{et
  al.}~\cite{Strong:2012,Strong:2014} applied model internal
correction in health-economic studies, and discussed information
measures for the model improvement. Internal model corrections were
also demonstrated in Reynolds-averaged Navier Stokes (RANS)
applications~\cite{Oliver:2011, Emory:2011}, large eddy simulation
(LES) computations~\cite{Huan:2017}, molecular
simulations\reb{~\cite{Pernot:2017}, particle-laden flows~\cite{Zio:2018}}, and chemical ignition
modeling~\cite{Hakim:2017,Sargsyan:2015}.

In our framework, developed in \cite{Sargsyan:2015}, we embed model error in
model components such as constitutive laws or phenomenological parameterizations
rather than as additive corrections to model outputs. Specifically, we focus on
an approach where corrections are added to select model input parameters,
therefore allowing the model to be treated as a black-box.  The framework is
developed in a general Bayesian context, where existing model parameters are
inferred simultaneously along with parameters that characterize model error.
Additionally, we employ Polynomial Chaos
(PC)~\cite{Ghanem:1991,Xiu:2002c,Najm:2009a,OlmOmk:2010} to represent the
augmented inputs, permitting effortless extraction and propagation of
uncertainties, eventually leading to efficient Bayesian computation. The
approach leads to a model error representation that is consistent with the
underlying model physics, and can be employed as a diagnostic device to enable
attribution of model errors to specific submodels, thus enabling targeted model
improvement and optimal experimental design.  The strength of the developed
framework is highlighted in model-to-model calibration studies where the
classical independent Gaussian discrepancy models are not defensible.

In this present work, we extend and enhance the probabilistic framework of
\cite{Sargsyan:2015} to enable calibration with respect to noisy observational
data, and to provide a principled way of attributing predictive uncertainties to
components due to data noise/error, model error, as well as any additional
errors associated with the lack-of-information. This generalization allows the
utilization of the construction with both computational and/or experimental
data, while retaining the non-intrusive PC construction and the facile
implementation of the method with black-box models. We detail a range of
options for construction of likelihoods for model error estimation, and
\reb{highlight some of them} in different scenarios. Overall, this work provides solid
foundations for the earlier developments, and extends them to more general
practical contexts.

The paper is organized as follows. Section~\ref{sec:merr} provides a
general overview of model error quantification and its associated
challenges. The embedded treatment is introduced in
Section~\ref{sec:de}, along with the Bayesian machinery, PC
representation, and predictive uncertainty attribution. The
methodology is then demonstrated on synthetic problems in
Section~\ref{sec:numer}, and a chemical ignition application in
Section~\ref{sec:ign}. The paper ends with discussions and future work
suggestions in Section~\ref{sec:disc}.

\section{The model error challenge} \label{sec:merr}

Consider a ``truth'' model $g(x_i)$ that generates datum $y_i$ at
operating condition $x_i$ (\eg spatial and temporal coordinates)
through the relationship $y_i=g(x_i)+\eid$ where $\ed_i$ is the
measurement noise, for a total of $i=1,\ldots,N$ data
measurements. While the data can be observed, the truth model is
unknown to us.  Instead, we have access to a different model
$f(x_i;\vlam)$, where $\vlam=(\lambda_1,\dots,\lambda_d)$ is a
$d$-dimensional parameter vector that can be estimated from experimental
data.
The model discrepancy, here defined to be the
difference between the truth model and our model of interest, is
$\delta(x):=g(x)-f(x;\vlam)$.
The fit model $f(x;\vlam)$ thus relates to the
observations $y_i$ via \be \label{eq:koh}
y_i=\underbrace{f(x_i;\vlam)+\delta(x_i)}_{\textrm{truth model }
  g(x_i)}+\eid. \ee
Conventional parameter estimation (\ie calibrating for $\vlam$)
often ignores model errors, effectively assuming that $f(x;\vlam)$ is able to
exactly replicate $g(x)$ for some $\vlam$.
However, this assumption typically
does not hold for complex physical systems, where simplified,
under-resolved, and poorly conceived models are often
unavoidable. Consequently, the estimated parameter values would be
biased, due to
neglected model deficiencies. Bayarri \emph{et al.}~\cite{Bayarri:2007} present
a clear illustration of this challenge. Moreover, ignoring model error, and
thereby treating discrepancies between model predictions and available
measurements as exclusively the result of uncorrelated data errors/noise
results in a calibrated model whose predictive uncertainty can exhibit
excessive overconfidence. This issue becomes quite evident as the number
of data points $N$ increases, as we illustrate through the following
example.

\begin{figure}[!tb]
\centering{
\subfigure[N=5]{ \includegraphics[width=0.31\textwidth]{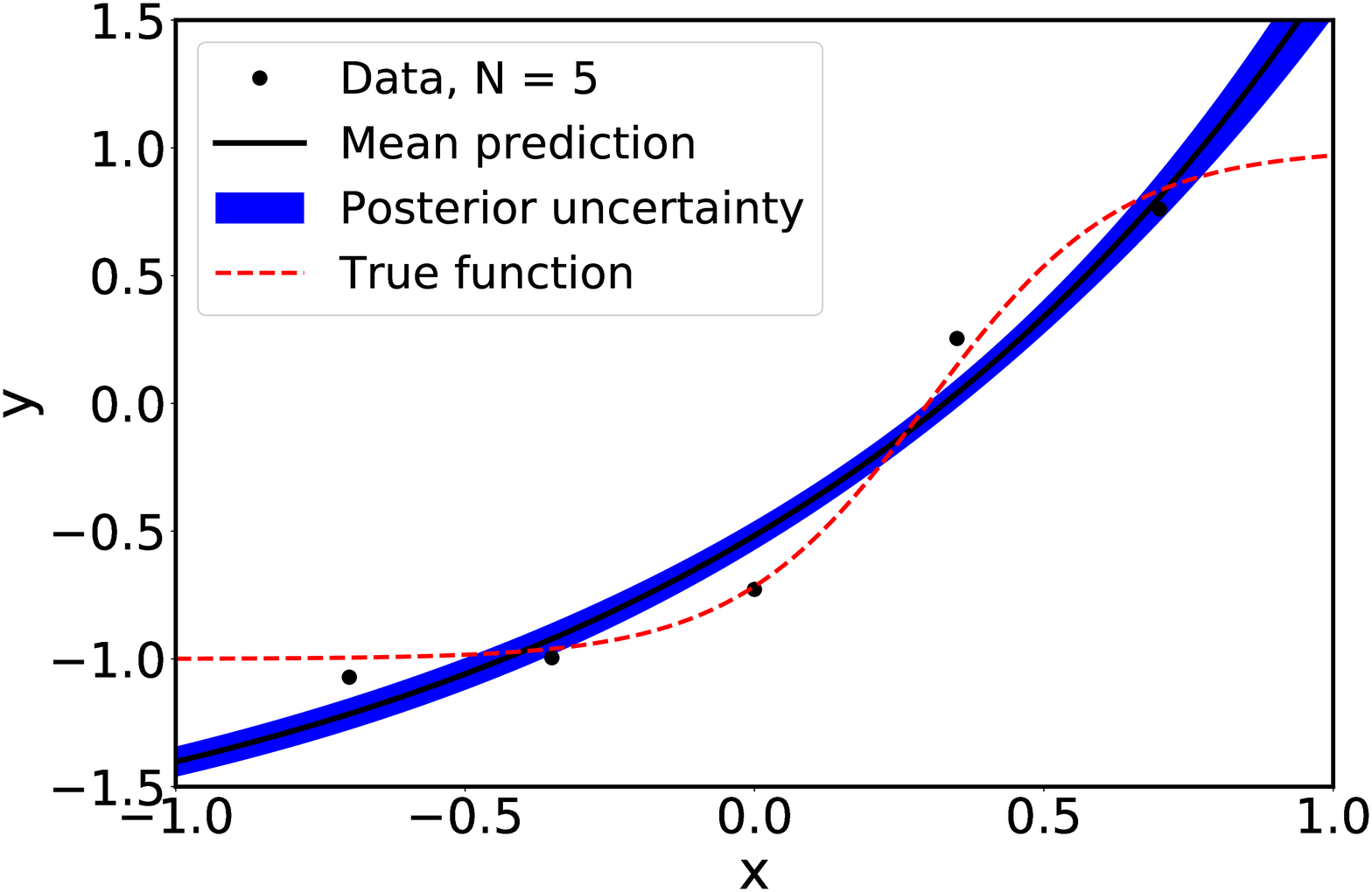}}
\subfigure[N=20]{ \includegraphics[width=0.31\textwidth]{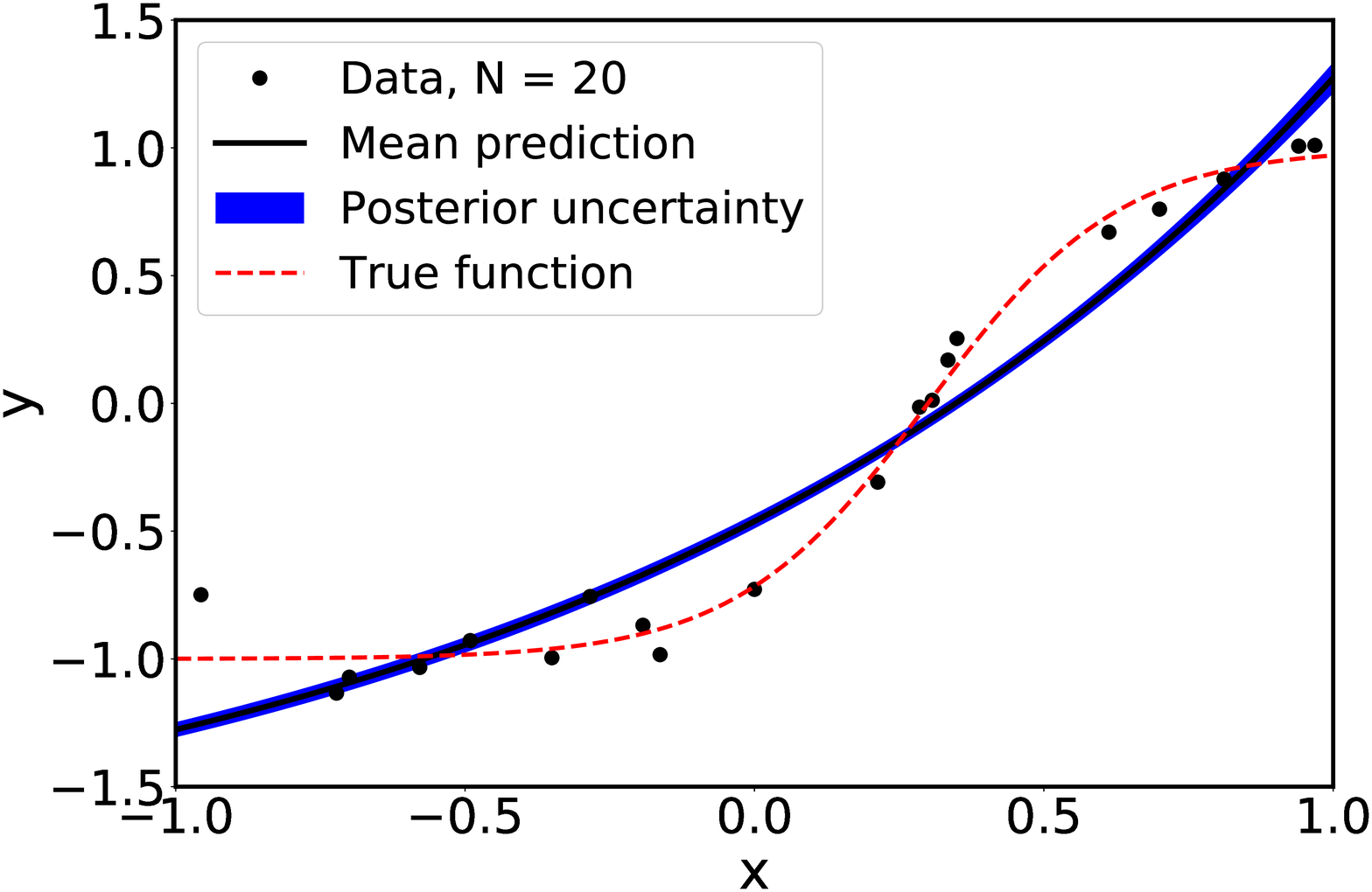}}
\subfigure[N=50]{ \includegraphics[width=0.32\textwidth]{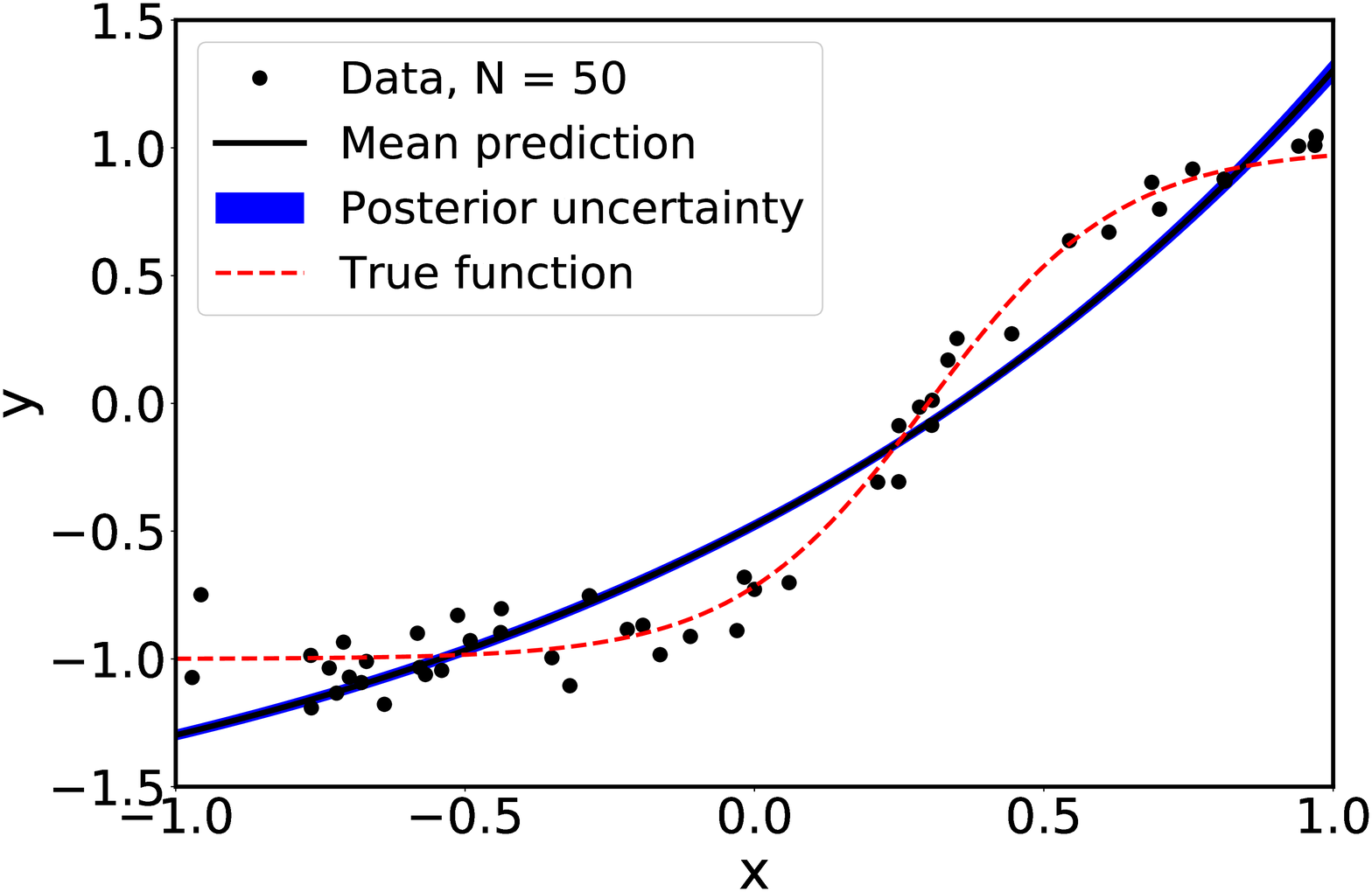}}\\
\subfigure{\includegraphics[width=0.32\textwidth]{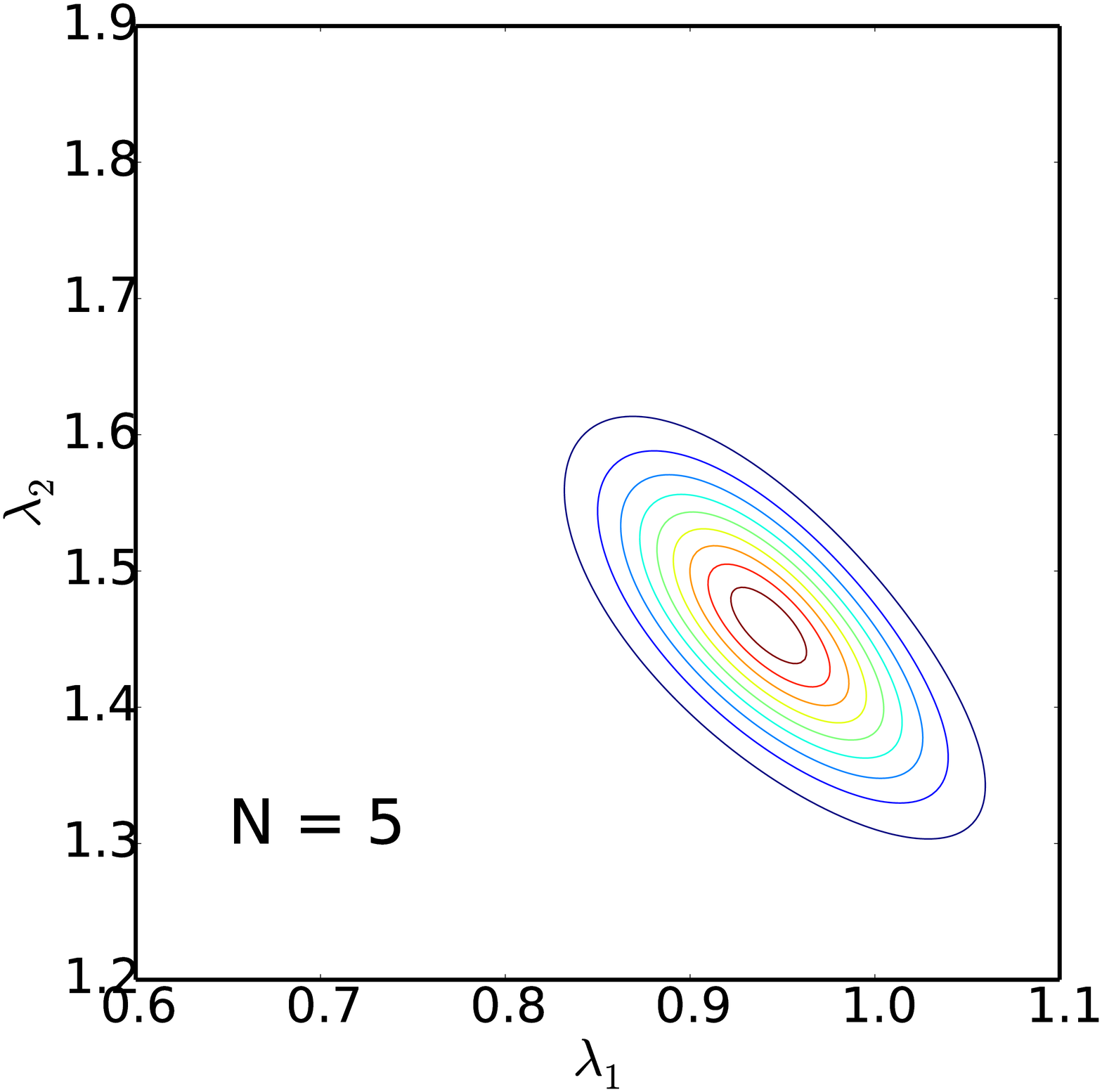}}
\subfigure{\includegraphics[width=0.32\textwidth]{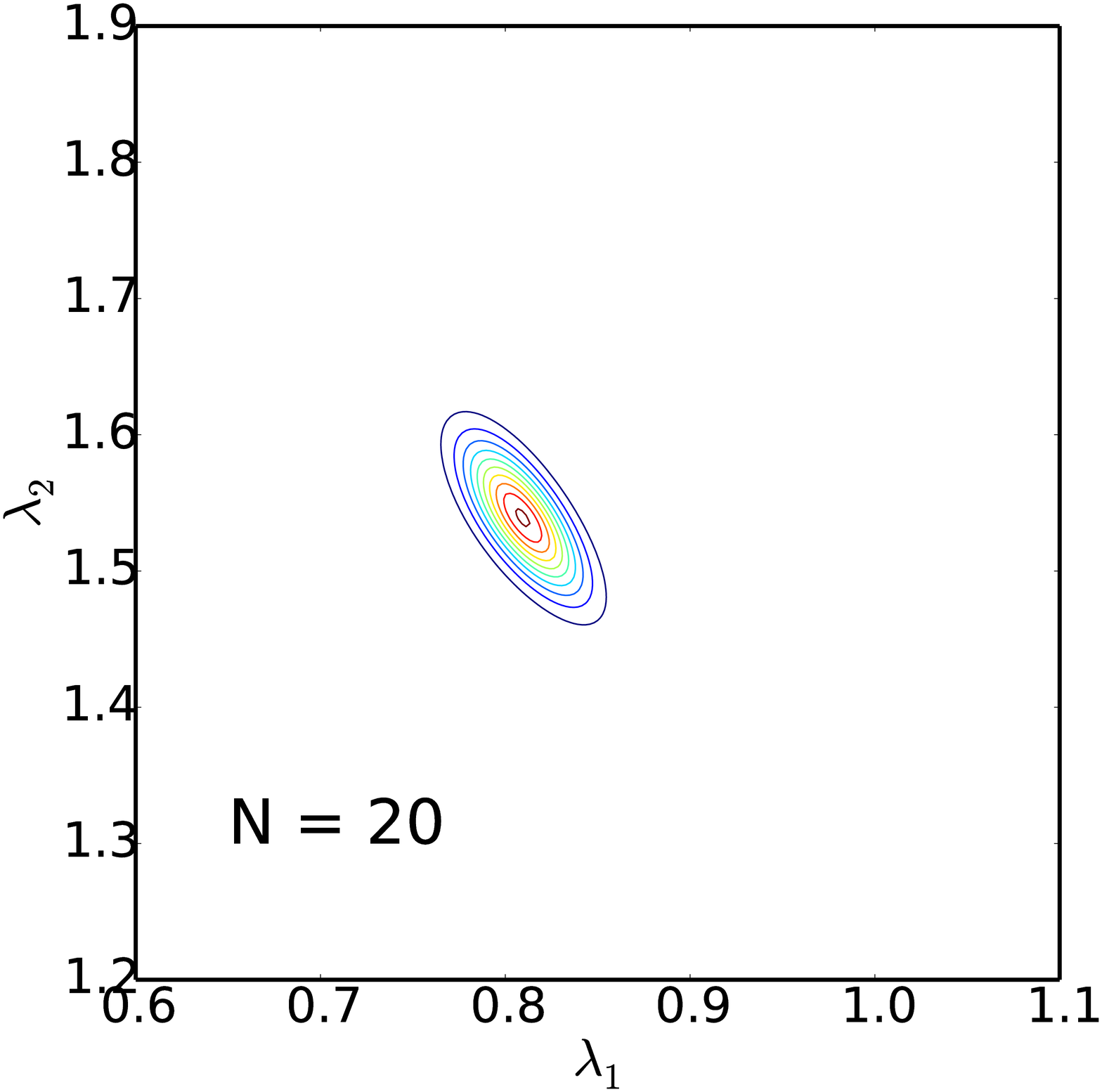}}
\subfigure{\includegraphics[width=0.32\textwidth]{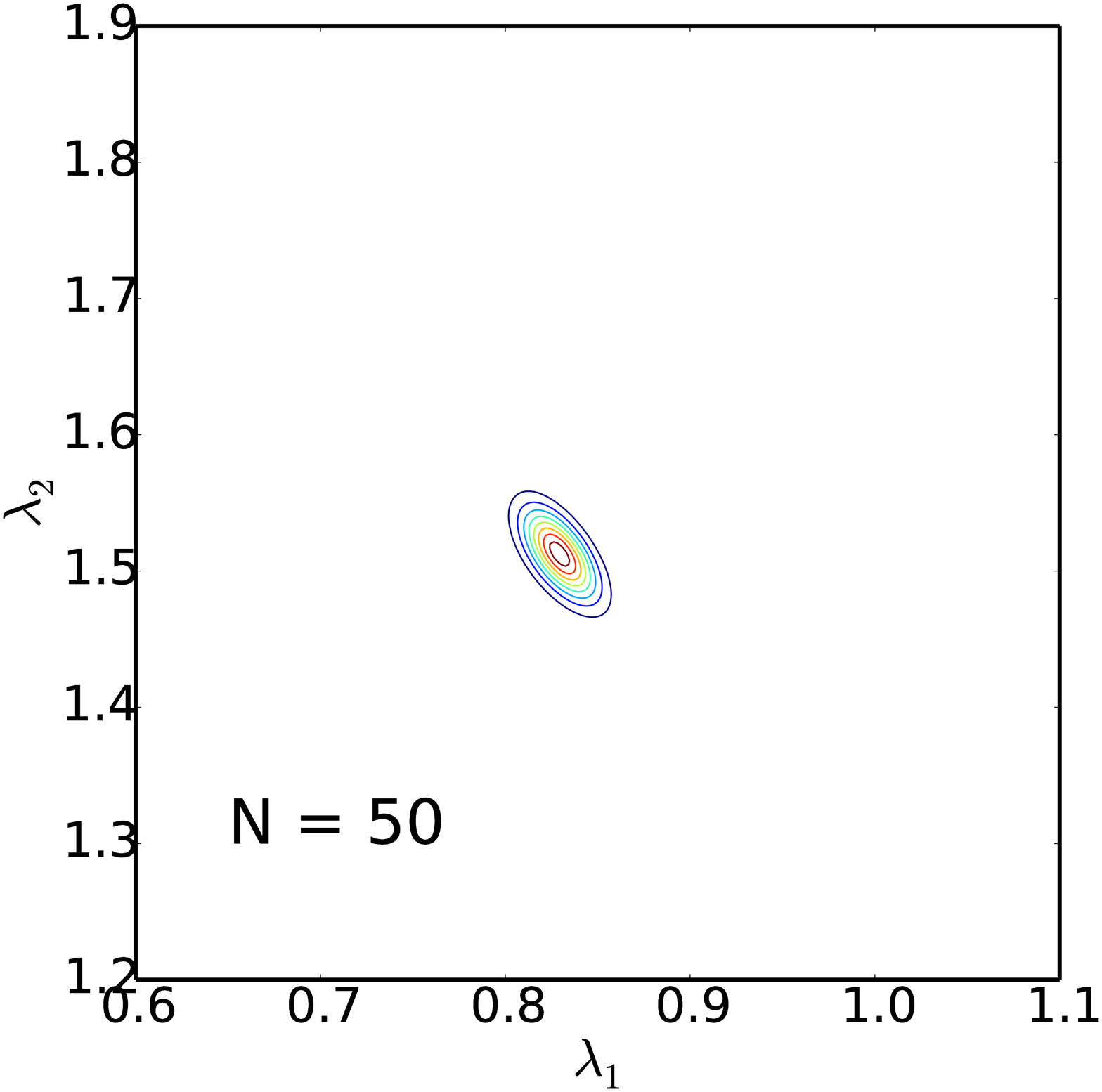}}}
\caption{\label{fig:ex1} Demonstration of a classical Bayesian calibration -- essentially a probabilistic least-squares fit --
with increasing amount of data. Top row
shows how pushed-forward (PF) \reb{standard deviation}, obtained by pushing posterior probability density functions (PDFs)
through the model, shrink and do not capture the true discrepancy between data and the model.
Correspondingly, the bottom row shows how calibrated parameter distributions shrink around potentially
wrong values of $\lambda_1$ and $\lambda_2$.
} \end{figure}

Consider an
exponential-growth model $f(x;\vlam)=\lambda_2 e^{\lambda_1 x}-2$,
while the truth model includes `saturation' and has the form
$g(x)=\tanh{3(x-0.3)}$.  Each synthetic datum is the truth model
corrupted with a measurement noise in the form of independent additive
Gaussian with zero mean and standard deviation $\sigma=0.1$. As seen
in Figure~\ref{fig:ex1}, as the number of data points increases,
Bayesian calibration
leads to shrinking uncertainty bands
in the model prediction that do not adequately capture the mismatch
between model prediction and data. In other words, the calibrated
results without model error become overconfident around values that
are, as illustrated elsewhere~\cite{Bayarri:2007}, potentially biased.
Model error needs to be taken into account in order to address these
challenges.

Moreover, and as already stated, the conventional Bayesian construction
for calibration with model error explicitly
represents the model error term $\delta(x)$ in~(\ref{eq:koh}), for
example, as a stochastic process (\emph{e.g.} a Gaussian process) with
a covariance structure that is optimized within a given
class~\cite{Kennedy:2001}.  However, this approach has potential
pitfalls when calibrating physical models for purposes of
prediction~\cite{Bayarri:2007a,Brynjarsdottir:2013,Oliver:2015,Pernot:2017a}. For
example, there is no mechanism for utilizing the QoI-specific additive
term $\delta(x)$ for predictions of other QoIs from the same model
$f(x;\vlam)$. A general statistical additive term without additional
treatment from physical knowledge can also violate requisite physical
constraints imposed by $g(x)$, and, presumably satisfied by $f(x;\vlam)$ by
construction. Lastly, the form~(\ref{eq:koh})
suffers identifiability difficulties since it is the sum of
$\eid$ and $\delta(x_i)$ that is relevant in the measurement magnitude
$y_i$.  While progress has been made to incorporate physics-backed
regularization for $\delta(x)$ through informative
priors~\cite{Brynjarsdottir:2013}, constructing them is often
\textit{ad hoc} and not feasible in general.

\section{Embedded model error representation} \label{sec:de}

We develop a calibration approach that accounts for model error via
probabilistic embedding \textit{within} the model, in elements of
$f(x;\vlam)$. For example, the embedding can be carried out in an
intrusive manner, via physics-backed phenomenological
parameterizations that are not yet part of the model, effectively
enhancing the model as $\tf(x;\vlam,\vtheta)$ to include new
parameters $\vtheta$. In this work, we demonstrate a simpler,
non-intrusive approach in which a subset of parameters in $\vlam$ are
selected and augmented by an additive discrepancy adjustment $\vdel$,
and aim to allow the embedded model $f(x;\vlam+\vdel)$ to produce
predictive uncertainties that are consistent with the
data\footnote{With some abuse of notation, we use vector notation
  $\vdel$ for the internal correction in contrast to the explicit
  model discrepancy $\delta(x)$.}.  This idea in its simplest form has
already been demonstrated in~\cite{Sargsyan:2015} for synthetic
examples and a simple chemical kinetics model, in which model
parameters $\vlam$ were cast as probabilistic quantities with a
requirement that their randomness, while respecting the physical
constraints \emph{by construction}, be consistent with
the data in some sense, \ie
$f(x_i;\vlam)\approx y_i$. \reb{Also, the authors in~\cite{Pernot:2017} investigated the embedded correction in some detail, considering various options in both synthetic studies and in a realistic model calibration setting.} The embedded model error representation is
generally parameterized by $\valpha$, and obeys a PDF $\pi_\vdel(\cdot;\valpha)$, or is explicitly written as
$\vdel(\valpha, \vxi)$ to highlight the stochastic dimension in
$\vxi$.  The task of estimating parameters $\vlam$ is thus
reformulated as an estimation of the augmented set of parameters
$\tvalpha=(\vlam,\valpha)$. Note that the resulting uncertainty
inflation, \reb{as introduced in~\cite{Pernot:2017a} and further demonstrated} in~\cite{Pernot:2017}, does not necessarily
improve the validity of the inadequate model, but rather allows for
meaningful calibrated predictions endowed with a degree of uncertainty
that captures model predictive inadequacy, while remaining consistent with the
physical constraints required by the physics.

The specific selection of model components
for embedding is a problem-dependent task. In principle, one can
select existing model components where simplifications, approximations, or
phenomenological modeling have been employed.
In~\cite{Huan:2017}, the authors employed
global sensitivity analysis (GSA) in order to isolate one or two most
impactful parameters for model error embedding. However, GSA only uses
model information and is not data-driven. A more rigorous and
comprehensive approach would involve Bayesian model selection based on
evidence computation to determine the best probabilistic
embedding. Note that it is also possible to envision more complicated
scenarios, \emph{e.g.} an $x$-dependent random process discrepancy
term $\vdel(x,\valpha,\vxi)$.

The overall data model with additive model error embedding can be rewritten as
\be \label{eq:datamodel}
y_i\approx H_i(\tvalpha)\equiv h_i(\vxi,\ved;\tvalpha) =f(x_i;\vlam+\vdel(\valpha,\vxi))+\epsilon_i \textrm{ for } i=1,\dots, N,
\ee
or, in vector notation,
 \be \label{eq:datamodelv}
 \vy\approx \vH(\tvalpha)\equiv\vh(\vxi,\ved;\tvalpha)=\vf(\vlam+\vdel(\valpha,\vxi))+\ved,
 \ee
where
$\vh(\vxi,\ved;\tvalpha)=(h_1(\vxi,\ved;\tvalpha),\dots,h_N(\vxi,\ved;\tvalpha))$
is a random vector, parameterized by $\tvalpha$, induced by $\vxi$ and
$\ved$, as a result of the model evaluations pushed-forward through
$f(x_i;\cdot)$, with the addition of data noise $\ved$. We define
corresponding upper case quantities $H_i(\tvalpha)$ and
$\vH(\tvalpha)$ to simplify the notation, with an understanding that
these are random quantities.  Without loss of generality, and for
ease of illustration, the components of the measurement error vector
$\ved$ will be modeled as independent, identically distributed normal
random variables with zero mean and fixed standard deviation $\sigma$.

The problem of calibrating model parameters $\vlam$ thus becomes
a \emph{density estimation} problem for the augmented stochastic input
$\vlam+\vdel(\valpha,\vxi)$, or a parameter estimation problem for
$\tvalpha=(\vlam,\valpha)$. \reb{Indeed, the estimation of a \emph{deterministic} $\vlam$ is replaced with the estimation of the PDF of \emph{stochastic}, augmented input $\vLam=\vlam+\vdel(\valpha,\vxi)$.} Uncertainty for epistemic parameters can
be reduced with increased data size, which translates to a
corresponding reduction of prediction uncertainty. While this is
sensible when the discrepancy between the model prediction
$f(x_i;\vlam)$ and the data $y_i$ is due to measurement noise only, it
is not desirable when this discrepancy includes model error (one would
not expect model error to disappear even with infinite data). The
uncertainty in the calibrated model prediction $f(x_i;\vlam)$ thus
needs to reflect this residual uncertainty due to model error, and the
embedded approach allows such an implementation via additional
stochastic dimensions encoded in $\vxi$; this will be described with
more details in Section~\ref{s:pred}.

\subsection{Bayesian inference of model error representation} \label{s:bayes}

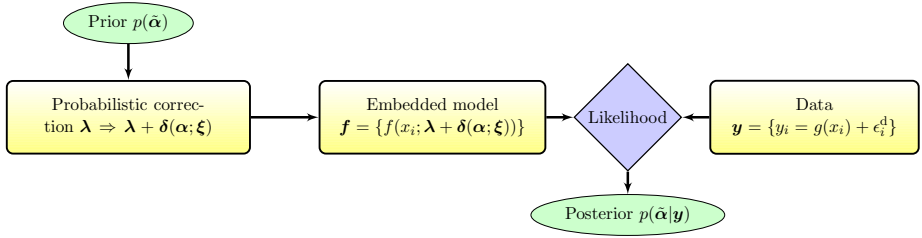
\begin{figure}[!tb]
\usetikzlibrary{positioning,arrows,shapes,shadows}
\tikzstyle{format} = [draw, very thick, fill=blue!40]
\tikzstyle{medium} = [ellipse, draw, thin, fill=green!20]
\tikzstyle{qoi}=[rectangle, draw=black, rounded corners, fill=green, drop shadow]
\tikzstyle{cblue}=[circle, draw, thin,fill=blue!10, scale=0.8]
\tikzstyle{decision} = [diamond, draw, fill=blue!20, 
    text width=5.5em, text badly centered, node distance=3cm, inner sep=0pt]
    \tikzstyle{biggray}=[rectangle, draw=black, thin, fill=gray!20,
                                 anchor=south east, text width=10em,
                                 text height=5em, minimum width=12.5em,
                                 minimum height=5.5em, rounded
                                 corners=20pt]
\centering
\scalebox{0.6}{
\begin{tikzpicture}[node distance=3cm, auto, thick]
\tikzset{
    mynode/.style={rectangle,rounded corners,draw=black, top color=white, bottom color=yellow!50,very thick, inner sep=1em, minimum size=3em, text width=10em, text centered},
    mynode2/.style={rectangle,rounded corners,draw=black, top color=white, bottom color=green!50,very thick, inner sep=1em, minimum size=3em, text centered},
    myarrow/.style={->, >=latex', shorten >=0pt, ultra thick},
    mylabel/.style={text width=7em, text centered} 
}  
    \path[use as bounding box] (0.5,0) rectangle (22,5);
    \path[->] node[biggray, opacity=0.0] at (4.0,1.5) (cmod) {};
     \path[->] node[mynode, text width=12em, right= -4.3 of cmod](alpha){Probabilistic correction $\vlam\Rightarrow \vlam+\vdel(\valpha;\vxi)$};
    \path[->] node[medium, above=0.8 of alpha] (prior) {Prior $p(\tvalpha)$}
                      (prior) edge [myarrow]  (alpha);
    \path[->] node[mynode, text width=11em, right=1.5 of alpha] (fi) {Embedded model \\$\vf=\{f(x_i;\vlam+\vdel(\valpha;\vxi))\}$}
                  (alpha) edge [myarrow] (fi);
    \path[->] node[decision, right=0.6 of fi] (xds) {Likelihood}
                  (fi) edge [myarrow]  (xds);
    \path[->] node[mynode, text width=10em, right=0.6 of xds] (gs) {Data \\$\vy=\{y_i=g(x_i)+\eid\}$}
                  (gs) edge [myarrow]  (xds);
   \path[->] node[medium, below=0.5 of xds] (ds) {Posterior $p(\tvalpha|\vy)$}
                  (xds) edge [myarrow] (ds);
    \end{tikzpicture}
}
\caption{Schematic of the Bayesian inference of $\tvalpha=(\vlam,\valpha)$, \ie estimation of model parameters ($\vlam)$ and stochastic model error correction parameters ($\valpha$).
\label{fig:sketch}}
\end{figure}

The reformulated parameter estimation problem for $\tvalpha=(\vlam,\valpha)$ can be thought of as a density estimation
problem for the embedded random input $\vLam(\tvalpha)=\vlam+\vdel(\valpha;\vxi)$, and is tackled as Bayesian parameter estimation of $\tvalpha$.
Bayesian methods are well-suited for dealing with uncertainties from
different sources, from intrinsic noise in the system to parametric
uncertainty and experimental errors. Although it is usually more
computationally involved compared to, for example, regularized
optimization techniques, the Bayesian framework provides a rigorous
platform for capturing the \emph{state-of-knowledge} about quantities
of interest before and after assimilating the data. Furthermore,
Bayesian techniques are very convenient for dealing with
\emph{nuisance} parameters, \ie parameters that are generally unknown
and are not of intrinsic interest, say, for prediction purposes.
The key relationship for Bayesian inference is Bayes formula~\cite{Bernardo:2000,Sivia:2006,Carlin:2011}, which
in this context reads as
\be \label{eq:bayes}
p(\tvalpha|\vy)=\frac{p(\vy|\tvalpha)p(\tvalpha)}{p(\vy)}.
\ee
The \emph{prior} probability density $p(\tvalpha)$ and the \emph{posterior} probability density $p(\tvalpha|\vy)$ represent degrees
of belief about model parameters $\tvalpha$ before and after the data $\vy$ is incorporated, respectively. The
\emph{evidence} $p(\vy)$ often  plays the role of a normalizing constant. However, it becomes a critical
measure when multiple models are compared against each other given the same
set of available data. The key factor in
Eq.~(\ref{eq:bayes}) is the \emph{likelihood}
$\LL_\vy(\tvalpha)=p(\vy|\tvalpha)$ that relates the data to the model
parameters. Figure~\ref{fig:sketch} demonstrates the schematic of the embedded
model error estimation framework within the Bayesian paradigm.

The posterior distribution is difficult to evaluate analytically for general likelihoods and priors, and is also
challenging to estimate numerically due to a usually high dimensional $\tvalpha$.
In such cases, Markov chain Monte Carlo (MCMC) methods are typically
used, in which one generates a Markov chain whose stationary
distribution is the posterior distribution~\cite{Gamerman:2006,
  Gilks:2005}. The posterior distribution $p(\tvalpha|\vy)$ provides
the current state-of-knowledge about the parameter of interest
$\tvalpha$, while the maximum a posteriori (MAP) value
\be
\tvalpha_{\textrm{MAP}}=\argmax_{\tvalpha} p(\tvalpha|\vy)
\ee
is also
of interest as the vector of most probable value $\tvalpha$ conditioned on measurements $\vy$. A common point estimate
for the density of the embedded random input
$\vLam(\tvalpha)=\vlam+\vdel(\valpha;\vxi)$ is characterized by
the MAP value of $\tvalpha$, which can either be extracted from the
full MCMC procedure, or computed via optimization techniques.
Whether optimization or MCMC is used, they all require multiple evaluations of the
likelihood and the prior in Eq.~(\ref{eq:bayes}). Construction of
appropriate likelihoods and, to a smaller extent, prior distributions are therefore crucial components
of the presented approach.

\color{black}

\subsubsection{Likelihood construction} \label{sec:lik}

The construction of a justifiable likelihood is perhaps the most
critical step for obtaining the posterior probability
distribution. From the data model~(\ref{eq:datamodel}) one can
write the \emph{true} likelihood as
\be \label{eq:lik_true}
\LL_\vy(\tvalpha)=\pi_{\vH(\tvalpha)}(\vy),
\ee
where $\pi_{\vH(\tvalpha)}(\vy)$ is the PDF of
the data prediction vector $\vH(\tvalpha)=\vh(\vxi,\ved;\tvalpha)$ defined in
Eq.~\eqref{eq:datamodelv}. Therefore,
the true likelihood evaluation~(\ref{eq:lik_true}) requires estimation
of an $N$-dimensional PDF as $\vy\in\RR^N$. This can be performed by sampling of both
$\vxi$ and $\ved$ for a given $\tvalpha$, followed by a Kernel Density
Estimation (KDE) to obtain a smooth estimate of
$\pi_{\vH(\tvalpha)}(\vy)$~\cite{Scott:1992,Silverman:1986}.
For example, with $R$ samples of $\vH(\tvalpha)$
the KDE-estimated likelihood is written as
\be \label{eq:kde}
\LL_\vy(\valpha)=\pi_{\vH(\tvalpha)}(\vy)\approx \frac{1}{R}\sum_{r=1}^R K(\vh^{(r)}-\vy).
\ee
Typically, Gaussian kernels $K(\vz)=(2\pi)^{-d/2}|\vW|^{-1/2}e^{-\vz^T\vW^{-1}\vz/2} $ are used. The bandwidth
matrix $\vW$, required to be symmetric and positive definite, is often selected as a diagonal matrix with entries
proportional to the marginal standard deviations of the samples. Note that this procedure has to be repeated at
each MCMC sample $\tvalpha$ and is computationally infeasible in most practical cases.

The true likelihood is degenerate for model-to-model calibration studies
when there is no measurement error, as recognized in~\cite{Sargsyan:2015} and, in a slightly different context,
in~\cite{Arnst:2010}. The degeneracy is a direct consequence of the fact that, if no data error is expected, there is
\emph{zero} probability that the model $f(x;\vlam)$ is able to replicate the `truth' $g(x)$ across all $x$ no matter
how well one tunes parameter $\vlam$, unless $g(\cdot)$ itself is a special
case of $f(\cdot;\vlam)$ for some fixed value of $\vlam$. With data noise, while the likelihood is not degenerate anymore, it may still lead to
posteriors that are hard to sample from\reb{~\cite{Pernot:2017,Pernot:2017a}}. Besides, as already indicated, the true likelihood computation requires a high-dimensional KDE at each
likelihood evaluation. Below, several approximating options for the true likelihood, that trade-off computational feasibility and accuracy, are listed.

\begin{itemize}
\item{\textbf{Independent-component approximation:}} The likelihood can be approximated as a product of its
marginal components as
\be \label{eq:ic}
\LL^{\textrm{IC}}_\vy(\tvalpha)=\prod_{i=1}^N \pi_{h_i}(y_i),
\ee
where each marginal PDF $\pi_{h_i}(y_i)$ is estimated via sampling and KDE. That is, having constructed $R$
samples of $\vh$, one approximates the marginal PDFs as
\be
\pi_{h_i}(y_i)\approx \frac{1}{R} \sum_{r=1}^R K(h^{(r)}_i-y_i)\textrm{ for all } i=1,\dots, N,
\ee
where, again, Gaussian kernels $K(z)=(2\pi)^{-1/2} w^{-1} e^{-z^2/(2w^2)}$ are typically employed. Note that KDE
estimates in this case are one-dimensional and computationally much more feasible compared to
Eq.~(\ref{eq:kde}). Nevertheless, the dependence on the KDE bandwidth $w$, sample size $R$, and the general
expense of sampling at each likelihood computation step, can make this approach computationally challenging,
albeit to a lesser degree than the computation of the true likelihood via Eq.~(\ref{eq:kde}).

\item{\textbf{Multivariate normal approximation:}} In this case, one can extract the mean and the covariance of the
random vector $\vh$ and, instead of estimating the true joint PDF $\pi_\vh(\vy)$, evaluate a multivariate normal
(MVN) corresponding to the mean and the covariance. If random sampling is employed
for extracting the first two moments of $\vh$, one has
\be \label{eq:mom}
\vmu^h\approx \frac{1}{R} \sum_{r=1}^R \vh^{(r)} \qquad \textrm{ and }\qquad \vC^h\approx\frac{1}{R-1} \sum_{r=1}
^R \left(\vh^{(r)}-\vmu^h\right) \left(\vh^{(r)}-\vmu^h\right)^T,
\ee
then one arrives at the likelihood
\be \label{eq:mvn}
\LL^{\textrm{MVN}}_\vy(\tvalpha)=(2\pi)^{-\frac{N}{2}}|\vC^h|^{-\frac{1}{2}} e^{-\frac{1}{2}(\vy-\vmu^h)^T {(\vC^h)}
^{-1}(\vy-\vmu^h)}.
\ee

\reb{Note that this approximation, while avoiding KDE, is not practical either since the covariance matrix is singular, inheriting the degeneracy of the true likelihood.}

\item{\textbf{Independent normal approximation:}} This option is a combination of the independent-component and
the multivariate normal approximations. With random sampling, the mean and the variance of the components of $\vh$ are
computed as
\be
\mu^h_i\approx \frac{1}{R} \sum_{r=1}^R h^{(r)}_i \qquad \textrm{ and }\qquad (\sigma^h_i)^2=\frac{1}
{R-1}\sum_{r=1}^R (h^{(r)}_i-\mu^h_i)^2,
\ee
 followed by the likelihood estimate
 \be \label{eq:in}
 \LL^{\textrm{IN}}_\vy(\tvalpha)=(2\pi)^{-\frac{N}{2}}\prod_{i=1}^N (\sigma^h_i)^{-1}e^{-\frac{(y_i-\mu^h_i)^2}
{2(\sigma^h_i)^2}}.
 \ee

\item{\textbf{Moment-matching approximation:}} To avoid computing expensive likelihoods, and in order to build
constraints driven by the purposes of the modeler, one can construct other forms of approximate likelihoods, inspired by
Approximate Bayesian Computation (ABC), otherwise called likelihood-free methods
~\cite{Beaumont:2002,Marjoram:2003,Sisson:2011}. This approach measures the discrepancy between a chosen set of statistics of
model outputs and the corresponding estimates from the data. In this regard, component-wise means and
variances are common choices. The ABC formulation relies on a kernel function, typically a Gaussian
$K(z)=e^{-z^2/2}/\sqrt{2\pi}$, and a chosen distance measure between statistics of interest on data prediction ($S_\vh$) and
observed data ($S_\vy$), as well as a pre-selected `tolerance' parameter $\epsilon$ to arrive at
\be \label{eq:abc}
\LL^{\textrm{ABC}}_\vy(\tvalpha)=\epsilon^{-1} K(\epsilon^{-1}\rho(S_\vh,S_\vy))=\frac{1}{\epsilon\sqrt{2\pi}}e^{-
\frac{\rho\left(S_\vh,S_\vy\right)^2}{2\epsilon^2}}.
\ee
It is of interest for predictive purposes to require agreement, on average,
between (a) the mean prediction and the data, and (b) the predictive standard
deviation and the discrepancy between the mean prediction and the data.
That is, besides matching the
means $\vmu^h\approx \vy$, one should additionally constrain the standard deviation as $\vsigma^h\approx
\gamma|\vmu^h-\vy|$, where the absolute value is understood to be taken element-wise, and $\gamma>0$ is an
additional parameter defined by modeler. In this case, the ABC likelihood takes the form
\be \label{eq:abc2}
\LL^{\textrm{ABC}}_\vy(\tvalpha)=\frac{1}{\epsilon\sqrt{2\pi}}\prod_{i=1}^N \exp\left(-\frac{(\mu^h_i-y_i)^2+
(\sigma^h_i-\gamma|\mu^h_i-y_i|)^2}{2\epsilon^2}\right).
\ee

Note that all PDFs and moments of $\vh$ or their components depend on $\tvalpha$, albeit dropped for simplicity in Eqs.~\eqref{eq:kde}-\eqref{eq:abc2}.

A proof-of-concept demonstration of such an ABC approach has been performed in~\cite{Sargsyan:2015} in the
context of calibrating a low-fidelity chemical kinetic model with respect to (noise free) simulation data from a higher-fidelity
model. As a direct consequence of the ABC likelihood, the resulting calibrated uncertain model outputs were shown
to be centered on the data, at the same time exhibiting a degree of uncertainty consistent with the average
discrepancy from the data.

\end{itemize}

\begin{table}[!t]
\begin{center}
\footnotesize
\begin{tabular}{ l | c | c }
  \hline
 Likelihood type & Formula & Eq. \\
  \hline
  \hline
   True likelihood & $\LL_\vy(\tvalpha)=\pi_\vh(\vy)$ & (\ref{eq:lik_true})\\ \hline
  Independent-component & $\LL^{\textrm{IC}}_\vy(\tvalpha)=\prod_{i=1}^N \pi_{h_i}(y_i)$ & (\ref{eq:ic})\\
  \hline
 Multivariate normal  & $\LL^{\textrm{MVN}}_\vy(\tvalpha)=(2\pi)^{-\frac{N}{2}}|\vC^h|^{-\frac{1}{2}} e^{-\frac{1}{2}
(\vy-\vmu^h)^T {\vC^h}^{-1}(\vy-\vmu^h)}$ & (\ref{eq:mvn})\\
  \hline
     Independent normal & $ \LL^{\textrm{IN}}_\vy(\tvalpha)=(2\pi)^{-\frac{N}{2}}\prod_{i=1}^N (\sigma^h_i)^{-1}\exp
\left(-\frac{(y_i-\mu^h_i)^2}{2(\sigma^h_i)^2}\right)$ & (\ref{eq:in})\\ \hline
   ABC; mean and stdev & $\LL^{\textrm{ABC}}_\vy(\tvalpha)=\frac{1}{\epsilon\sqrt{2\pi}}\prod_{i=1}^N \exp\left(-
\frac{(\mu^h_i-y_i)^2+(\sigma^h_i-\gamma|\mu^h_i-y_i|)^2}{2\epsilon^2}\right)$
& (\ref{eq:abc})\\
  \hline
\end{tabular}
\end{center}
\caption {A summary of potential options for likelihood construction, including the corresponding equation number.
Note that all PDFs and moments of $\vh$ and their components depend on $\tvalpha$, but are dropped for simplicity.
}
\label{tab:liks}
\end{table}

\noindent Table~\ref{tab:liks} summarizes the likelihood options in a concise way. Note that, in this section, the PDF of $
\vh$, as well as, when necessary, computation of its moments, are estimated via random sampling, requiring a
potentially large number $R$ of model evaluations. With the non-intrusive polynomial chaos (PC) approach
described further in Section~\ref{s:nisppc}, one can (a) reduce the sampling cost by replacing model
evaluations with PC evaluations, if KDE estimates are necessary, and (b) evaluate moments exactly from PC
representations, without additional cost. \reb{In this work we explore moment-based likelihoods only, \ie independent-normal
and ABC likelihoods, where the first two moments are efficiently computed via PC representation. The KDE-requiring likelihoods
need substantially more sampling for PDF estimates to reduce the noise in likelihood computation, while the multivariate normal likelihood
often fails due to the rank-deficiency of the covariance matrix, a direct consequence of the degeneracy of the true likelihood.}

\subsubsection{Prior construction}
\label{sec:prior}

Another major ingredient of Bayes formula~(\ref{eq:bayes}) is the prior PDF $p(\tvalpha)$ on
parameters $\tvalpha=(\vlam,\valpha)$. The prior, in the absence of information to the contrary, is often
separated for convenience, \ie $p(\tvalpha)=p(\vlam)p(\valpha)$.
Prior selection is a known conceptual challenge for any Bayesian
method. In this context, while one selects $p(\vlam)$ according to some physical considerations, \eg expert knowledge,
or a result of a previous calibration, the additional difficulty arises from
the fact that $\valpha$ are not physically-meaningful model parameters, but
rather parameters that define the discrepancy term $\vdel(\valpha,\vxi)$. At the same time, prior selection offers
several opportunities for imposing specific constraints or regularization. It is important to select
appropriate priors that capture the initial knowledge on $\valpha$ before any
observational or simulation data is available. For bounded embeddings this would typically entail priors with
specific irregular support, while for unbounded embeddings, such as a zero-mean multivariate normal (MVN)
$\vdel(\valpha,\vxi)$ the constraints can be derived from the requisite constraints on the covariance structure.
The latter is a reasonably mature area in Bayesian hierarchical
modeling~\cite{Leonard:1992,Alvarez:2014,Daniels:1999}, with several options available, such as both
Wishart~\cite{Chung:2015} and inverse-Wishart priors~\cite{Kass:2006}, Cholesky-factor
priors~\cite{SmithKohn:2002}, reference priors~\cite{Yang:1994}, shrinkage priors~\cite{Wang:2013},
separation-based priors that separate standard deviation and correlation
coefficients~\cite{Huang:2013,Gelman:2006,Barnard:2000}, and matrix logarithm~\cite{Hsu:2012}. The prior selection
will not be made the focus of the current paper; unless otherwise specified, we will restrict ourselves to uniform priors, potentially with physics-based support constraints.

\subsubsection{Posterior sampling and MCMC challenges}

Sampling of the posterior distribution via MCMC is a challenging task
due to the \reb{potential} high-dimensionality of $\tvalpha$ and the near-degeneracy discussed in Section~\ref{sec:lik}. One
approach of mitigating the high-dimensionality is to embed the
model error representation in a few parameters at a time, instead of the full input $\vlam$. This enables attribution of
model errors to specific parameters and provides grounds for attributing model errors to certain submodels.
Another way of mitigating the high-dimensionality is to resort to simple posterior
maximization, via standard optimization methods, instead of the full MCMC sampling. This is a key advantage of
the embedded representation -- even if one resorts to the best value $\tvalpha_{\textrm{MAP}}$, it still
leads to prediction with uncertainties that capture model errors, albeit without a component relevant to the posterior
distribution.

Nevertheless, sampling of the posterior distribution via MCMC or even
simply maximizing the posterior distribution may be a difficult task
in general, due to the structure of the posterior distribution
itself. It may be multimodal as well as peaked along lower-dimensional
manifolds, making MCMC sampling extremely challenging. Structural
improvements to the posterior distribution may be achieved by
appropriate regularization via prior selection.

\subsection{Prediction with model error} \label{s:pred}

The key advantage of the developed method is highlighted when one uses the calibrated model for
prediction purposes. Explicit additive representation of the model discrepancy function $\delta(x)$ in~\cite{Kennedy:2001}
makes the predictions challenging since it may violate physical constraints and is
specific to the observable quantities used for the calibration. On the contrary, in our approach, model
discrepancy is captured \emph{within} the model, which makes definitions of predictive quantities relatively
straightforward and meaningful. Each fixed value of the vector $\tvalpha=(\vlam,\valpha)$ induces a random input vector
$\vLam(\tvalpha)=\vlam+\vdel(\valpha,\vxi)$
which in turn leads to a stochastic process $F(x;\tvalpha)$ that is the result of pushing forward the stochastic
germ $\vxi$
 through the model $f(x;\vLam(\tvalpha))$. The mean and covariance of this process are denoted by $\mu^f(x;\tvalpha)
$ and $C^f(x,x';\tvalpha)$, respectively, where
\bea \label{eq:ppfmeanxi}
\mu^f(x;\tvalpha) &=& \EE_\vxi [F(x;\tvalpha)] \\
C^f(x,x';\tvalpha) &=& \CC_\vxi [F(x,x';\tvalpha)].
\eea

The MAP value $\tvalpha_{\textrm{MAP}}$ can be used to arrive at the best PDF for $\vLam$ that encompasses the augmented model error
representation. This input can subsequently be
pushed forward through the model $f(x;\vLam)$ to obtain the \emph{MAP pushed-forward} process $\fmp(x)= F(x;\tvalpha_{\textrm{MAP}})$,
with moments denoted as $\mu_{\textrm{MPF}}(x)$ and $C_{\textrm{MPF}}(x,x')$ which can be computed via
sampling similar to~(\ref{eq:mom}) for general $x$, or via non-intrusive spectral projection with polynomial chaos
described further in Section~\ref{s:nisppc}.

Furthermore, accounting for the full posterior distribution of $\tvalpha$, one can compute the moments of \emph{posterior pushed-forward}
process $F(x;\tvalpha)$ for any value of $x$ as a combination of two stochastic
dimensions (model-error $\vxi$ and posterior range sampled by MCMC samples
$\tvalpha$):
\be \label{eq:ppfmean}
\mu_{\textrm{PF}}(x)=\EE_\vxi \EE_{\tvalpha}[F(x;\tvalpha)] =\EE_{\tvalpha}[\mu^f(x;\tvalpha)],
\ee
and, due to variance decomposition formula,
\be \label{eq:ppfcor}
C_{\textrm{PF}}(x,x')=\EE_{\tvalpha}[C^f(x,x';\tvalpha)]+\CC_{\tvalpha}[\mu^f(x;\tvalpha)],
\ee
where $\EE_{\tvalpha}$ and $\CC_{\tvalpha}$ denote the mean and covariance with respect to the posterior distribution
$p(\tvalpha|\vy)$, respectively, and can be computed with MCMC samples and standard estimators.
Pushed-forward variance for each fixed $x$ can be computed simply by $\sigma^2_{\textrm{PF}}(x)=C_{\textrm{PF}}(x,x)$, and is decomposed as
\be \label{eq:predvar}
\sigma^2_{\textrm{PF}}(x)=\underbrace{\EE_\tvalpha[\sigma^f(x;\tvalpha)^2]}_{\textrm{model error}}+
\underbrace{\VV_\tvalpha[\mu^f(x;\tvalpha)]}_{\textrm{posterior uncertainty}}
\ee
which consists of contributions due to the embedded model error as well as due to posterior uncertainty stemming from the quality and amount of data.
The key feature of the developed approach -- and one to be illustrated in numerical examples below --
is that, with increasing amount of data, the model error component
$\EE_\tvalpha[\sigma^f(x;\tvalpha)^2]$ does not decrease - rather it saturates to a fixed value. At the same time, the
data error encoded into the posterior variance term $\VV_\tvalpha[\mu^f(x;\tvalpha)]$ typically
reduces with increasing amount of data.

 While the pushed-forward process can be useful
for prediction purposes for general values of $x$, in order to compare with data, it is also important to consider the \emph{posterior predictive}
distribution~\cite{Gelman:1996}, which is by definition the combination of the pushed-forward process at the design conditions $x_i$
and the data noise model $\epsilon_i$. The $i$-th component of the posterior predictive random vector is then
$F(x_i;\tvalpha)+\epsilon_i$
with mean and covariance
\be \label{eq:ppmean}
\mu_{\textrm{PP}}(x_i)=\EE_{\tvalpha}[\mu^f(x_i;\tvalpha)]=\mu_{\textrm{PF}}(x_i),
\ee
\be \label{eq:ppcov}
C_{\textrm{PP}}(x_i,x_j)=\EE_{\tvalpha}[C^f(x_i,x_j;\tvalpha)+\sigma^2 \delta_{ij}]+\CC_{\tvalpha}[\mu^f(x_i;
\tvalpha)]=C_{\textrm{PF}}(x_i,x_j)+\sigma^2 \delta_{ij}.
\ee

The posterior predictive covariance~(\ref{eq:ppcov}) should be
contrasted with an imposed parameterized covariance -- typically a square exponential -- when modeling the model
discrepancy term explicitly in an additive fashion~\cite{Kennedy:2001}.
The major conceptual difference is that the covariance obtained via model error embedding is
directly informed by the model structure, rather than parameterized in a model-independent way.

Note that often $\sigma$ can be viewed as a hyperparameter to be inferred together with $\tvalpha$, in which case
the last term $\sigma^2$ in Eq.~(\ref{eq:ppcov}) is replaced with its posterior average $\EE_{\tvalpha}[\sigma^2]$. This is demonstrated
in the chemical kinetics example in Section~\ref{sec:ign_vasu}.

Both likelihood computation via any of the options from
Table~\ref{tab:liks}, and
predictive/pushed-forward quantity computations, require sampling of model input $\vLam(\tvalpha)=\vlam+\vdel(\valpha,\vxi)$ and subsequent
evaluations of the model $f(x,\vLam(\tvalpha))$ with given $\tvalpha$, or ability to compute
the pushed-forward moments $\mu^f(x;\tvalpha)$ and $C^f(x,x';\tvalpha)$. Polynomial Chaos (PC) machinery
provides a flexible mechanism of representing random variables allowing computationally inexpensive sampling,
moment evaluation, as well as, generally, convenient tools for forward uncertainty propagation
~\cite{Ghanem:1991, Xiu:2002c, OlmOmk:2010}.

\subsubsection{Polynomial chaos representation of augmented inputs}
\label{sec:pcin}

\reb{The stochastic discrepancy term $\vdel(\valpha,\vxi)$ can be parameterized in different ways to incorporate a potentially large class of PDFs, such as Gaussian mixture models or parametrized PDF families.}
While the present framework remains valid with any \reb{such} parameterization,
the focus of this paper is specifically on Polynomial Chaos (PC) representations, since they allow
flexible parameterization of a wide class of random variables, with efficient sampling, moment evaluation and
forward propagation machinery. In order to avoid crowding the notation, and for demonstrating proof-of-concept,
assume that model error representation is embedded in \emph{all} components of $\vdel$\reb{, and the number of stochastic dimensions, \ie the dimensionality of $\vxi$, is the same as the number of input parameters $d$}. In such case, the
components of the model input are written as an expansion
\be \label{eq:inpc}
\Lambda_j(\tvalpha,\vxi)=\lambda_j+\delta_j(\valpha,\vxi)=\lambda_j+\sum_{k=1}^{K_j-1} \alpha_{kj} \Psi_k(\vxi), \textrm{ for } j=1,\dots,d
\ee
with respect to standard orthogonal polynomials $\Psi_k(\vxi)$ of an independent-component standard random
vector $\vxi$. The two most commonly used PC expansions are a) Gauss-Hermite, \ie
the \emph{germ} $\vxi$ is a vector of \emph{i.i.d.} Gaussian random variables,
and, correspondingly, the $\Psi_k(\vxi)$ are Hermite polynomials that are orthogonal with respect to the PDF of $\vxi
$, $\pi_\vxi(\vxi)=\frac{e^{-\vxi^2/2}}{\sqrt{2\pi}}$, and b) Legendre-Uniform, in which $\vxi$ is a vector of
\emph{i.i.d.} standard uniform random variables on $[-1,1]$,
and the $\Psi_k(\vxi)$ are Legendre orthogonal polynomials. The latter is a more meaningful parameterization for model inputs that are
bounded. The orthogonality of polynomial bases is written as
\be
\int_\vxi \Psi_k(\vxi) \Psi_j(\vxi) \pi_\vxi(\vxi) d\vxi = \begin{cases}||\Psi_k||^2, \textrm{ if } k=j\\0, \qquad\textrm{ otherwise} \end{cases}
\ee
where the norm $||\Psi_k||=\left(\int_\vxi \Psi_k^2(\vxi) \pi_\vxi(\vxi) d\vxi
\right)^{1/2}$ is usually precomputed and stored a priori. Typically, with an understanding that each probabilistically
cast $\delta_j$ corresponds to a stochastic dimension, one chooses $\vxi$ to have the same size as $\vdel$.
The order of each expansion in Eq.~(\ref{eq:inpc}) is selected a priori, denoted by $p_j$, leading to $K_j=(d+p_j)!/
(d!p_j!)$ terms. Note that formulating model input adjustments as PC expansions~(\ref{eq:inpc}), the model-error estimation is
essentially reformulated as a parameter estimation for deterministic quantities $\alpha_{kj}$ that comprise
the set $\tvalpha=\{\alpha_{kj}\}_{j=1,\dots,d}^{k=0,\dots, K_j-1}$, in which, out of convenience, we denoted $\lambda_j=\alpha_{0,j}$ for all $j=1,\dots,d$.

Below we discuss two commonly used embeddings that allow efficient Bayesian calibration.

\bi
\item \emph{Multivariate normal input:} A multivariate normal assumption for
$\vdel(\valpha,\vxi)$ is a convenient way of parameterizing the internal model-error correction, particularly in cases when there is no restriction on the range of input parameters.
It, as does the uniform embedding discussed below, also has the practical
advantage of being first-order in $\vxi$, thus eliminating difficulties associated with
inferring high-order PC coefficients~(\ref{eq:inpc}). Indeed, the first-order PC expansion for $\vdel(\valpha,\vxi)$,
corresponding to an MVN, which leads to
\be \label{eq:inpc1}
\Lambda_j(\tvalpha,\vxi)=\lambda_j+\delta_j(\valpha,\vxi)=\lambda_j+\sum_{k=1}^d \alpha_{kj} \xi_k, \textrm{ for } j=1,\dots, d
\ee
is a special case of Eq.~(\ref{eq:inpc}) for $p_j=1$ for all $j=1,\dots,d$. It highlights one key challenge of such PC
representations -- many coefficient sets $\valpha=\{\alpha_{kj}\}_{j=1,\dots,d}^{k=0,\dots,d}$ may correspond to the
same joint PDF for $\vLam$ and therefore will lead to the same
likelihood value and, consequently, to a multimodal posterior distribution or a posterior distribution that is constant along certain low-dimensional manifolds. Barring judicious choice of priors
for $\valpha$, this poses practical difficulties for optimization or MCMC methods. Nevertheless, for the linear
case~(\ref{eq:inpc1}), a simple re-parameterization resolves this ambiguity. Specifically, one can propose a
`triangular' linear PC form
\be \label{eq:inpct}
\delta_j(\xi_1,\dots,\xi_j)=\sum_{k=1}^j \alpha_{kj} \xi_k, \textrm{ for } j=1,\dots, d
\ee
with an additional constraint $\alpha_{kd}>0$ for $k=1,\dots,d$ enforced via Bayesian priors. In this case, there is
a one-to-one correspondence between any MVN PDF of $\vLam$ and the set of coefficients
$\tvalpha=\{\alpha_{kj}\}_{j=0,\dots,d}^{k=0,\dots,j}$, effectively ruling out a large set of potentially challenging, degenerate posteriors. \reb{The `triangular' parameterization in fact corresponds to entries of
Cholesky factorization matrix of the covariance of the underlying MVN.}

\item \emph{Uniform i.i.d. input:} In physical models, one often has bounds for input parameters that cannot be exceeded
under any circumstances. In such cases, simple Gaussian additive adjustments to the model inputs may not be the best option.
We employ Legendre-Uniform PC expansions for such cases that are bounded. However, in general this may require prior constraints on $\tvalpha$ that are technically challenging to implement. To simplify, and for the sake of demonstration,
a linear Legendre-Uniform PC with \emph{i.i.d.} embedding can be used,
\be\label{eq:inpc2}
\Lambda_j(\tvalpha,\vxi)=\lambda_j+\delta_j(\valpha,\vxi)=\lambda_j+\alpha_{1j} \xi_j, \textrm{ for } j=1,\dots, d
\ee
corresponding to model-error correction $\delta_j(\xi_j)=\alpha_{1j}\xi_j$, for $\xi_j\in$U$[-1,1]$. The \emph{i.i.d.} embedding allows
simple prior constraints
\be \label{eq:luprior}
 \left\{
 \begin{array}{l}
\lambda_{j}+\alpha_{1j}\le a_j\\
\lambda_{j}-\alpha_{1j}\ge b_j.
 \end{array}
 \right.
\ee
on $\tvalpha=(\vlam,\valpha)$ to enforce parameters within given
ranges $\Lambda_j=\lambda_j+\alpha_{1j}\xi_j\in[a_j,b_j]$. The
prior~\eqref{eq:luprior}, together with $\alpha_{1j}\ge0$ (added due
to invariance, and to avoid bimodal posteriors) corresponds to a
triangular region in the $(\lambda_j,\alpha_{1j})$ space, a subset of
$\RR^2$.

\ei

\subsubsection{Uncertainty propagation and predictive moment estimation} \label{s:nisppc}

Representing inputs of the function $f(x;\vLam(\tvalpha,\vxi))$ with PC expansions~(\ref{eq:inpc}), or its simpler forms \eqref{eq:inpc1} and~\eqref{eq:inpc2}, allows efficient propagation of
the uncertainties through $f(x;\cdot)$ via non-intrusive spectral projection (NISP), as follows. One constructs a PC expansion for the output for each
$x$ as
\be \label{eq:fpc}
f(x;\vLam(\tvalpha,\vxi))\simeq\sum_{k=0}^{K-1} f_k(x;\tvalpha) \Psi_k(\vxi),
\ee
where the coefficients $f_k(x;\tvalpha)$ are found via orthogonal projection that is numerically computed with
quadrature integration, using a precomputed quadrature point-weight pairs $(\vxi^{(q)},w_q)$ for
$q=1,\dots,Q$,
\bea
f_k(x;\tvalpha)&=&\frac{1}{||\Psi_k||^2} \int_\vxi f(x; \vLam(\tvalpha,\vxi)) \Psi_k(\vxi) \pi_\vxi(\vxi) d\vxi \approx \nonumber \\
&\approx& \frac{1}{||\Psi_k||
^2} \sum_{q=1}^Q w_q f(x;\vLam(\tvalpha,\vxi^{(q)})) \Psi_k(\vxi^{(q)}).
\label{eq:fpcq}
\eea
The PC coefficients depend on $\tvalpha$ since the relationship $\vLam(\tvalpha,\vxi)$ encapsulated in the input PC
expansion~(\ref{eq:inpc}) is parameterized by $\tvalpha$. It is also important to recognize that this PC propagation
approach essentially relies on \emph{sampling}, too, since the model evaluations are driven by the underlying
quadrature sampling of $\vxi$. Such sampling is, however, much more efficient and accurate for a large class of
models, compared to Monte-Carlo based uncertainty propagation approaches~\cite{Eldred:2009, Sargsyan:2016}.

Note that the classical calibration strategy that ignores the model error and
assumes $f(x;\vlam)$ replicates the `truth' $g(x)$, is merely a special case of the developed model-error embedding
framework for $0$-th order PC expansion in~(\ref{eq:inpc}), \ie $\Lambda_j(\tvalpha,\vxi)=\lambda_j$, with no explicit model-error
dimension ($\vxi$) dependence, and $\tvalpha\equiv\vlam$.

Next, we return to the predictive process $F(x;\tvalpha)=f(x;\vLam(\tvalpha,\vxi))$ described in Section~\ref{s:pred} in terms of PC representations. Both generic predictions for arbitrary $x$ and likelihood computations for $x_i$'s require either KDE estimation or moment computation. This
entails extensive
sampling of the function outputs and may become prohibitively expensive.  For sampling, the PC
representation~(\ref{eq:fpc}) can serve as a `surrogate' approximation of the function and be sampled instead of it. However, the crucial advantage of the PC representation is highlighted in moment-based likelihood cases,
Eqs.~(\ref{eq:mvn}),~(\ref{eq:in}) and~(\ref{eq:abc}). In these cases, only moments are required to compute the
likelihoods, and the PC expansion~(\ref{eq:fpc}) offers analytical, closed
forms for them, circumventing sampling.

The predictive stochastic process $F(x;
\tvalpha)$ is now written as a PC expansion
\be \label{eq:pcf}
F(x;\tvalpha) \simeq \sum_{k=0}^{K-1} f_k(x;\tvalpha) \Psi_k(\vxi)
\ee
from which moments can be easily extracted as
\be \label{eq:predmom}
\mu^f(x;\tvalpha)\simeq f_0(x;\tvalpha) \qquad\textrm{ and }\qquad C^f(x,x';\tvalpha)\simeq \sum_{k=1}^{K-1} f_k(x;
\tvalpha)f_k(x';\tvalpha) ||\Psi_k||^2.
\ee
The overall pushed-forward moments are written as
\bea
\mu_{\textrm{PF}}(x)&=&\EE_\tvalpha[\mu^f(x;\tvalpha)]\simeq\EE_\tvalpha[f_0(x;\tvalpha)] \label{eq:mpf}\\
C_{\textrm{PF}}(x,x')&=&\EE_\tvalpha[C^f(x,x';\tvalpha)]+\CC_\tvalpha[\mu^f(x;\tvalpha)] \simeq \nonumber \\
&\simeq&  \underbrace{\sum_{k=1}^{K-1} \EE_\tvalpha[f_k(x;\tvalpha)f_k(x';\tvalpha)] ||\Psi_k||^2}_{\textrm{Model error}}+\underbrace{\CC_\tvalpha[f_0(x;\tvalpha)]}_{\textrm{Posterior uncertainty}}. \label{eq:cpf}
\eea

Table~\ref{tab:pred} summarizes the predictive process $F(x;\tvalpha)$, both for
fixed $\tvalpha$ and posterior average with respect to $\tvalpha$. It also lists its moments, and provides formulae
specific for the PC representation.

Correspondingly, the data model~(\ref{eq:datamodel}) now takes the form, as a combination of pushed-forward
process evaluated at $x_i$'s and measurement errors,
\be \label{eq:datamodelpc}
y_i\approx \sum_{k=0}^{K-1} f_k(x_i;\tvalpha) \Psi_k(\vxi)+\sigma \xi_{d+i}, \textrm{ for } i=1,\dots, N,
\ee
from which one can extract moments as in~(\ref{eq:predmom}). The
right-hand-side of Eq.~\eqref{eq:datamodelpc} includes all components of
predictive uncertainty (model error $\vxi$ and data error $\xi_{i+d}$) in a PC
form for given $\tvalpha$. One can further build a PC expansion given posterior
samples of $\tvalpha$ arriving at an overall PC representation that also
includes posterior uncertainty. Details of this procedure are given in
Appendix~\ref{sec:app}. Further in this text, we will keep $\tvalpha$ as the
uncertain dimension due to the posterior PDF, and the associated variance
contributions will be accounted for via estimators that are based on MCMC
samples.

When the predictive process $F(x;\tvalpha)$ is evaluated at specific locations $x_i$ at
which the data is taken, it allows direct comparison with the data points $y_i$, thus facilitating likelihood
computations. The corresponding moments can be written as
\bea
\mu_i(\tvalpha)&=&\EE_\vxi f(x_i;\vLam(\tvalpha,\vxi))\simeq f_0(x_i;\tvalpha),\label{eq:pcmu}\\
C_{ij}(\tvalpha)&=&\EE_\vxi \left[\left(f(x_i;\vLam(\tvalpha,\vxi))-\mu_i(\tvalpha)\right) \left(f(x_j;\vLam(\tvalpha,\vxi))-
\mu_j(\tvalpha)\right)\right]\simeq\nonumber\\
&\simeq&\sum_{k=1}^{K-1}f_k(x_i;\tvalpha) f_k(x_j;\tvalpha) ||\Psi_k||^2 + \delta_{ij} \sigma^2. \label{eq:pccov}\\
\sigma^2_i(\tvalpha)&=&\EE_\vxi \left[f(x_i;\vLam(\tvalpha,\vxi))-\mu_i(\tvalpha)\right]^2\simeq\sum_{k=1}^{K-1}
f^2_k(x_i;\tvalpha) ||\Psi_k||^2+\sigma^2,\label{eq:pcsig}
\eea
\reb{These moments feed directly into likelihood computation as $\mu_i^h$ and $\sigma_i^h$ in Eqs.~\eqref{eq:in} or Eq.~\eqref{eq:abc2} in which we had dropped $\tvalpha$ for simplicity.}
Note that in the equations above, as well as before in this text, the symbol $\simeq$ highlights the fact that the
result is based on a PC expansion which approximates equality and extra care is needed, \emph{e.g.}, in terms of
appropriate truncation of the expansion.
Quadrature point sampling and consequent function evaluations are still present during the
construction of the PC expansion, but for reasonably smooth forward functions they are much more efficient than
Monte-Carlo based approaches for moment computation. Besides, as shown in the next section, one can pre-construct and employ a polynomial surrogate \emph{before} the inference, allowing efficient MCMC without expensive model evaluations for each $\tvalpha$, \ie at each MCMC step.

\begin{table}[!t]
\begin{center}
\begin{footnotesize}
\begin{tabular}{| c | c | c | c |}
  \hline
- & Process & Mean & Covariance \\
  \hline
  \hline
 Fixed $\tvalpha$ &   $F(x;\tvalpha)$ & $\mu^f(x;\tvalpha)$ & $C^f(x,x';\tvalpha)$\\ \hline
  PC apprx. &   $\sum_{k=0}^{K-1} f_k(x;\tvalpha) \Psi_k(\vxi)$ & $f_0(x;\tvalpha)$ & $\sum_{k=1}^{K-1} f_k(x;\tvalpha)f_k(x';\tvalpha) ||\Psi_k||^2$\\ \hline
PF, $\EE_\tvalpha$ & $\EE_\tvalpha[F(x;\tvalpha)]$  & $\EE_\tvalpha[\mu^f(x;\tvalpha)]$ & $\EE_\tvalpha[C^f(x,x';\tvalpha)]+\CC_\tvalpha[\mu^f(x;\tvalpha)]$\\
  \hline
     \end{tabular}
\end{footnotesize}
\end{center}
\caption {A summary of the predictive process and its moments, \reb{in a PC approximate form and in general pushed-forward, \ie posterior average form. Note that the posterior predictive (PP), as discussed after Eqs.~\eqref{eq:ppmean} and~\eqref{eq:ppcov}, adds data variance, $\sigma^2$ or its posterior average $\EE_\tvalpha[\sigma^2]$ in case when $\sigma^2$ is also an object of inference and part of $\tvalpha$}. }
\label{tab:pred}
\end{table}

\subsubsection{Surrogate construction and overall algorithm} \label{s:algo}

For most physical application problems, the model $f(x;\vLam)$ is quite expensive, or difficult to efficiently
evaluate during MCMC. Here we describe a procedure to pre-construct a surrogate $f_s(x;\vlam)\approx f(x;\vlam)$,
to alleviate the cost. The surrogate $f_s(x;\vlam)$ can be constructed as an approximation of the function
$f(x;\vlam)$ over the joint space $(x,\vlam)$, or over $\vlam$ only, for a select set of $x_i$
that have been utilized during calibration and prediction. Here, we will describe and employ the
latter approach for simplicity. For each $x_i$, consider a surrogate form
\be \label{eq:pfit}
f(x_i;\vlam)\approx f_s(x_i;\vlam)=\sum_{k=0}^{K-1} s_{ik} L_k(\vlam),
\ee
where $L_k(\vlam)$'s are scaled multivariate Legendre polynomials, with inputs scaled to their respective ranges $\lambda_j\in[a_j,b_j]$ for $j=1,\dots,d$. Similar to the general PC construction described earlier, the truncation, \ie the number of polynomial bases $K$, is driven by a predefined rule, \eg according to a total degree of retained polynomials.

The polynomial fit~\eqref{eq:pfit} is constructed via least-squares regression, using an arbitrary set of $R$ model evaluations $\{f(x_i,\vlam_r)\}_{r=1}^R$. In other words, the coefficients $c_{ik}$ are the solution of the minimization problem for each $i$
\be
\argmin_{c_{ik}} \sum_{r=1}^R \left(f(x_i,\vlam_r)-\sum_{k=0}^{K-1} c_{ik} L_k(\vlam_r)\right)^2.
\ee
Note that the simple least-squares solution is analytically available and does not require an optimization engine. Namely, the solution vector $\vc_i=(c_{i0},\dots,c_{iK})^T$ of size $K\times 1$ is written as
\be \label{eq:lsqsol}
\vc_i=(P^TP)^{-1} P^T \vf_i,
\ee
where $\vf_i=(f(x_i,\vlam_1), \dots, f(x_i,\vlam_R))^T$ is the model evaluations' vector of size $R\times 1$, and $P$ is the basis evaluation matrix of size $R\times K$ with entries $P_{rk}=L_k(\vlam_r)$. Note that the least-squares solution~\eqref{eq:lsqsol} is employed instead of NISP-like orthogonal projection, since it enables a surrogate error metric, leave-one-out error which can in principle be added to the predictive uncertainty budget. Besides, the regression-based surrogate is more general and does not require basis orthogonality.

Now, when we construct the PC propagation~\eqref{eq:fpc} via NISP, the surrogate $f_s(x;\vlam)$ replaces the actual forward model $f(x;\vlam)$ in the quadrature integration~\eqref{eq:fpcq}. Moreover, if the surrogate is built as a polynomial expansion~\eqref{eq:pfit} with the same order as the NISP in Eq.~\eqref{eq:fpc}, then no extra approximation error is induced due to NISP. The only additional error is due to the surrogate approximation itself which can be estimated, \eg via leave-one-out cross-validation as demonstrated in~\cite{Huan:2017}. In this work, for the numerical demonstrations in Section~\ref{sec:numer}, no surrogate is employed as the functions are synthetic and easy-to-evaluate, while for the chemical kinetics application in Section~\ref{sec:ign} the surrogate errors are negligible and are skipped for the clarity of presentation. The overall mechanism of surrogate-enhanced model error inference and prediction is listed in Algorithm~\ref{algo}.

\begin{algorithm}[htb]
\begin{scriptsize}
\SetAlgoLined
\textbf{Surrogate construction:}
\bi
\item Note: this step is skipped for the demo problems, and only applied to the chemistry problem in Section~\ref{sec:ign}.
\item Choose $R$ input samples $\vlam_r$, for $r=1,\dots,R$:
\bi
\item we chose uniform random set of points distributed in given parameter ranges. $\lambda_{ir}\in[a_i,b_i]$.
\ei
\item Evaluate forward model $R$ times and extract $f(x_i;\vlam_r)$ at each design condition $x_i$, for $i=1,\dots,N$ and $r=1,\dots,R$.
\item Build polynomial surrogate~\eqref{eq:pfit}:
\bi
\item we employed least-squares solution~\eqref{eq:lsqsol}.
\ei
\ei
\textbf{Model error inference:}
\bi
\item Select embedding type (see Section~\ref{sec:pcin}):
\bi
\item we chose MVN input~\eqref{eq:inpc1} for demo problems with no parameter range constraints, and the uniform i.i.d. inputs~\eqref{eq:inpc2} otherwise.
\ei
\item Select a prior (see Sections~\ref{sec:prior} and \ref{sec:pcin}).
\item Select a feasible likelihood form from Table~\ref{tab:liks}:
\bi
\item for most studies here, we chose independent-normal, or ABC.
\ei
\item Run MCMC inference:
\bi
\item we used adaptive MCMC (AMCMC) algorithm, introduced in~\cite{Haario:2001}.
\item at each MCMC step, to compute the likelihood, the PC propagation~\eqref{eq:datamodelpc} and moment estimation~\eqref{eq:pcmu}-\eqref{eq:pcsig} are completed via integration by quadrature~\eqref{eq:fpcq}.
\item for the chemistry problem in Section~\ref{sec:ign}, the expensive model evaluations required by \eqref{eq:fpcq} are replaced by the surrogate evaluations~\eqref{eq:pfit}.
\ei
\ei

\textbf{Prediction:}

\bi
\item Note: the predictions can be made for general design conditions $x$, even if data is collected at specific values $x_i$, for $i=1,\dots,N$.
\item Predictive moments of $f(x;\vlam)$ are computed as given in Eqs.~\eqref{eq:mpf} and~\eqref{eq:cpf}, and are summarized in Table~\ref{tab:pred}.
\ei

\caption{The full workflow of the surrogate-enhanced model error inference algorithm.}
\label{algo}
\end{scriptsize}
\end{algorithm}

\section{Numerical demonstrations} \label{sec:numer}

In this section, we illustrate the strengths of the embedded model-error
strategy with a few numerical examples. Note that
we skip the surrogate construction step as listed in Algorithm~\ref{algo},
since the synthetic model evaluations are not costly.

\begin{figure}[!h]
\centering{
\subfigure[PF Prediction with ABC]{ \includegraphics[height=0.47\textwidth]{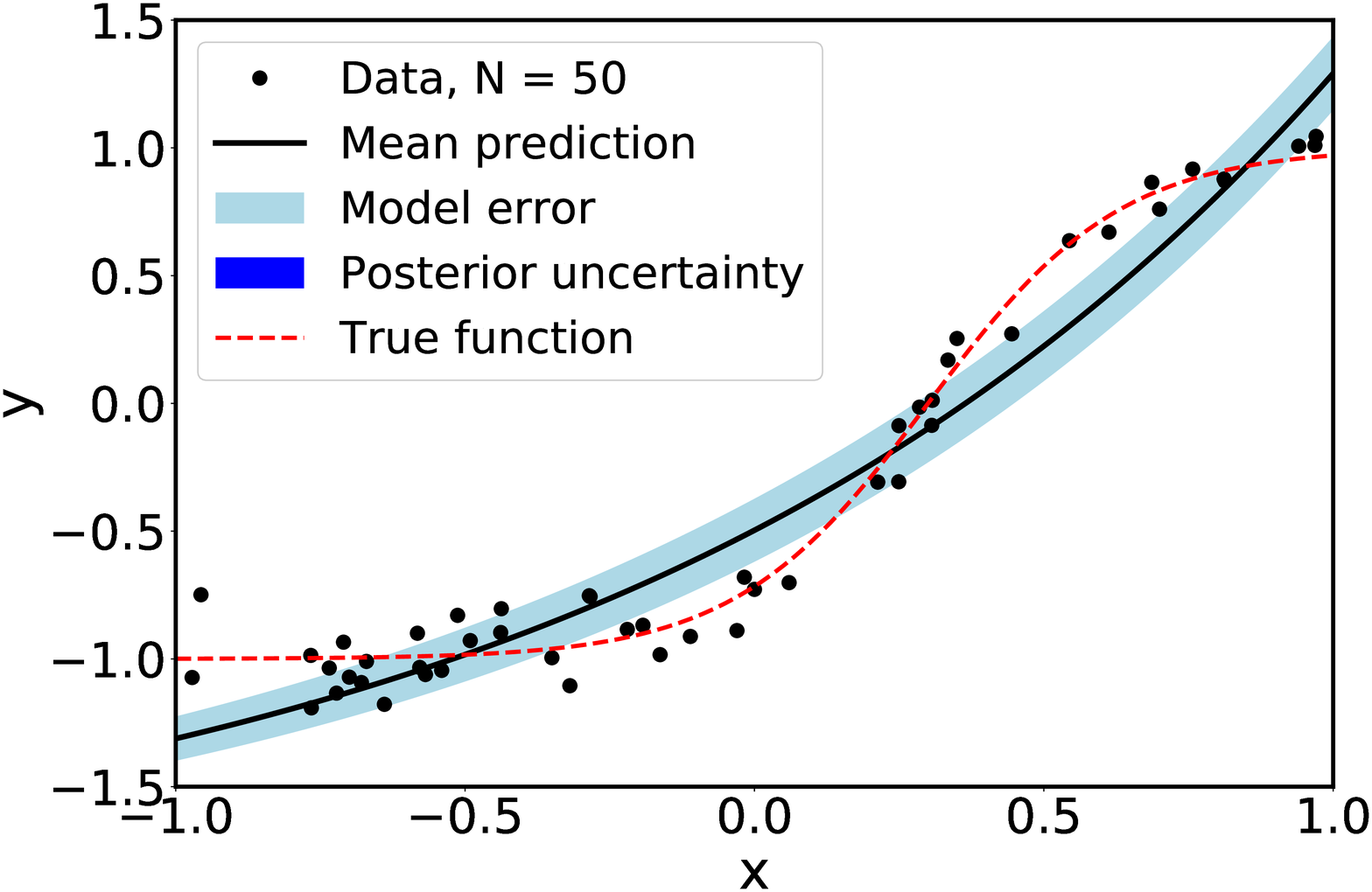}}\\
\subfigure[ABC Posterior]{ \includegraphics[height=0.47\textwidth]{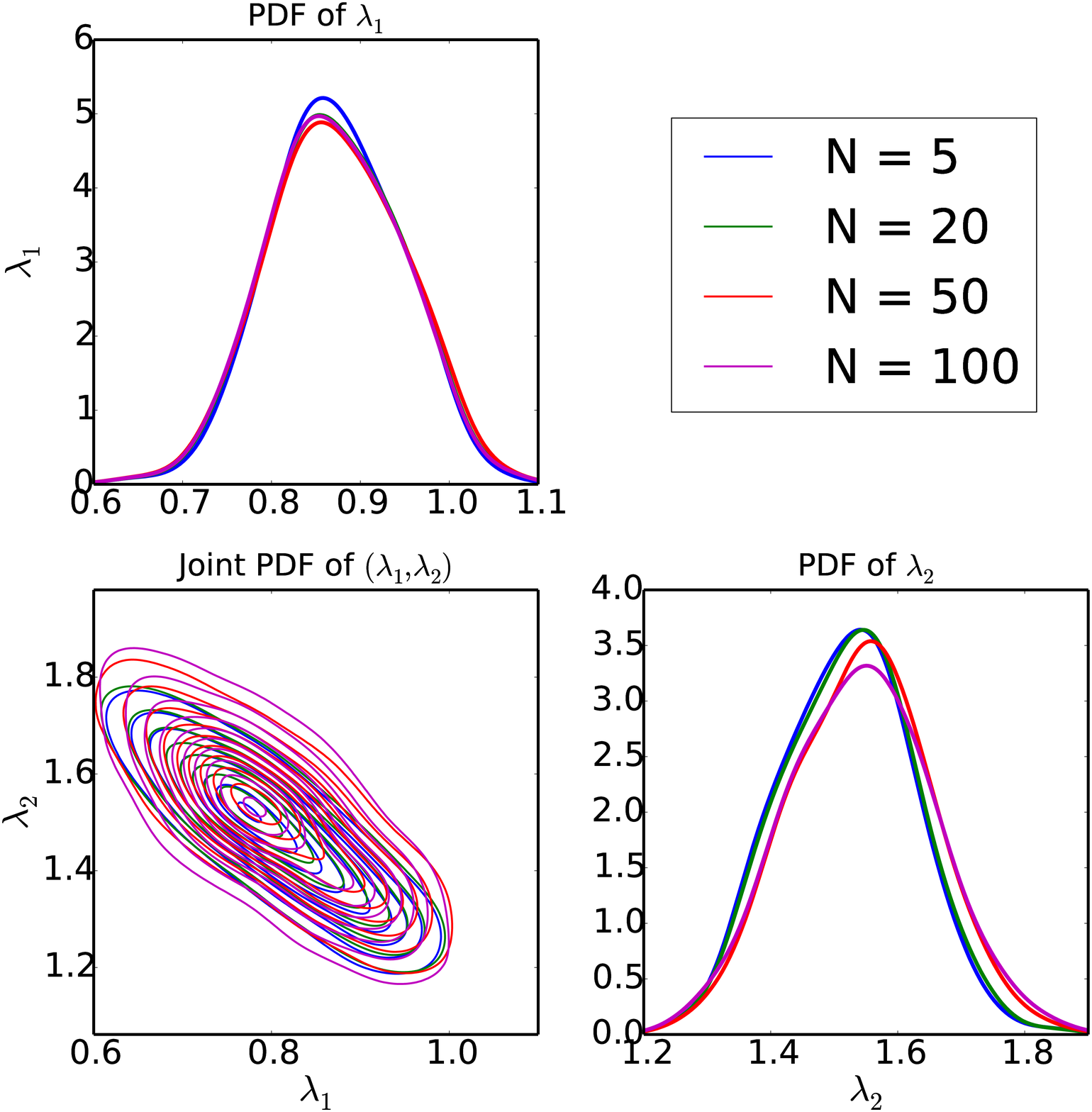}}
\subfigure[Classical Posterior]{ \includegraphics[height=0.47\textwidth]{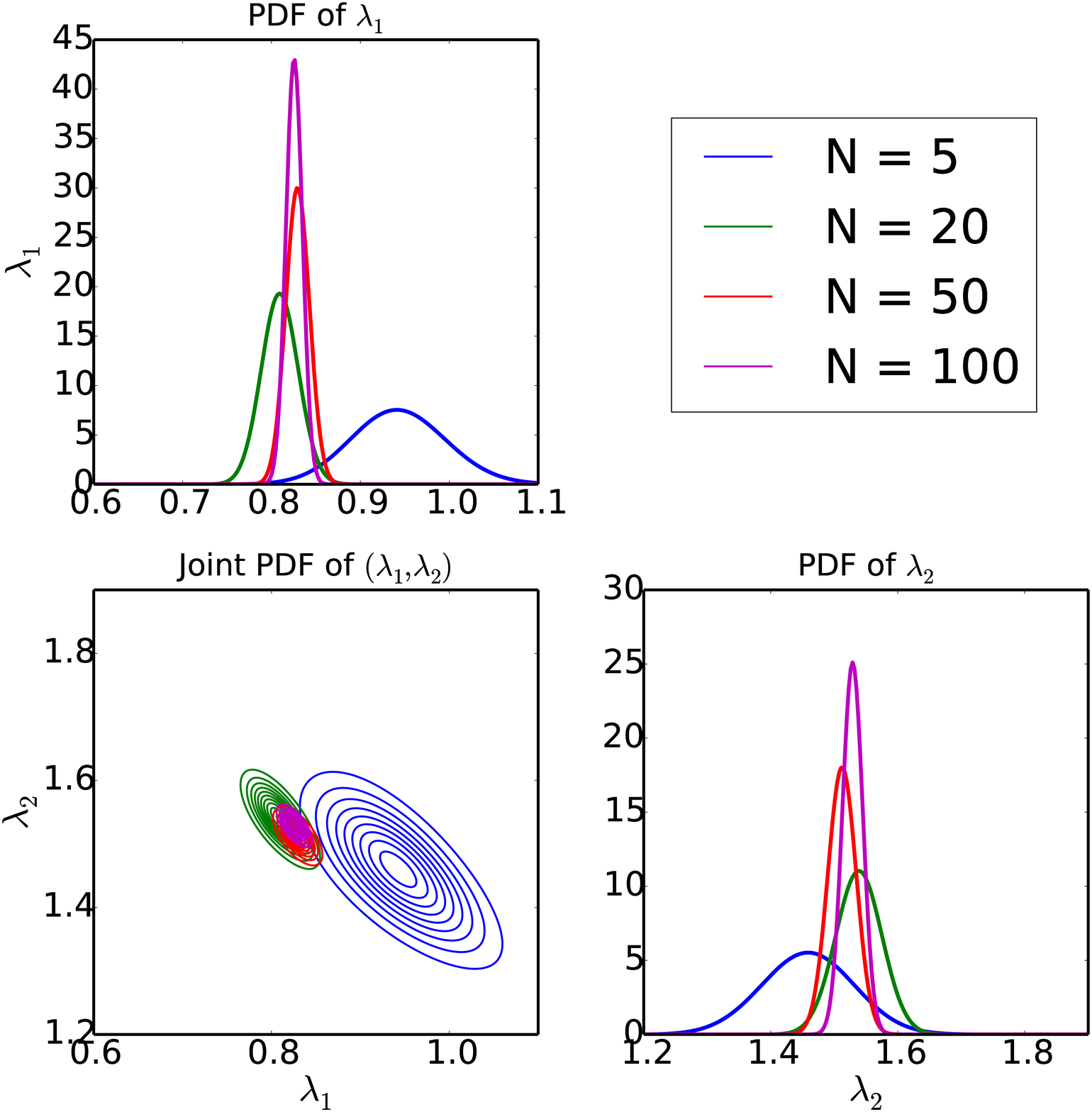}}}
\caption{\label{fig:ex1_new} Illustration of the pushed-forward predictions with ABC likelihood, as well as comparison of the posterior PDFs as $N$ grows for classical case and the case with model error augmentation with ABC likelihood. \reb{The uncertainty bands in the first plot correspond to
one PF standard deviation from the PF means, as defined in Eqs.~\eqref{eq:mpf} and~\eqref{eq:cpf}. Note that the posterior uncertainty component is small enough to be invisible on the given scale in the first subplot.}
} \end{figure}

\paragraph{Demo 1 (consistent predictive errorbars)}
Let us revisit the motivational example from Section~\ref{sec:merr}.
That is, we calibrate the model $f(x;\vlam)=\lambda_2 e^{\lambda_1x}-2$, with noisy data measured \reb{at uniformly random $x$-locations} from an
underlying `truth' model $g(x)=\tanh 3(x-0.3)$. As Figure~\ref{fig:ex1}
highlighted, the conventional calibration approach results
in posterior PF uncertainty which is not
representative of the correct discrepancy between the
calibrated model and the underlying `truth' model. Now, we employ the embedded model error calibration as follows.
Consider a `triangular' MVN additive term for the parameter
$\vlam=(\lambda_1,\lambda_2)$ as in Eq.~\eqref{eq:inpct}, and employ
the ABC likelihood \reb{\eqref{eq:abc2}} with $\gamma=1$ and
$\epsilon=0.0001$. The choice of $\gamma$ specifies the requirement that,
on average, the $1\sigma$ range of the predictive uncertainty is equal to
$\gamma\times$ the \reb{absolute} discrepancy between the data and the mean prediction. The choice of
$\epsilon \ll 1$ specifies the stringency with which the ABC kernel density
enforces the desired constraints on the predictive mean and standard deviation. \reb{The data noise magnitude is assumed known and fixed as $\sigma=0.1$ both in the likelihood and in data generation.}
Figure~\ref{fig:ex1_new}(a) clearly demonstrates
that the resulting pushed-forward prediction errorbars are representative of the
true discrepancy, unlike the classical case, shown in the $N=50$ subplot
of Figure~\ref{fig:ex1}\footnote{\reb{In this figure, as well as in several others later in the text, the multi-color bands are plotted in the following way: if the total variance is $v=v_1+v_2$, we plot $\sqrt{v_2}$ as the half-width of the first band, and $\sqrt{v_1+v_2}$ as the half-width of the second band.}}. \reb{Furthermore}, Figures~\ref{fig:ex1_new}(b) and
(c) demonstrate posterior predictive PDFs of
$\vlam=(\lambda_1,\lambda_2)$ for embedded and
classical cases, respectively, as the number of data points
increases. \reb{Note that in the embedded case $\vlam$ is effectively cast as a random variable $\vLam$, and the plotted PDF is the posterior predictive of $\vLam$, induced by both posterior \emph{and} variability of $\vLam$ due to $\vxi$.} It can be seen that model-error embedding guards against
posterior shrinkage.

\begin{figure}[!t]
\vspace*{-6mm}
\centerline{
\includegraphics[height=50mm]{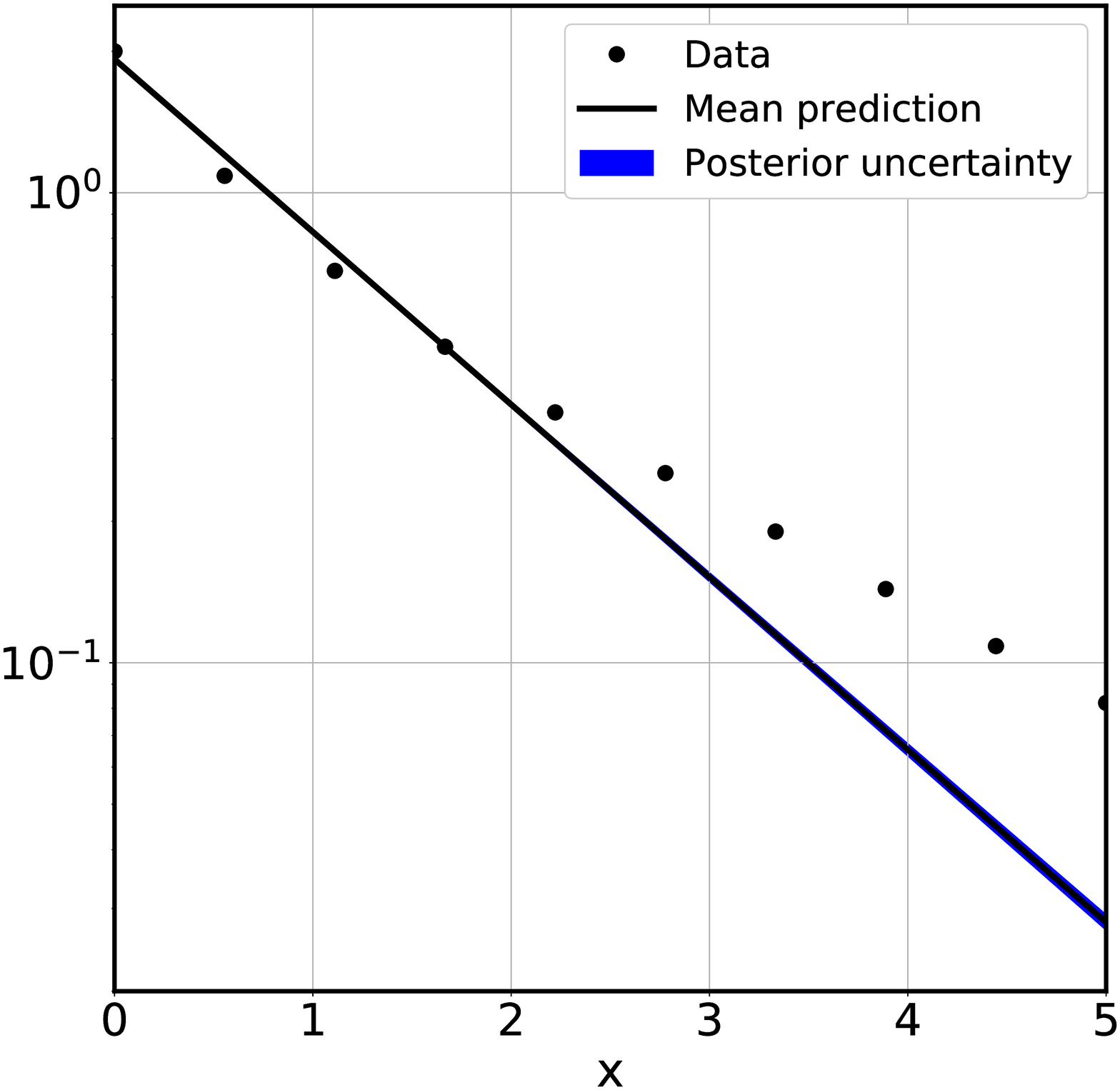}\hspace*{-0.5cm}\hfill
\includegraphics[height=50mm]{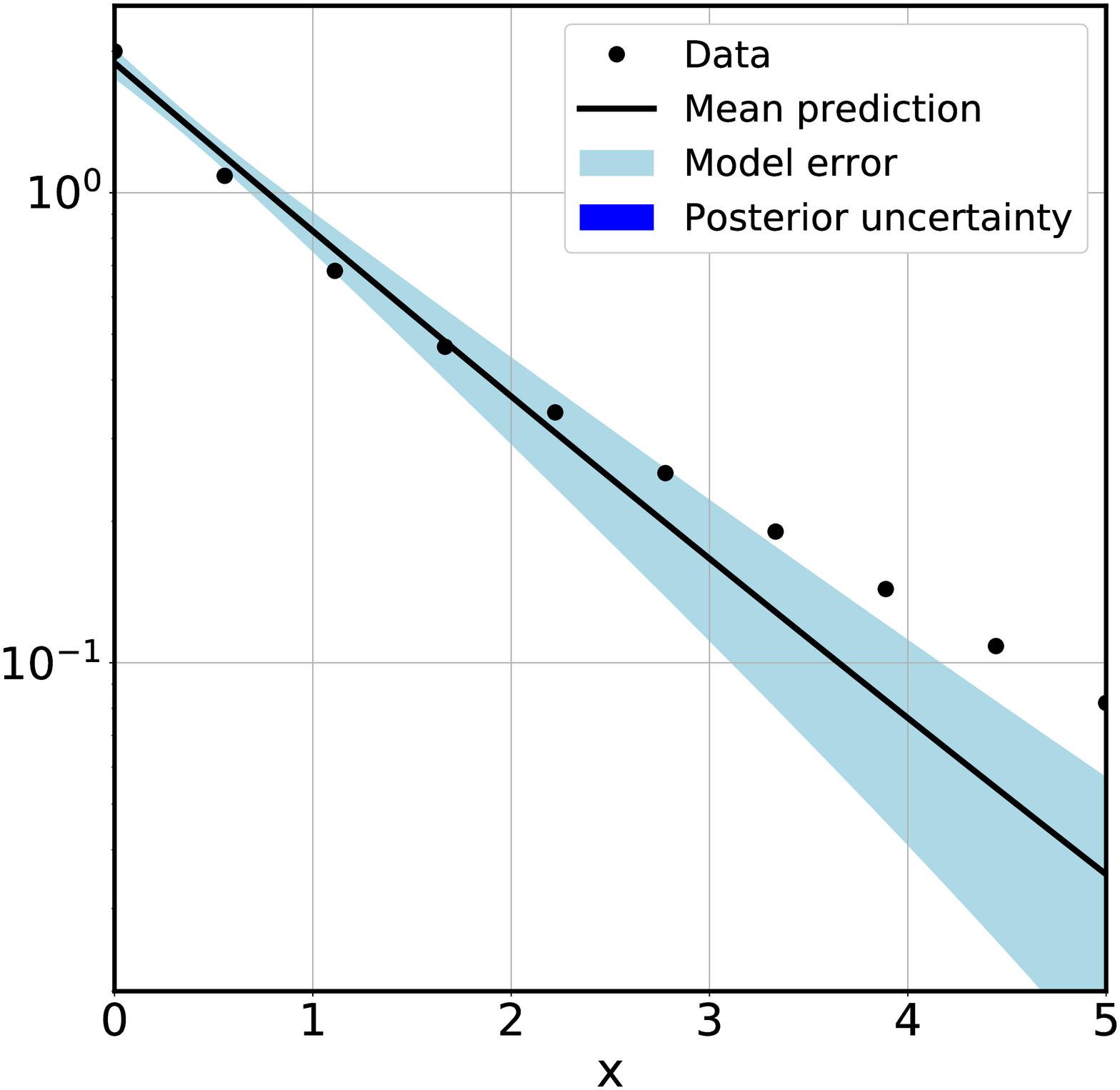}\hspace*{-0.5cm}\hfill
\includegraphics[height=50mm]{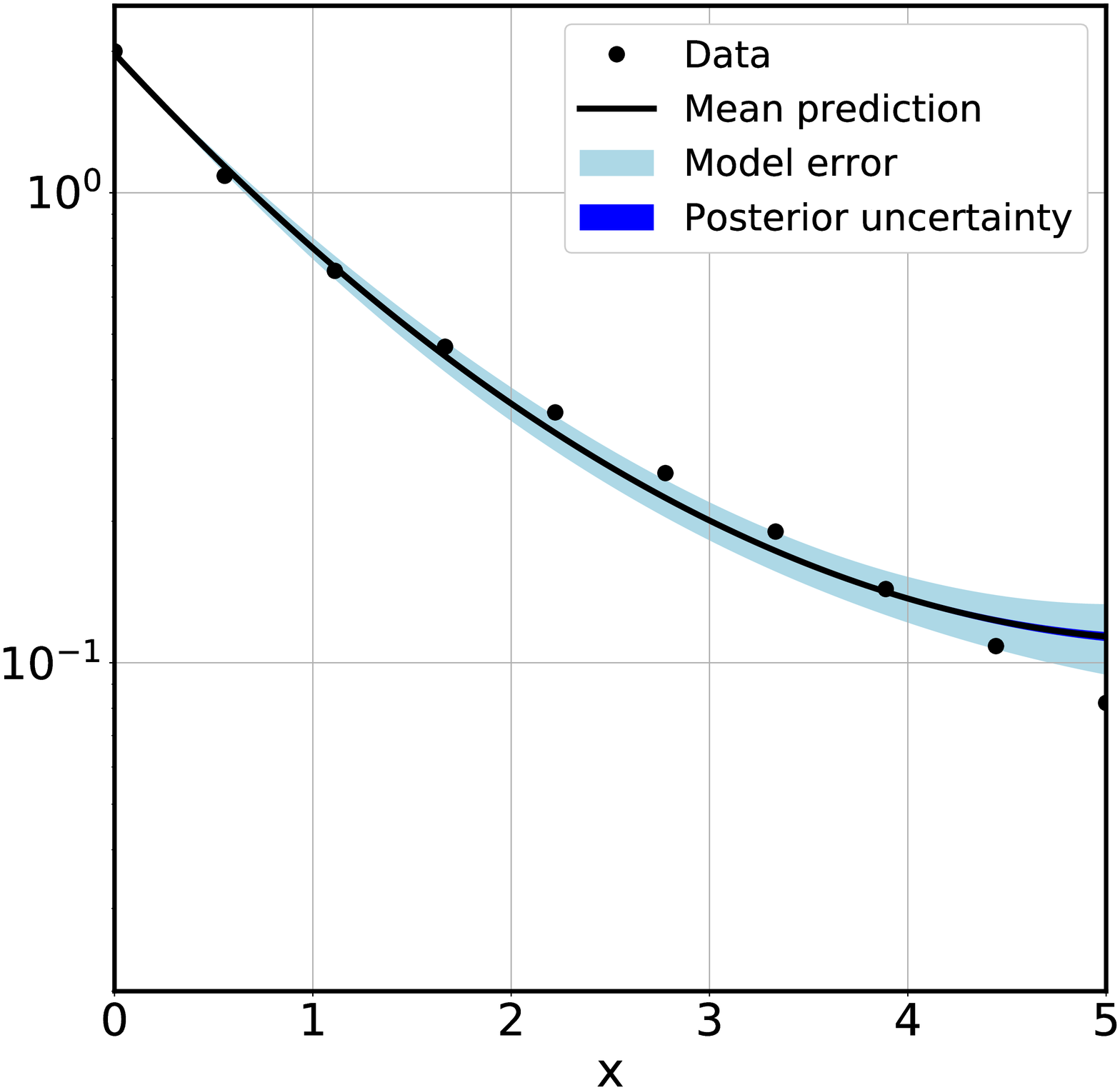}
}
\vspace*{-4mm}
\caption{\label{fig:fitexp} Calibration of a single-exponent model
$f(x;\vlam)=e^{-(\lambda_1+\lambda_2x)}$ with additive Gaussian noise
(left) fails to predict the truth model $g(x)=e^{-0.5x}+e^{-2x}$, while the
calibration with embedded model error leads to uncertain predictions that
capture the truth model (middle). A quadratic-exponential model
$f_2(x;\vlam)=e^{-(\lambda_1+\lambda_2x+\lambda_3x^2)}$ captures the truth
better with smaller uncertainties (right). The uncertainty bands correspond to
one PF standard deviation from the PF means, as defined in
Eqs.~\eqref{eq:mpf} and~\eqref{eq:cpf}. \reb{Note that the posterior
  uncertainty component is relatively small, and barely visible on the given scale.}
}
\end{figure}

\paragraph{Demo 2 (targeted model improvement)}
For another illustration, consider a `truth' model $g(x)=e^{-0.5x}+e^{-2x}$ mimicking
a physical system that involves two decay scales. Assuming no measurement noise, the
data at $10$ equidistant locations in $x\in[0,5]$ come from this model
exactly. \reb{Note that unlike the other demonstrations, in this case we have set both $x$- and $y$-values of the data to be noiseless.} Now, consider a single-decay model
$f(x;\vlam)=e^{-(\lambda_1+\lambda_2x)}$, as if the analyst is unaware of the
second decay source. A classical calibration for $\vlam=(\lambda_1,\lambda_2)$ leads to
wrong values for the decay rate, and fails to capture the double-exponential
dynamics as shown in the left plot of Figure~\ref{fig:fitexp} -- clearly the uncertain
prediction is not consistent with the range of discrepancy from the
truth. \reb{We used a conventional, Gaussian \iid noise assumption with a fixed standard deviation $\sigma=0.01$.}
On the other hand, augmenting with an MVN discrepancy term to $\vlam$ and calibrating
for the best values of the MVN parameters
$\valpha$, via the ABC likelihood~\reb{(\ref{eq:abc2})}, leads
to an uncertain model prediction that captures the discrepancy from the
data better (middle plot). \reb{The ABC likelihood parameters are set to $\epsilon=0.001$ and $\gamma=1$, while we used $3$-rd order NISP for uncertainty propagation within the likelihood computation.} One can envision a next step, \emph{e.g.}
embed more complicated `physics', albeit still not the true double-exponential.
For example, one can keep the single-exponential description and propose a
quadratic-exponential model,
$f_2(x;\vlam)=e^{-(\lambda_1+\lambda_2x+\lambda_3x^2)}$ with an additional parameter $\lambda_3$. As the right plot
suggests, this allows better predictions with smaller model errors.
The present approach allows such embedding of model error terms in targeted model components, \reb{at the same time facilitating} the evaluation of the resulting uncertain predictions for model comparison/selection studies.

\begin{figure}[!t]
  \vspace{0pt}
  \centering
  \includegraphics[width=5.0in]{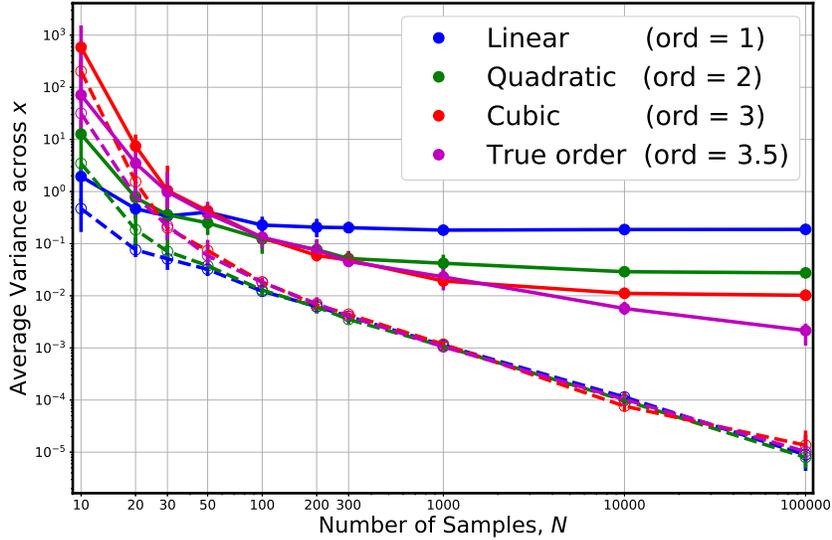}
  \caption{Demonstration of the decay of predictive variance components as the number of data points increases. The solid lines \reb{and filled markers} correspond
  to model error, while the dashed lines \reb{and hollow markers} correspond \reb{to posterior uncertainty}, according to the pushed-forward variance decomposition in~\eqref{eq:predvar}.
  Besides the true fractional-power model, three simple models with varying
  degree of accuracy are used, clearly demonstrating the saturation of the model error
component to a limiting value, while the \reb{pushed-forward posterior uncertainty} due to $N$ keeps reducing. The points in the plot indicate \reb{averaged variances} across all values of $x_i$, and $100$ replica simulations \reb{(with newly sampled data sets)} have been performed, reporting only median
results with 25\%-75\% quantiles.
}
  \label{fig:varconv}
 \end{figure}

\paragraph{Demo 3 (convergence with increasing $N$)}
\reb{This exercise} demonstrates in more detail how the data error reduces with increasing $N$ while the
model error term does not. This
illustration uses a fractional-power polynomial `truth' function $g(x)=6+x^2-\reb{0.5}(x+1)^{3.5}$ and observations
that are \reb{sampled as uniformly random in $x\in[-1,1]$ and} corrupted \reb{in $y$} by a Gaussian, \emph{i.i.d.} noise term of standard deviation $\sigma=0.5$. \reb{This data noise magnitude is assumed to be known and is also included in the likelihood computation.} Four simple
models are calibrated with respect to their parameter vector $\vlam$, (a) linear, $f(x;\vlam)=\lambda_0+\lambda_1 x
$, (b) quadratic, $f(x;\vlam)=\lambda_0+\lambda_1 x+\lambda_2 x^2$,
(c) cubic, $f(x;\vlam)=\lambda_0+\lambda_1 x+\lambda_2 x^2 + \lambda_3 x^3$, and
(d) true \reb{order}, $f(x;\vlam)=\lambda_0+\lambda_1 x+\lambda_2x^2+\lambda_3 (x+1)^{3.5}$. \reb{In this demonstration an independent normal form is used for the likelihood as given in Eq.~\eqref{eq:in}. Note that normality here is not an approximation, since the MVN embedding is used and the models are linear in $\vlam$. For the same reason, a first-order \emph{exact} NISP is used. Also, we used this form in place of ABC to avoid having to tune an extra parameter $\epsilon$ that can affect posterior width.}
\reb{As Figure~\ref{fig:varconv} shows,} while the \reb{posterior uncertainty} component (as given in the decomposition~\eqref{eq:predvar})
reduces with increasing amount of data, the model error `saturates' at
a limiting value driven by the quality of the model, for all the approximate models.

\begin{figure}[!tb]
\centering{
\subfigure[]{ \includegraphics[width=0.48\textwidth]{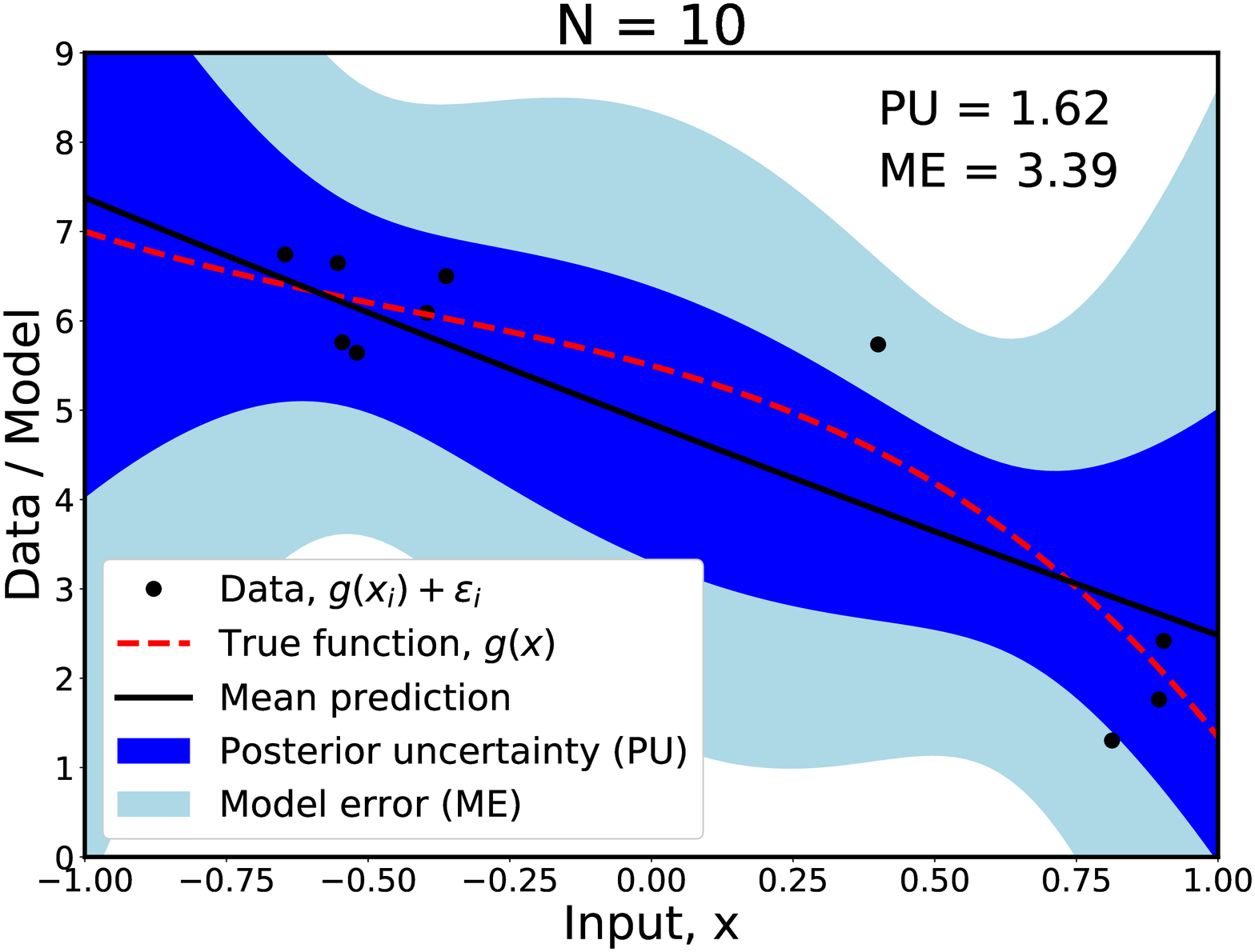}}
\subfigure[]{ \includegraphics[width=0.48\textwidth]{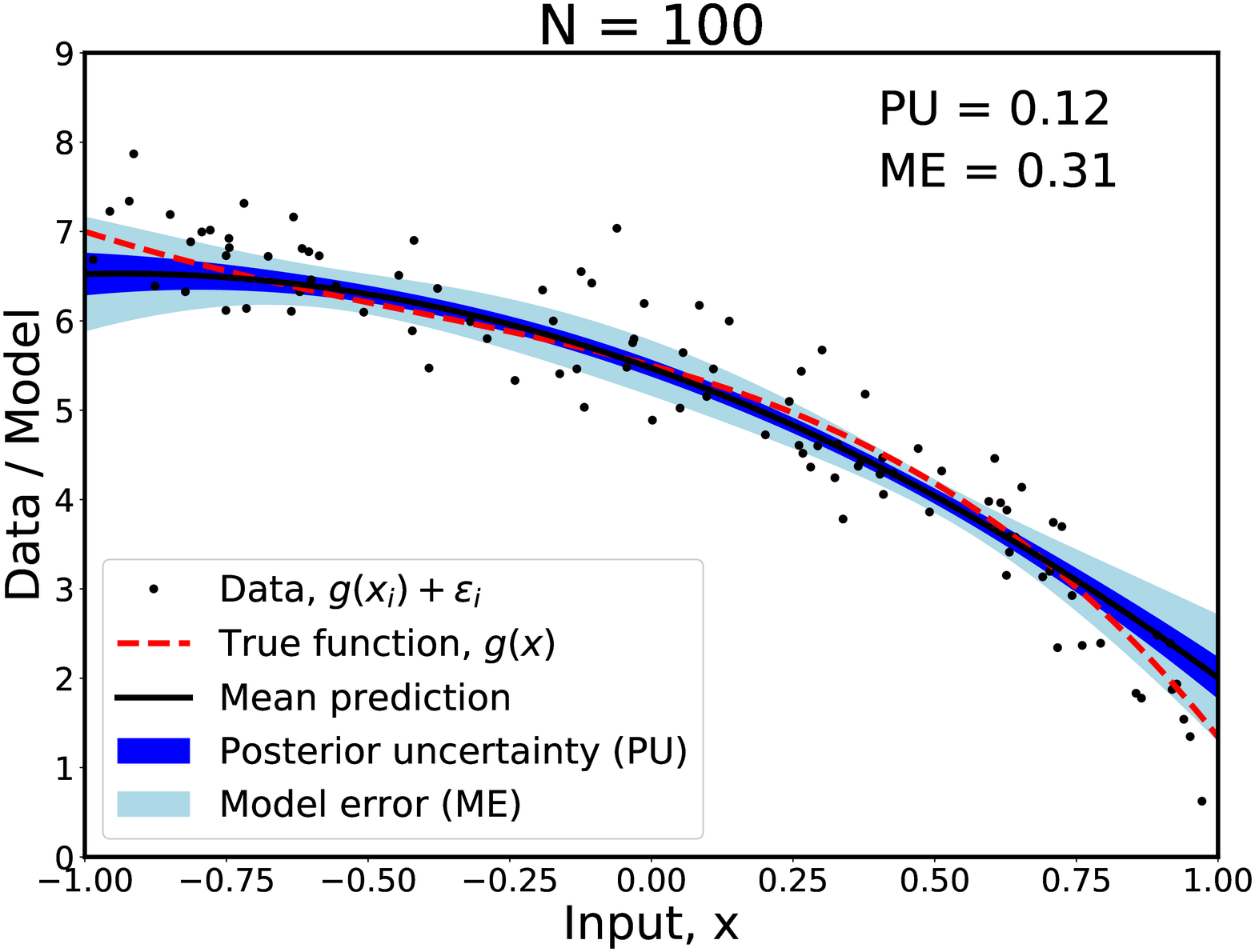}}\\
\subfigure[]{ \includegraphics[width=0.48\textwidth]{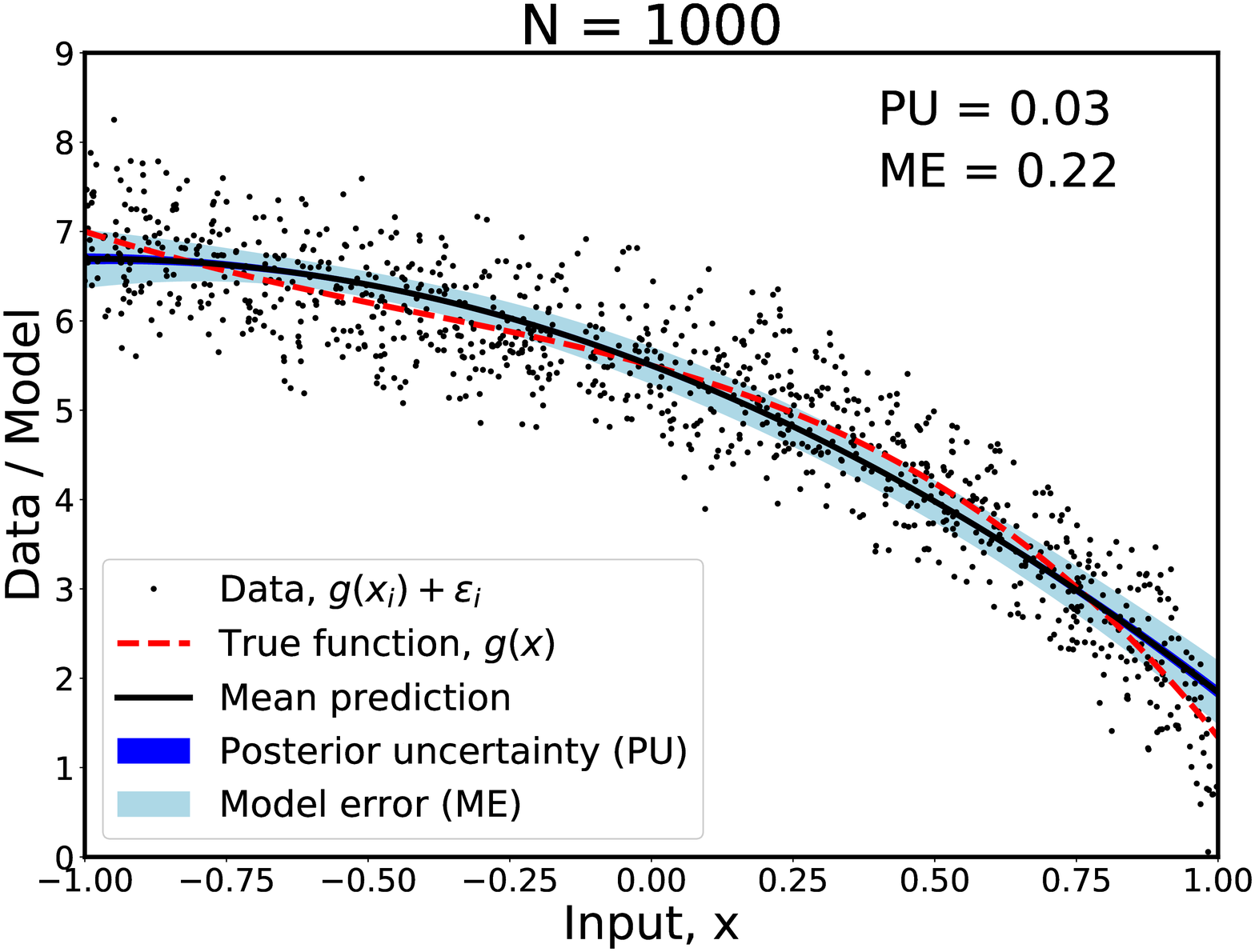}}
\subfigure[]{ \includegraphics[width=0.48\textwidth]{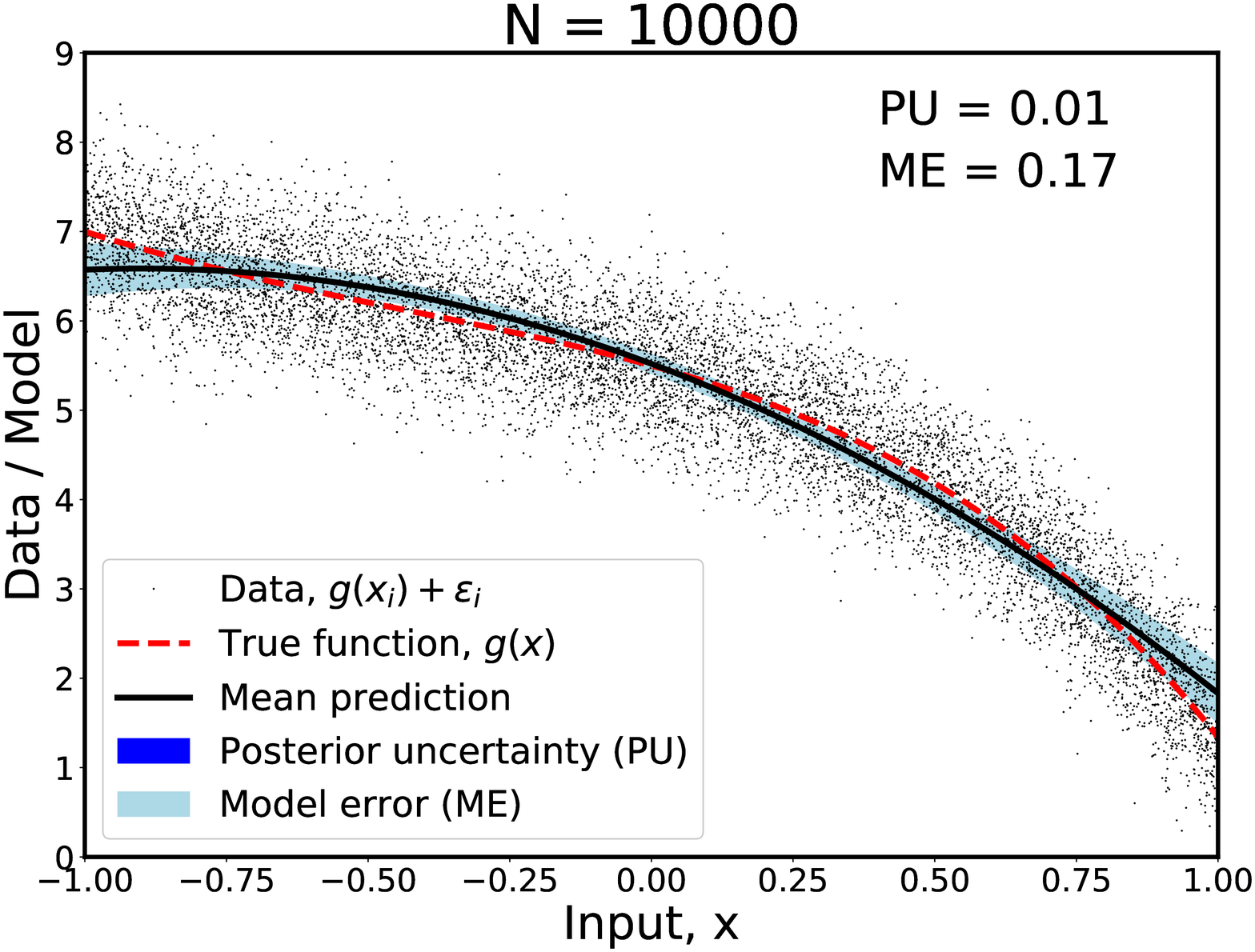}}}
\caption{\label{fig:fits} Demonstration of the pushed-forward predictions given noisy observations of a `truth'
function $g(x)=6+x^2-\reb{0.5}(x+1)^{3.5}$ being calibrated by a quadratic model $f(x;\vlam)=\lambda_0+\lambda_1 x+
\lambda_2 x^2$ via MVN model error embedding in all parameters $\vlam$. The
four frames correspond to the four $N$ values
$(10,10^2,10^3,10^4)$, as indicated, respectively. The value of model error component (light blue region) of
the pushed-forward variance decomposition in~\eqref{eq:predvar} \reb{starts saturating} after
enough data is collected, while posterior uncertainty keeps reducing.
\reb{The posterior uncertainty (PU) and model error (ME) labels in the plots indicate the square root of average variance across $x$}. The actual average fit variances correspond to the green convergence curves for the quadratic case in Figure~\ref{fig:varconv}.
} \end{figure}

Figure~\ref{fig:fits} illustrates this effect further for the quadratic
model case, with four different values of $N$, demonstrating the fact that the model error component of the
variance remains relatively constant with increasing $N$.

\section{Ignition model} \label{sec:ign}

\begin{figure}[!tp]
  \vspace{0pt}
  \centering
  \includegraphics[width=4.0in]{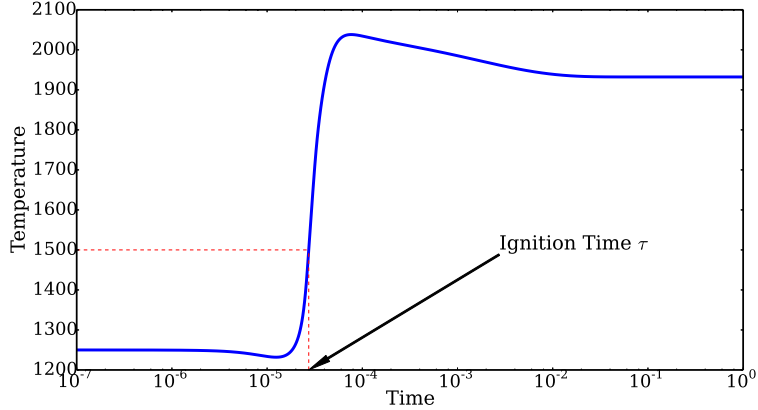}
  \caption{An example of temperature profile and ignition time definition.}
  \label{fig:ign}
 \end{figure}

This section illustrates the method on an example problem taken from chemical kinetics. Consider the chemical model with the classical two-step mechanism described by \cite{Westbrook:1981}, which accounts for the incomplete oxidation of n-dodecane
\begin{eqnarray}
\textrm{C}_{12}\textrm{H}_{26} + \tfrac{25}{2} \textrm{O}_2 &\stackrel{k_1}{\rightarrow}& 12 \textrm{CO} + 13 \textrm{H}_2\textrm{O}
\label{eq:2step_mech1}\\
\textrm{CO} + \tfrac{1}{2} \textrm{O}_2 &\underset{k_{2b}}{\stackrel{k_{2f}}{\rightleftharpoons}}& \textrm{CO}_2 \mbox{.}
\label{eq:2step_mech2}
\end{eqnarray}
The reaction rates are modeled using Arrhenius laws, with the parameters taken from~\cite{Westbrook:1981,Dryer:1973}
\begin{eqnarray}
k_1 & = &  A e^{\left(-\frac{E}{RT}\right)}[\textrm{C}_{12}\textrm{H}_{26}]^{0.25}[\textrm{O}_2]^{1.25}\\
k_{2f} & = & 3.98 \times 10^{14} e^{\left(-\frac{40000}{RT}\right)}[\textrm{CO}][\textrm{H}_2\textrm{O}]^{0.5}[\textrm{O}_2]^{0.25}\\
k_{2b} & = & 5 \times 10^{8} e^{\left(-\frac{40000}{RT}\right)}[\textrm{CO}_2] \mbox{.}
\label{eq:2step_rates}
\end{eqnarray}
where $R$ is the ideal gas constant, $T$ is the temperature in Kelvin, and $[X]$ the molar concentration of species $X$ in units of mol/cm$^{3}$. The units
of time and energy are sec and cal, respectively. Rate $k_1$ is the
rate of progress of fuel oxidation. The pre-exponential factor $A$ is parameterized as
\be
\ln A = a_1 + a_2e^{a_3\phi} + a_4\tanh((a_5+a_6\phi)T_0+a_7),
\label{eq:tanh7}
\ee
in which it is modeled as a function of initial temperature, $T_0$, and equivalence ratio, $\phi$.
Such a special form allows accounting for the desired behavior of the autoignition
delay time in ($T_0$, $\phi$) space; e.g., in the Negative Temperature
Coefficient (NTC) region~\cite{Leenson:1984}.
Varying the Arrhenius parameters with mixture state is a well
known technique that has been applied elsewhere with good success~\cite{Fernandez-Tarrazo:2006,Franzelli:2010,Misdariis:2015}.
Together with the activation energy $E$, the parameter $a_1$ is of interest, \ie $\vlam=(E,a_1)$, while for the purposes of the demonstrations in this work, we fix the rest of the parameters at their nominal values $(a_2,a_3,a_4,a_5,a_6,a_7)=(-2.13,-2.05,1.89,-0.01,2.87\cdot 10^{-4},8.43)$, found in~\cite{Hakim:2016a}. The operating conditions
for a given ignition simulation are the pressure (in atm), the equivalence
ratio, and the scaled inverse of the initial temperature. In terms of the notation in this work, the operating conditions are $x=(P,\phi,1000/T_0)$, and the model output of
interest $f(x;\vlam)$ is the logarithm of the ignition time, $\ln(\tau)$. The ignition time is defined as the time when the temperature reaches $1500$ K, as demonstrated in Figure~\ref{fig:ign}. The simulations of the 2-step mechanism are performed using TChem~\cite{Safta:2011}, a software toolkit for the analysis of complex kinetic models.

We demonstrate calibration of this 2-step model in two case studies: (a) experimental shock-tube data from~\cite{Vasu:2009}, and (b) simulation data generated from the skeletal mechanism by \cite{Narayanaswamy:2014} with 255 species and 2289 reactions.

\subsection{Calibration with experimental data} \label{sec:ign_vasu}
Here we employ experimental shock-tube ignition data from~\cite{Vasu:2009}
collected in a heated, high-pressure shock tube for two fixed
values of equivalence ratio ($\phi=0.5$ and $\phi=1.0$), over a range of
variation of the two other
operating conditions, pressure ($P$) and initial temperature ($T_0$).
As such, the logarithm of ignition time, $\ln(\tau)$, collected under varying
conditions $x=(P,\phi,1000/T_0)$, is the data generation model $g(x)$. Overall, there are $42$ data samples, or quartets, $\{(P,\phi,1000/T_0)_i, \ln(\tau_i)\}_{i=1}^{42}$, $16$ of which correspond to $\phi=0.5$, while the other $26$ are collected under equivalence ratio $\phi=1.0$.
We note that
the two other operating conditions, $P$ and $T_0$, are measured rather than strictly enforced, and the authors in~\cite{Vasu:2009}
somewhat heuristically evaluate the overall noise in ignition time measurement as $\pm 10\%$. Besides,
the definition of ignition time in the experiments differs somewhat from the definition employed in the
model-to-be-calibrated. However, the difference is of little consequence. Given the sharp temperature increase and
overall shape of the temperature evolution as in Figure~\ref{fig:ign}, the minor
discrepancy in the definition of ignition time can essentially be regarded as data
noise. Overall, given $\ln(1.1 \tau)\approx \ln(\tau)+0.1$ and $\ln(0.9 \tau)\approx \ln(\tau)-0.1$, we note that the nominal standard deviation is $\sigma=0.1$. For the inference, we assume an \emph{i.i.d.} additive data noise model on $\ln(\tau)$, distributed as a normal random variable with vanishing mean and constant standard deviation to be inferred.

\begin{figure}[!tp]
  \vspace{0pt}
  \centering{
\subfigure[No model error]{\includegraphics[width=0.98\textwidth]{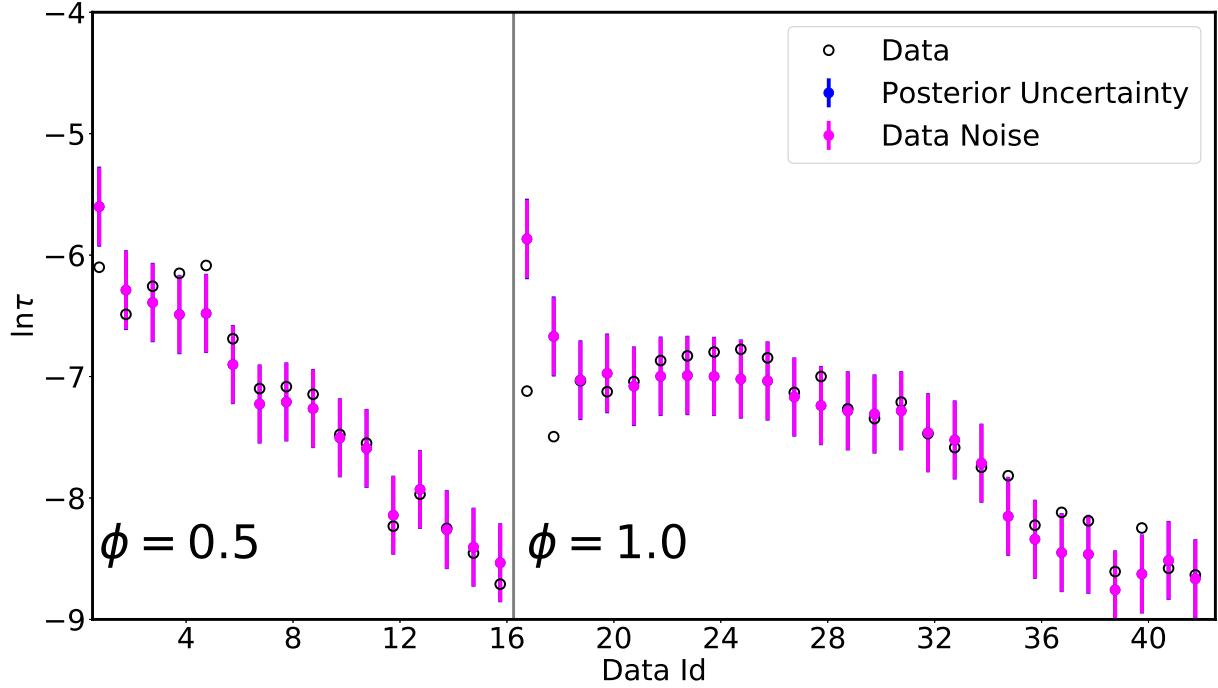}}\\

\subfigure[Model error]{  \includegraphics[width=0.98\textwidth]{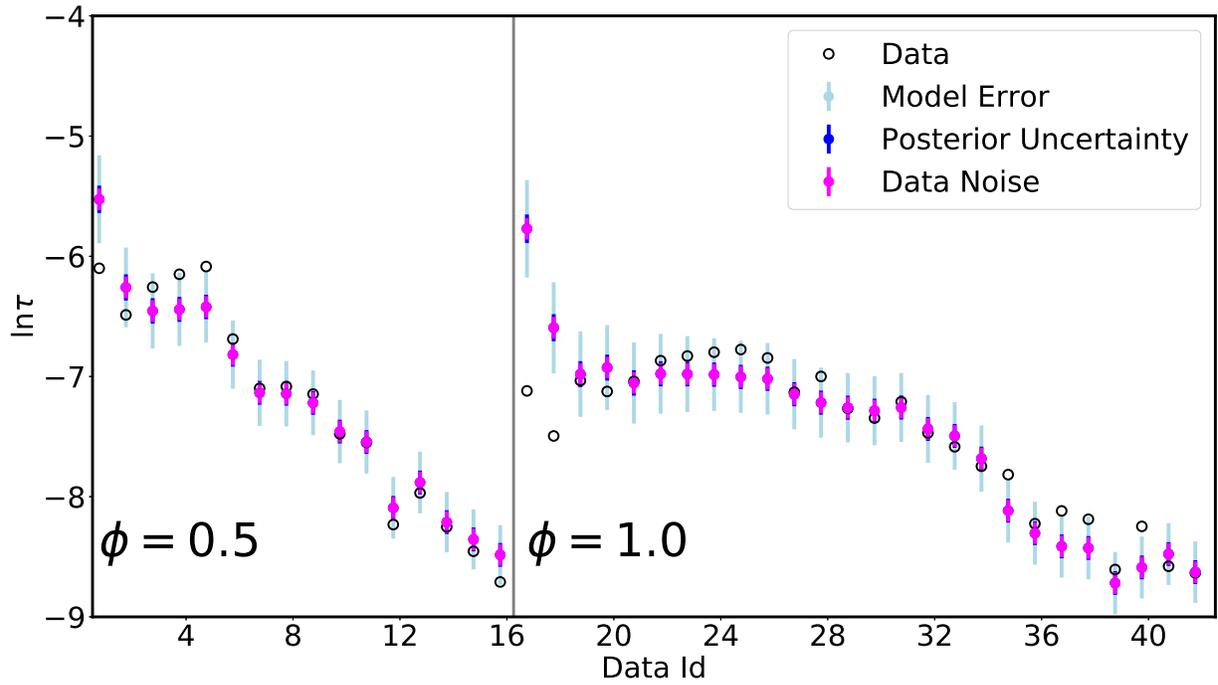}}
}
  \caption{Demonstration of uncertainty decomposition of the predictions, according to Eq.~\eqref{eq:pprd}. The classical Bayesian calibration, with
  inferred constant data noise $\sigma$ is shown in the first plot, while the results
  with the embedded model error approach are shown in the second plot. Note that in both cases, the posterior uncertainty contribution is relatively small.
  }
  \label{fig:vasucal}
 \end{figure}

\begin{figure}[!h]
\includegraphics[width=0.98\textwidth]{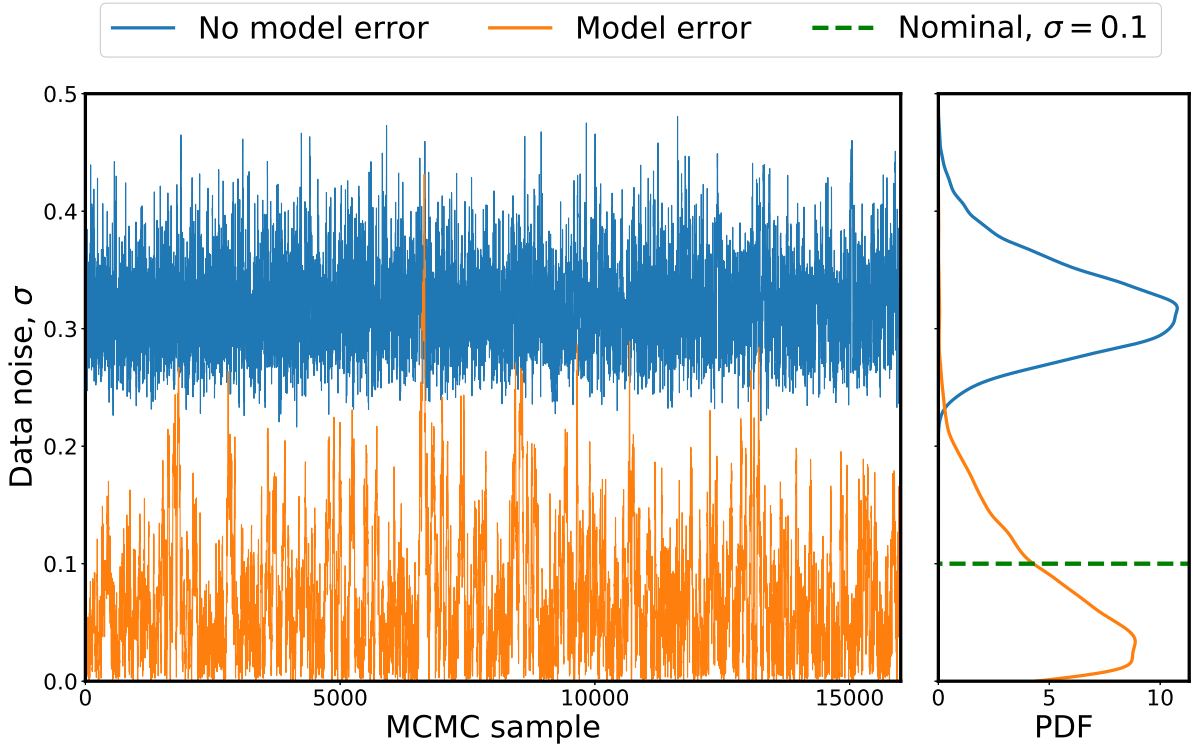}
  \caption{Demonstration of posterior samples of data noise standard deviation with or without including model error.
  }
  \label{fig:sigmapdf}
 \end{figure}

Following Algorithm~\ref{algo}, we build a polynomial surrogate for the output $\ln(\tau)$ of the
two-step mechanism~\eqref{eq:2step_mech1}-\eqref{eq:2step_mech2} at all
operating conditions $x_i$, for $i=1,\dots, N$ and with $N=42$.
The polynomial surrogate is chosen to be $3$-rd order, with the coefficients
found by regression described in Section~\ref{s:algo}, using $R=100$ input parameters $\vlam=(E,a_1)$, sampled uniformly over ranges $E\in[10000,40000]$
and $a_1\in [15,37]$. The polynomial regression allows analytical extraction of leave-one-out (LOO) error as a measure of its accuracy~\cite{Christensen:2002}, and can subsequently augment predictive uncertainty budget, as demonstrated in~\cite{Huan:2018b}.
For the examples shown, the LOO errors are on the order of $10^{-2}$, while the observable $\ln{\tau}$ varies in the range between $-11$ and $-6$, therefore we ignore surrogate error in the present discussion. Moreover, as we employ
a third-order polynomial surrogate model, the PC-based NISP propagation with third-order truncation provides an exact extraction of mean and variance of the PF predictions. The data noise is assumed constant, independent across operating conditions $i=1,\dots,42$, and is inferred together with the model parameters.
The goal of the demonstration is to highlight the ability of the model error approach to differentiate model error from data noise appropriately. In other words, we demonstrate that including embedded model error term allows estimating data noise standard deviation much more accurately, while the classical inference without accounting for model error overestimates the data standard deviation, compensating for unaccounted model error.
Figure~\ref{fig:vasucal} demonstrates the posterior predictive uncertainty that can be derived similar to Eq.~\eqref{eq:ppcov} as
\be
\sigma^2_{\textrm{PP}}(x_i)=\underbrace{\EE_\tvalpha[\sigma^f(x_i;\tvalpha)^2]}_{\textrm{model error}}+
\underbrace{\VV_\tvalpha[\mu^f(x_i;\tvalpha)]}_{\textrm{posterior uncertainty}}+\underbrace{\mathbb{E}_\sigma[\sigma^2]}_{\textrm{data noise}}.
\label{eq:pprd}
\ee
This equation is similar to the pushed forward uncertainty~\eqref{eq:predvar}, with the additional data noise term. The classical calibration, as Figure~\ref{fig:vasucal}(a) shows, captures the experimental data well \emph{on
average}, with posterior uncertainties, due to low amount of the
data, being consistent with the data. However, it includes no model error by construction, and attributes most of the posterior predictive uncertainty to data noise, thus overestimating the latter. When including model error, as shown in Figure~\ref{fig:vasucal}(b), with MVN embedding~\eqref{eq:mvn} in the
two main parameters $\vlam=(E,a_1)$ of the 2-step model, and independent-normal likelihood~\eqref{eq:in}, one achieves similar posterior predictive variance, but with an estimate of data noise size that is closer to the nominal value of $\sigma=0.1$. This more accurate estimation of $\sigma$ is highlighted in Figure~\ref{fig:sigmapdf}, in which the MCMC samples from both cases are shown, together with their respective PDFs and the nominal value of $\sigma$. \reb{Note that the nominal value itself is a heuristic estimate from the processed data~\cite{Vasu:2009}, and the posterior PDF by no means is expected to peak at the nominal value: the best one could hope for is a well-identified posterior PDF that includes the nominal value with a realistic probability.}

\subsection{Calibration with complex-model data}
Consider a more complex mechanism (albeit still a ``skeletal" mechanism for this fuel) with 255 species and 2289 reactions defined in
\cite{Narayanaswamy:2014}. We simulate ignition with this mechanism, using TChem~\cite{Safta:2011}, at arbitrary operating conditions. We select a uniform
$11\times 11\times 11$ grid in the $x=(P,\phi,1000/T_0)$ space varying these
conditions in the ranges $x\in[11,61]$~atm, $\phi\in[0.5,2.5]$,
$1000/T_0\in[0.8,1.3]$~K$^{-1}$.  The logarithm of the ignition time for this
mechanism is the data generation model $g(x)$ in the notation of this
work. The ignition time is defined as the time when temperature reaches $1500$~K. As in the previous case, the model-to-be-calibrated is the two-step mechanism from~\eqref{eq:2step_mech1}-\eqref{eq:2step_mech2}, for which we prebuild $3$-rd order polynomial surrogates for all $N=1331$ operating conditions, using $R=100$ random parameter samples of $\vlam=(E,a_1)$, sampled from ranges $E\in[10000,40000]$ and $a_1\in [15,37]$. Once again, the PC-based NISP propagation provides exact values for the PF moments. Unlike the experimental data from the previous case, the complex-model simulation produces data with no noise, and the discrepancy between the two-step mechanism and the complex model is purely due to model error. As such, the main assumption underlying the classical calibration, i.e. the \emph{i.i.d.} nature of data noise is unfounded. Still, we proceed with the classical calibration for demonstration, assuming constant data noise $\sigma$ that is inferred together with the model parameters $\vlam=(E,a_1)$. What we have found is that the posterior variance $\mathbb{V}_\vlam[f(x;\vlam)]$, being the only contributor to the PF uncertainty, is negligibly small due to abundance of data for calibration.

\begin{figure}[!tp]
  \vspace{0pt}
      \includegraphics[width=0.98\textwidth]{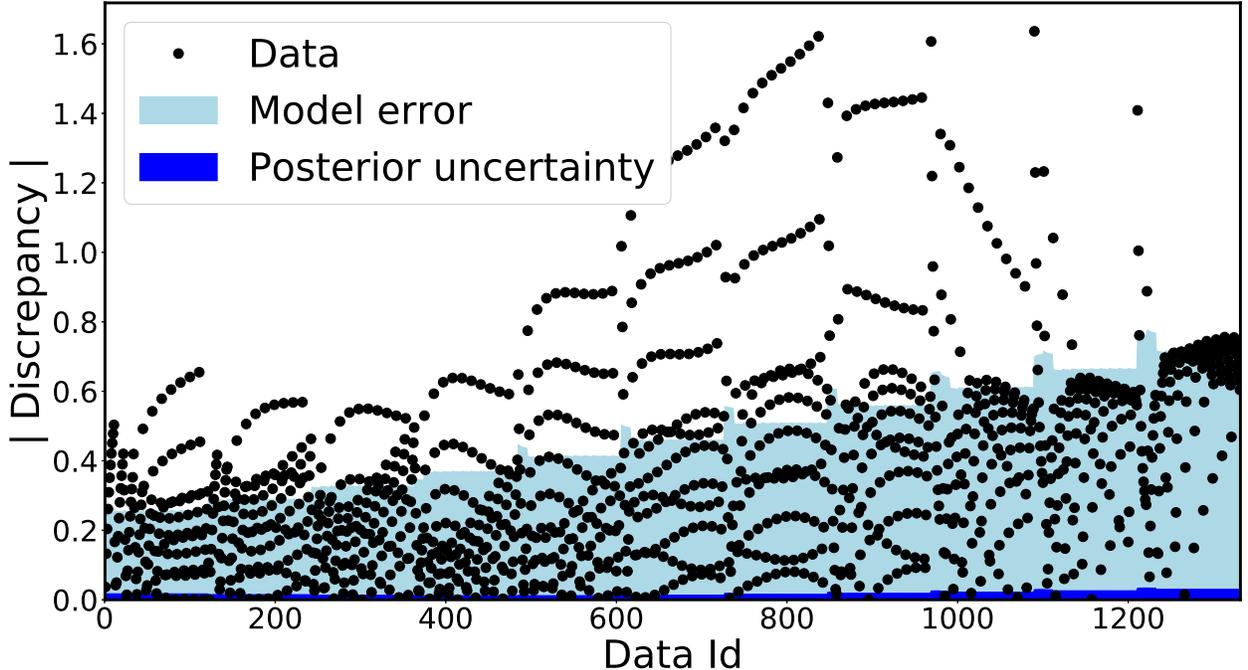}
  \caption{Demonstration of the 2-step mechanism $f(x,\vlam)$ fit to simulation data from the detailed mechanism
  $g(x)$, across $1331$ operating conditions for varying $x=(P,\phi,1000/T_0)$ on a uniform $11^3$ grid. The data points correspond to the actual discrepancy between complex-model data and two-step mechanism mean predictions, while the colored cloud demonstrates the PF standard deviations according to the variance decomposition~\reb{\eqref{eq:pprd}}. The MVN-embedded model-error calibration leads to PF uncertainties that are on average representative of the
  actual model discrepancies, unlike the classical calibration (results not shown), which includes negligibly small PF uncertainty.}

  \label{fig:naracal}
   \end{figure}

\begin{figure}[!bp]
  \vspace{0pt}
  \includegraphics[width=2.5in]{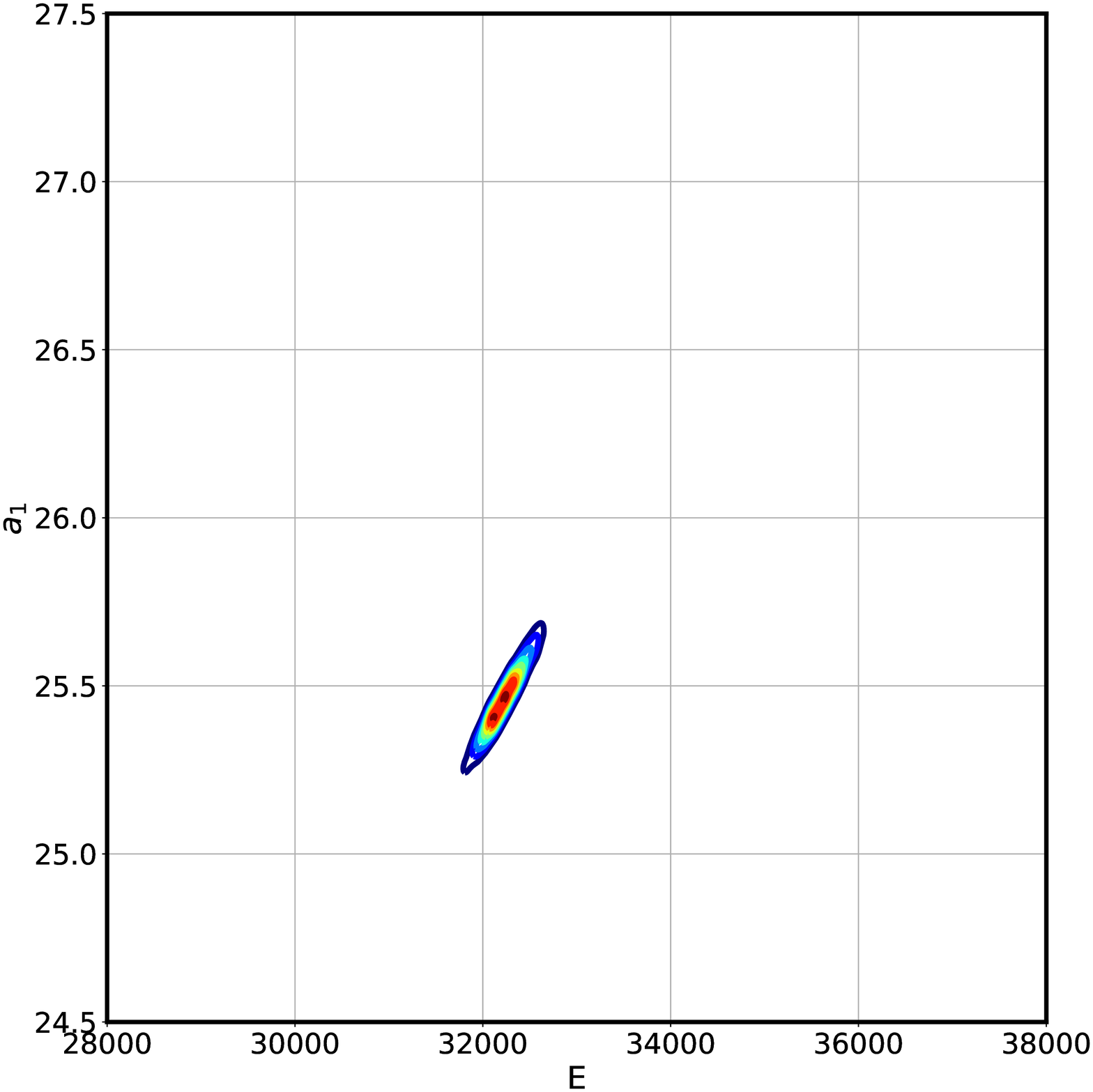}\hfill
  \includegraphics[width=2.5in]{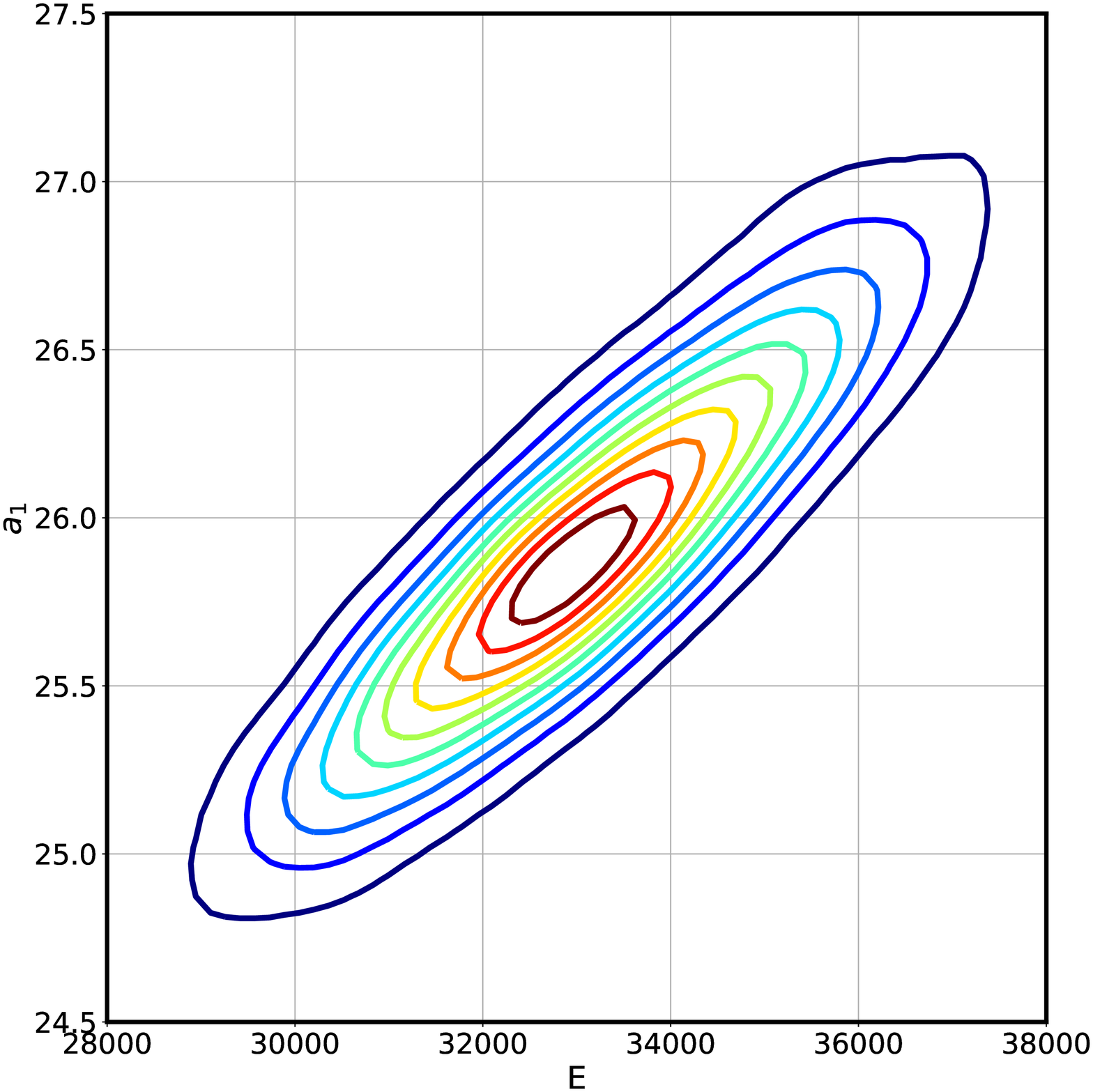}
  \caption{Joint posterior PDFs for $(E,a_1)$ for (left) classical calibration,
  and (right) model-error embedded calibration. The classical approach leads to overconfident posterior PDFs as predictive
  errorbars have shown in Figure~\ref{fig:naracal}.}

  \label{fig:Ea}
 \end{figure}

\begin{figure}[!h]
  \vspace{0pt}
  \includegraphics[width=0.45\textwidth]{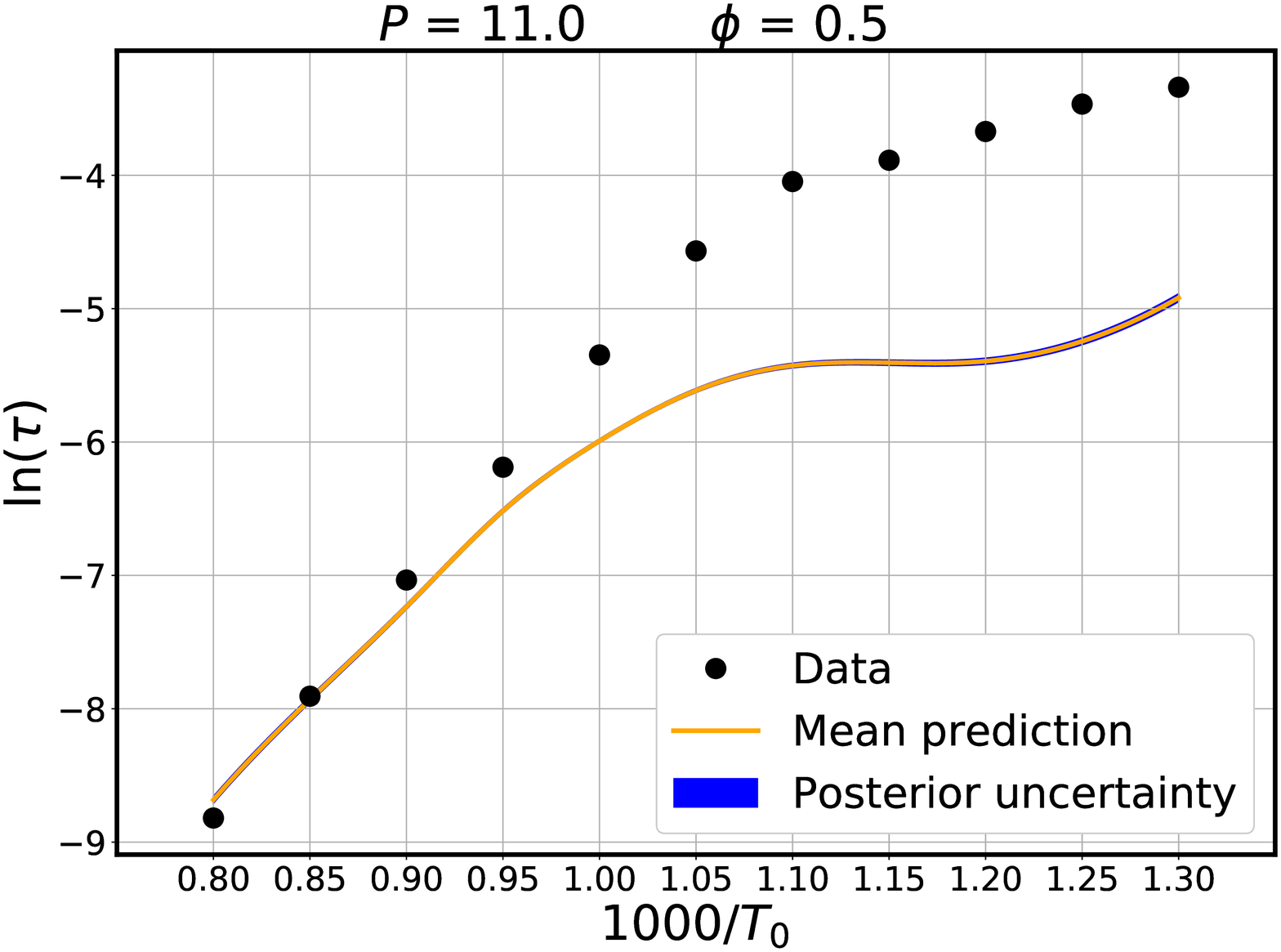}\hfill
  \includegraphics[width=0.45\textwidth]{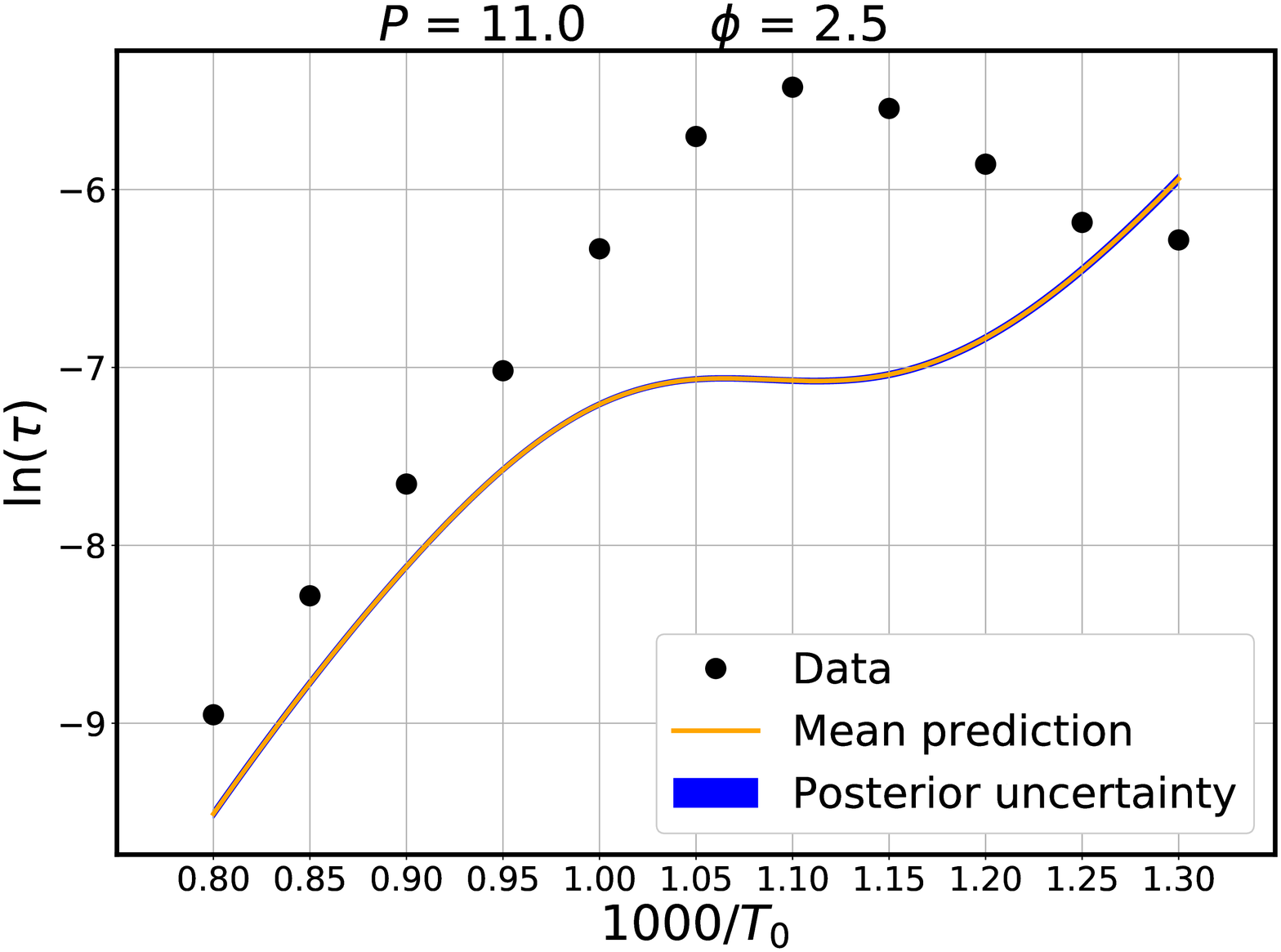}\\
  \includegraphics[width=0.45\textwidth]{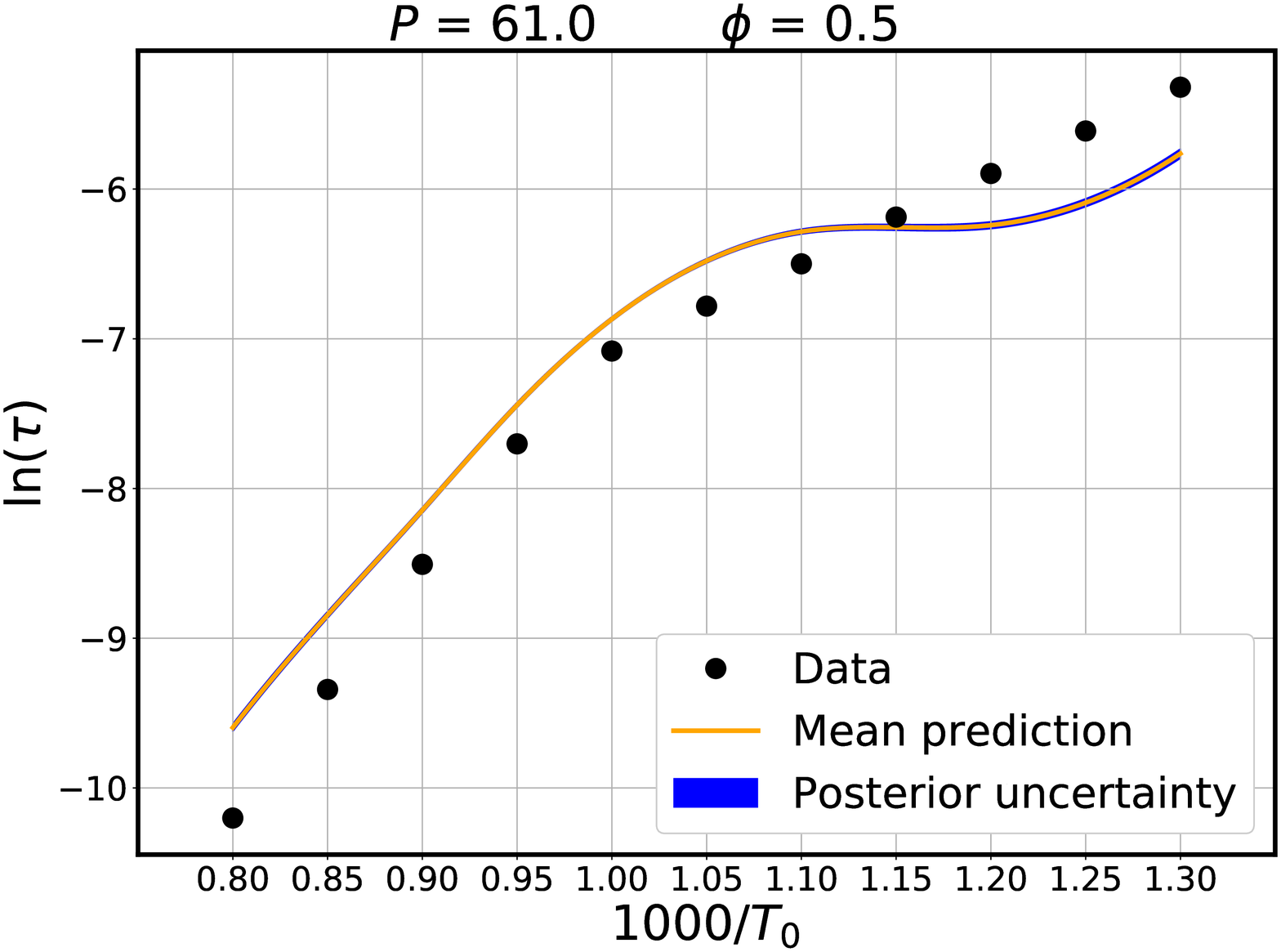}\hfill
  \includegraphics[width=0.45\textwidth]{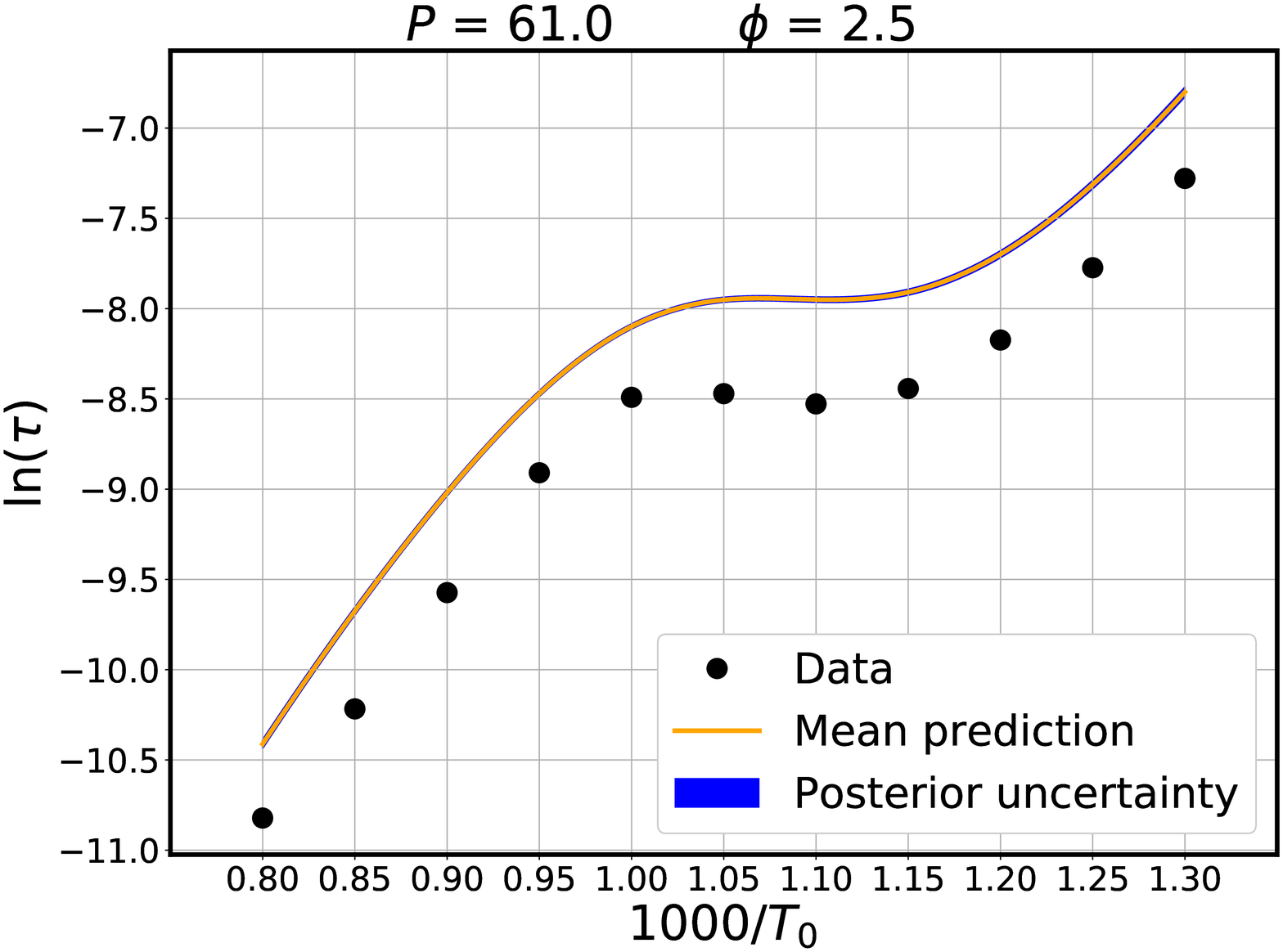}
  \caption{Demonstration of PF uncertainties of classical
    calibration. The output of interest, $\ln(\tau)$ is shown as a
    function of the inverse initial temperature, for $4$ `corner'
    cases of the two other operating conditions, the equivalence ratio
    $\phi$ and the pressure $P$. \reb{Note that the posterior
      uncertainty component is relatively small, and barely visible on the given scale.}}
  \label{fig:narafits_cl}
 \end{figure}

\begin{figure}[!h]
  \vspace{0pt}
   \includegraphics[width=0.45\textwidth]{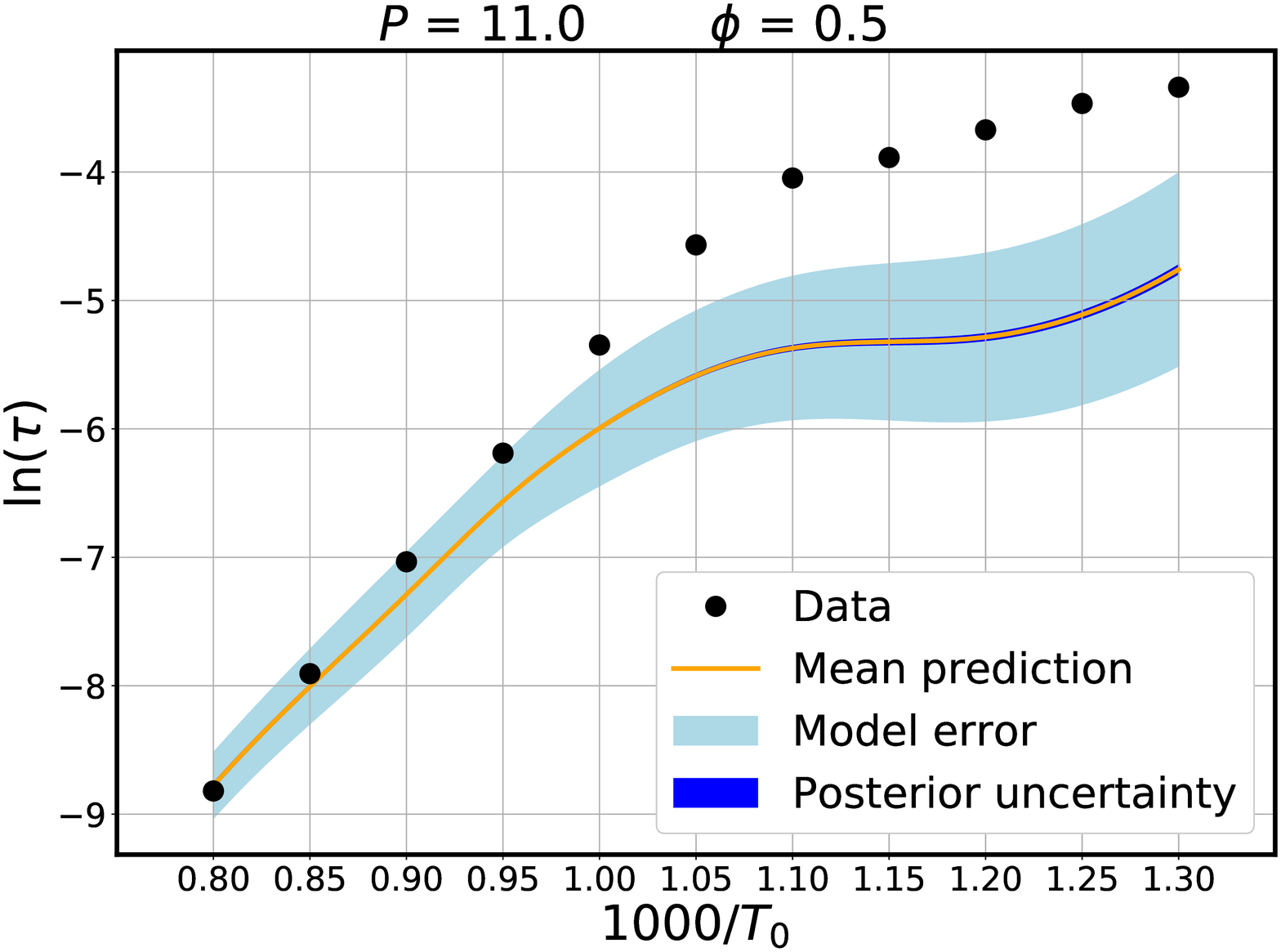}\hfill
  \includegraphics[width=0.45\textwidth]{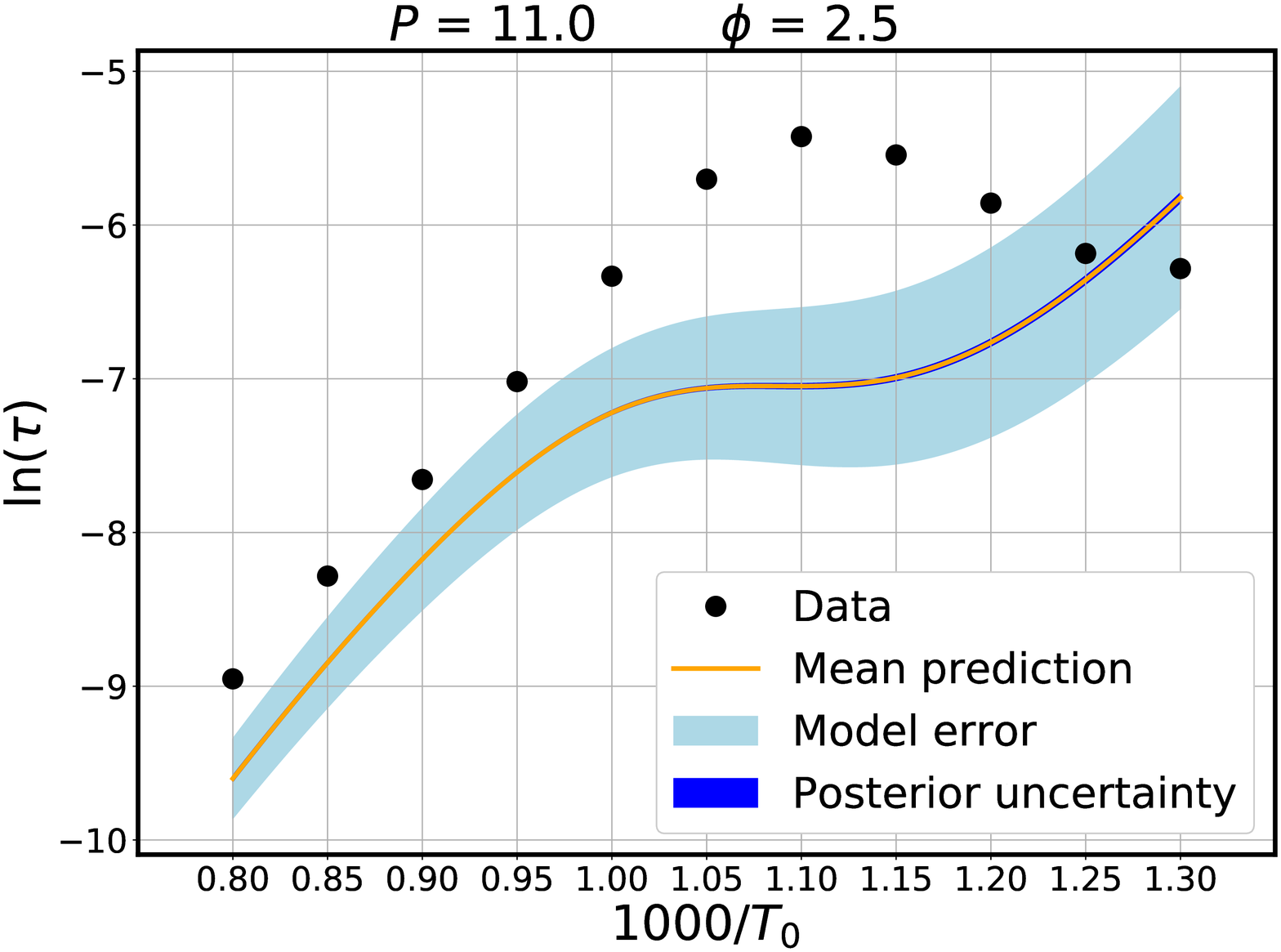}\\
  \includegraphics[width=0.45\textwidth]{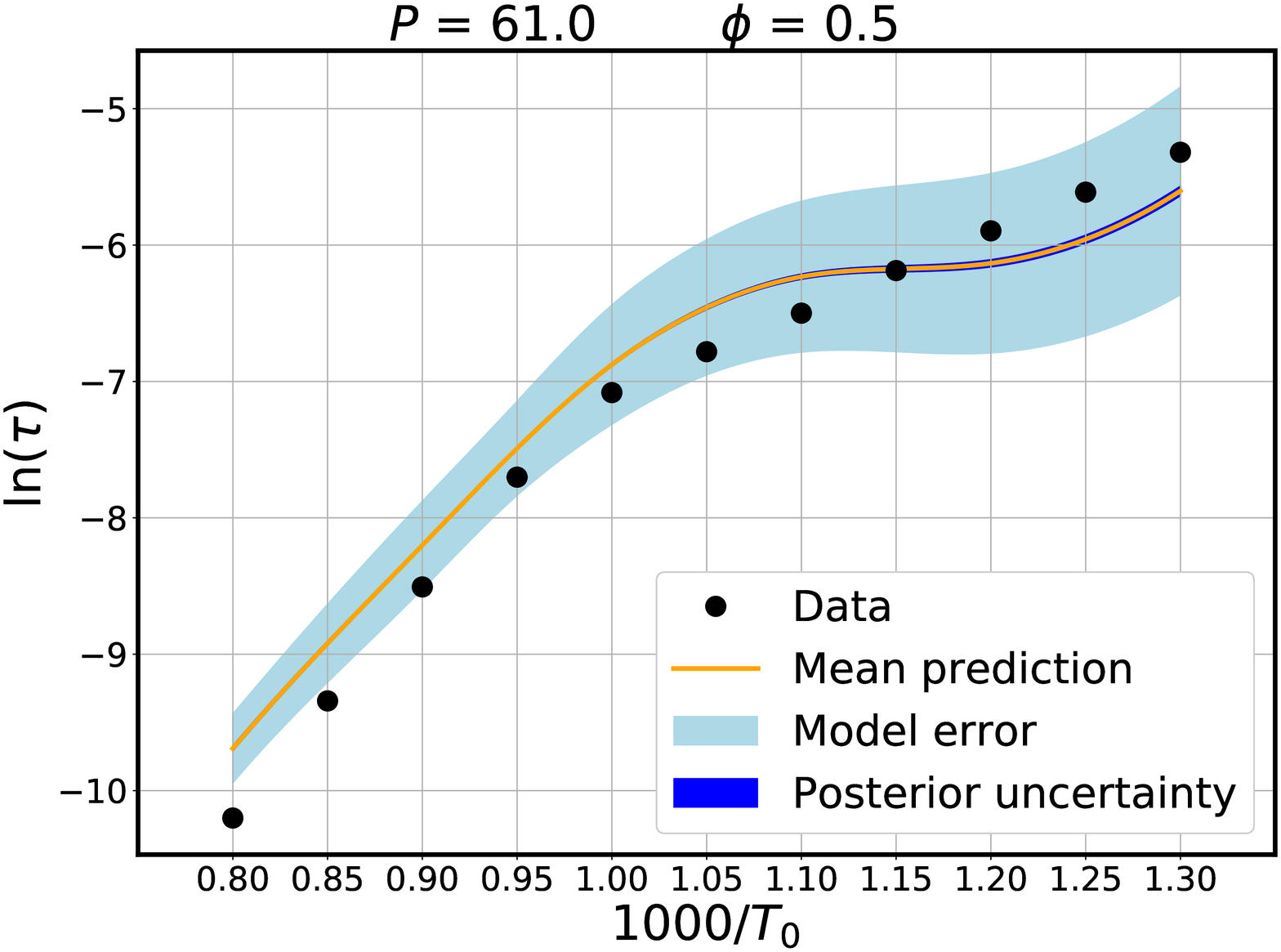}\hfill
  \includegraphics[width=0.45\textwidth]{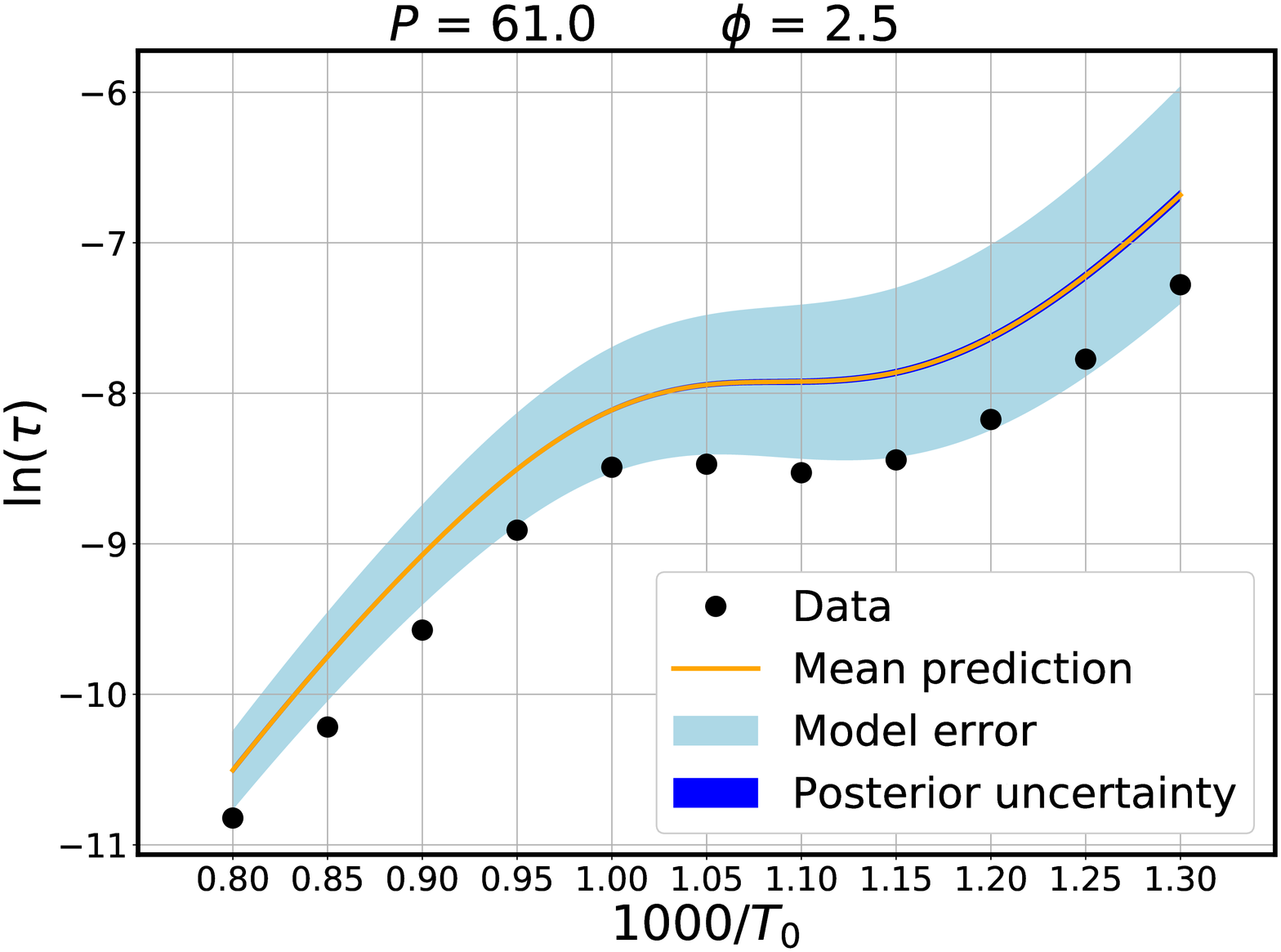}
  \caption{Demonstration of PF uncertainties of calibration with model
    error embedding. The output of interest, $\ln(\tau)$ is shown as a
    function of the inverse initial temperature, for $4$ `corner'
    cases of the two other operating conditions, the equivalence ratio
    $\phi$ and the pressure $P$. \reb{Note that the posterior
      uncertainty component is relatively small compared to model
      error, and barely visible on the given scale.}}
  \label{fig:narafits_gm}
 \end{figure}

Indeed, we have found that while the calibrated 2-step mechanism captures the detailed mechanism reasonably
well \emph{on average}, the PF uncertainty is by no means representative of the true \reb{magnitude of the} discrepancy. \reb{In contrast}, as Figure~\ref{fig:naracal} shows, the embedded model
error calibration (again, we employed MVN embedding~\eqref{eq:mvn} and
independent-normal likelihood~\eqref{eq:in}) leads to a predictive uncertainty component that fairly
describes the true discrepancy across all $1331$ data points. This figure shows the true discrepancy from complex-model data to the mean prediction of the two-step mechanism, with a  rrorbars behind it corresponding to the PF variance components according to the decomposition~\reb{\eqref{eq:pprd}}. Further, Figure~\ref{fig:Ea} illustrates the posterior joint PDFs for
$(E,a_1)$. While the classical calibration (on the left) leads to
an over-confident PDF, the
model-error embedded approach leads to posterior PDFs that are representative of actual
model discrepancies, as its PF predictions show in Figures~\ref{fig:narafits_cl} and~\ref{fig:narafits_gm}. These results demonstrate the model-to-model calibration for a small subset of the $1331$ operating conditions. Namely, they illustrate the ignition time dependence on initial temperature
for $4$ `corner' cases of pressure and equivalence ratio, $(11,0.5)$, $(11,2.5)$,
$(61,0.5)$ and $(61,2.5)$. Again, it is clear that the classical calibration (Figure~\ref{fig:narafits_cl})
underestimates the predictive uncertainties, while the model-error embedding (Figure~\ref{fig:narafits_gm}) allows predictive uncertainties to capture the discrepancy across a wide range
of conditions.

\section{Discussion} \label{sec:disc}

The framework presented in this work takes essential steps towards
representation and quantification of model errors.
Specifically, we present a principled approach for
representing model structural error by embedding it in the model itself.
The Bayesian machinery is employed for
estimation of embedded model error parameterization together with original model
parameters. \reb{The true likelihood, however, is degenerate or near-degenerate, leading to posteriors that are hard to sample from. In this work, we highlighted and employed moment-based likelihoods primarily. For such likelihoods, } polynomial chaos expansions with pre-built surrogates allow efficient
likelihood computation and uncertainty propagation. \reb{Specifically, we employed independent-normal approximation to the likelihood function and ABC-inspired likelihoods that compare the first two moments of the data and the embedded model. In principle, both are found to be viable options, and the choice should mostly be driven by the goals of the modeler (see, \eg \cite{Sargsyan:2015,Hakim:2017,Pernot:2017} for ABC, and~\cite{Huan:2018b,Zio:2018,Pernot:2017} for independent-normal). Different approaches do typically lead to different posterior distributions. This paper does not intend to provide comparison between the two -- interested readers are referred to~\cite{Pernot:2017} for an in-detail comparison between those and other approaches. The ABC likelihood includes a `tolerance' parameter $\epsilon$ which has a direct impact on posterior width, while independent-normal approximation does not require any additional hyperparameters. However, we have found that for small data sizes, independent normal likelihood may have stronger prior dependence and may suffer from identifiability more than ABC-based likelihoods do.}
The approach shows promising results in various contexts/scenarios
that are well-recognized to highlight fundamental challenges associated with
model error assessment. For example, model-to-model calibration is a special
case of the developed methodology. In such cases,  there is no observational
data, \ie $\epsilon_i=0$ in Eq.~(\ref{eq:koh}), one strives to calibrate a
lower-fidelity model $f(x;\vlam)$ with respect to \emph{simulation} data from a
higher-fidelity model that is effectively assumed to be the `truth'
$g(x)$. This scenario particularly highlights the deficiency of the conventional
calibration approaches that assume \emph{i.i.d} error between the high- and
low-fidelity models, while the embedded model error calibration allows a
principled way to construct model-to-model discrepancy terms that are directly
propagated through the physical model and satisfy physical constraints by definition.
This work extends this no-noise approach,
developed in~\cite{Sargsyan:2015}, to incorporate experimental/observational
data leading to full attribution of predictive uncertainties to various
components, including uncertainties associated with both structural errors and
data amount/quality.

Another advantage of the present approach is highlighted when the
predictive quantities are computed for \emph{another} QoI that relies on
$\vlam$ or a
subset thereof. Two different extrapolation modes for prediction are
implied, (a) extrapolating $f(x;\vlam)$ to design conditions $x$ that are outside
the range of $x_i$'s used for calibration, and (b) extrapolating to a different
observable or model, \eg $h(x';\vlam)$, potentially at qualitatively different
design conditions $x'$. The latter in particular highlights the strength of the
proposed approach compared to conventional statistical approaches of explicit
modeling of model discrepancy as an additive correction on a specific
observable~\cite{Kennedy:2001}. Such additive corrections do not have any
predictive meaning when applied to other observable QoIs, while the present
approach of embedding model error representations inside the models allows
predictions, with quantified uncertainties, for categorically different models
or model scenarios.  It is worth noting that extrapolation in \emph{any} sense
is a dangerous task -- and this work is not guaranteeing accurate
extrapolative predictions for general models of interest. Rather, by
``correcting" the model in physically consistent ways, the approach retains the
predictive strength of the physical model intact, whether for prediction of
other QoIs, or, to the extent that the physical model is ``valid" over a broad
set of operating conditions, for extrapolation. The approach lays a
foundation for making such predictions that are at least meaningful and informed
by the inferred model error magnitude.

As a future extension, one can tackle hierarchical models, as
hierarchical structure is quite common in physical models.
There are also potential opportunities for exploring connections between this
embedded model error approach and multi-fidelity model analysis.

While, generally throughout the paper, it is assumed that both the model
$f(x;\vlam)$ and the `truth' or high-fidelity model $g(x)$ are deterministic,
the developed machinery \reb{can be generalized} to stochastic models that
exhibit internal variability. Such internal variability will simply add an extra
uncertainty component to the pushed-forward process $F(x;\tvalpha)$ (if the
model-to-be-calibrated $f(x;\vlam)$ is stochastic) in Eq.~(\ref{eq:predvar}), or
to the posterior predictive random variable (if the high-fidelity model $g(x)$
is stochastic), effectively playing a role of observational error. In a PC
context, one can envision extra elements in the augmented
germ~(\ref{eq:fullgerm}) to capture internal stochasticity of $f(x;\vlam)$ or
$g(x)$.

Finally, it is also important to highlight some major challenges for the embedded model error
construction that are only partially addressed in this paper or relegated to
future work. For example, the dimensionality increase in the associated
calibration problem renders the MCMC sampling quite challenging, suggesting the
use of advanced MCMC sampling schemes that are well-tuned to sample in
data-informed lower-dimensional manifolds or nearly-degenerate posteriors. Also, the Bayesian
problem can often be prior-dominated, making appropriate prior selection a
crucial part of the methodology. Further, in general, identifiability is an
issue for inferring both model structural error parameterization and physical
parameters at the same time, particularly in the limit of low amount of
high-fidelity or observational data~\cite{Arendt:2012}. \reb{Having said that, in this work, a part of the output variance comes from propagating embedded uncertainties
    through the model, with its attendant nonlinearities, making the
    embedded model error approach less prone to identifiability challenges associated with explicit model output corrections.}

The calibration with model error embedding, including a variety of likelihood,
prior, and embedding form options, is implemented as a part of UQTk
v3.0~\cite{uqtk:web,Debusschere:2016}, a lightweight C++/Python library for a
range of basic UQ tasks developed at Sandia National Laboratories in Livermore,
CA.

\section*{Acknowledgments}
The authors would like to thank Jason Bender, Chi Feng, Youssef Marzouk and Cosmin Safta for comments and suggestions throughout this work.
The authors acknowledge the support by the Defense Advanced Research Projects Agency (DARPA)
program on Enabling Quantification of Uncertainty in Physical Systems (EQUiPS).
Also, KS and HNN acknowledge support for this work through the
Scientific Discovery through Advanced Computing (SciDAC) program funded by the U.S. Department of Energy (DOE),
Office of Science, Advanced Scientific Computing Research (ASCR).
HNN also acknowledges support for this work through the DOE, Office of Basic Energy Sciences
(BES), Division of Chemical Sciences, Geosciences, and Biosciences.
Sandia National Laboratories is a multimission laboratory managed and operated by
National Technology and Engineering Solutions of Sandia, LLC.,
a wholly owned subsidiary of Honeywell International, Inc.,
for the U.S. Department of Energy's National Nuclear Security Administration under contract DE-NA-0003525.

\begin{appendices}
\section{Attribution of predictive uncertainties} \label{sec:app}
\end{appendices}

Eq.~(\ref{eq:datamodelpc}) can be viewed as a $(d+N)$-dimensional PC expansion
with respect to the augmented PC germ $\hat{\vxi}=(\xi_1,\dots,\xi_d, \xi_{d+1},\dots,\xi_{d+N})$ in which the first
$d$ stochastic dimensions correspond to the embedded model error, and the last $N$ correspond to the error
associated with data collection.

One can proceed further with the PC paradigm and represent the posterior random vector $\tvalpha$ as a PC
expansion of dimension $d_\tvalpha$, the number of entries in $\tvalpha$,
\be \label{eq:pcalfa}
\tilde{\alpha}_j\simeq \sum_{k=0}^{K_\tvalpha-1} a_{jk} \Psi_k(\xi_{d+N+1},\dots,\xi_{d+N+j}), \qquad \textrm{ for }
j=1,\dots,d_\tvalpha.
\ee
This can be accomplished via the Rosenblatt transformation, using posterior samples obtained from MCMC. As
before, one will need to truncate at an appropriate PC order $p_\tvalpha$, leading to a number of terms $K_
\tvalpha=(d_\tvalpha+p_\tvalpha)!/(d_\tvalpha!p_\tvalpha!)$. Then, similar to the PC propagation of $\vLam$ through
the model $f(x;\vLam)$, one can propagate the PCE of $\tvalpha$ in Eq.~(\ref{eq:pcalfa}), through each of the
coefficients $f_k(x_i;\tvalpha)$ to obtain
\be
f_k(x_i;\tvalpha)\simeq \sum_{m=0}^{\tilde{K}-1} f_{kim}\Psi_m(\xi_{d+N+1},\dots,\xi_{d+N+d_\valpha})
\ee
for some truncation number $\tilde{K}$. Plugging this into the data model~(\ref{eq:datamodelpc}), one obtains the
fully expanded PC representation
\be \label{eq:fullpc}
y_i\approx \sum_{k=0}^{K-1} \sum_{m=0}^{\tilde{K}-1} f_{kim}\Psi_m(\xi_{d+N+1},\dots,\xi_{d+N+d_\valpha})
\Psi_k(\vxi)+\sigma \xi_{d+i} \qquad\textrm{ for } i=1,\dots, N.
\ee
This is a generic PC representation with respect to the fully augmented PC germ
\be \label{eq:fullgerm}
\hat{\hat{\vxi}}=(\underbrace{\xi_1,\dots,\xi_d}_{\textrm{Model error}},\underbrace{\xi_{d+1}, \dots,\xi_{d+N}}
_{\textrm{Measurement error}},\underbrace{\xi_{d+N+1},\dots,\xi_{d+N+d_\valpha}}_{\textrm{Posterior
uncertainty}}).
\ee
Even if the resulting PC expansion~(\ref{eq:fullpc}) has a very large stochastic dimensionality of $d_{full}=d+N+d_
\valpha$, one can employ generic PC tools for variance attribution -- without additional cost -- in order to attribute
the overall uncertainty of the data model into specific components/dimensionalities, such as model error,
measurement error and posterior uncertainty. Namely, the right-hand side of Eq.~(\ref{eq:fullpc}) can be written in a
general PC form
\be \label{eq:fullpcgen}
y_i\approx \sum_{\vp \in \SSS} f_\vp \Psi_\vp(\hat{\hat{\vxi}}),
\ee
where $\SSS$ denotes the set of multiindices, \ie dimension-wise orders, $\vp=(p_1,\dots,p_{d_{full}})$, induced
by the specific form of the expansion~(\ref{eq:fullpc}). Now the total variance of this expansion $V_{total}=
\sum_{\vp\in\SSS} f^2_\vp ||\Psi_\vp||^2$ can be attributed to fractional contributions corresponding to specific
subsets of the germ, via Sobol sensitivity indices~\cite{Sobol:2003, Saltelli:2004}. For example, the main Sobol
sensitivity index corresponding to a subset of dimensions $\hat{\hat{\vxi}}\_\subset \hat{\hat{\vxi}}$ is
\be \label{eq:fullpcvar}
S_{\hat{\hat{\vxi}}\_}=\frac{1}{V_{total}} \sum_{\vp\in\SSS\_} f^2_\vp ||\Psi_\vp||^2,
\ee
where $\SSS\_\subset\SSS$ is the subset of multiindices that involve \emph{only} dimensions in $\hat{\hat{\vxi}}\_$. When specific subsets of the full germ, highlighted in Eq.~(\ref{eq:fullgerm}), are taken, one recovers a
decomposition that is essentially the PC-based counterpart of the general variance
decomposition~(\ref{eq:predvar}). Such uncertainty attribution decomposes the overall uncertainty and allows informed
decision-making in selecting the submodels/parameterizations that contribute most to it.

\bibliography{local}
\bibliographystyle{abbrv}

\end{document}